\documentclass[fleqn,usenatbib]{mnras}

\usepackage[T1]{fontenc}
\usepackage{ae,aecompl}

\usepackage{esvect}
\usepackage{natbib}
\usepackage{efbox,graphicx}
\efboxsetup{linecolor=red,linewidth=0.5pt}
\usepackage{url}
\usepackage{epstopdf}
\usepackage{color}
\usepackage{textcomp}
\usepackage{gensymb}
\usepackage{multirow}
\usepackage{ragged2e}
\usepackage{hyperref}
\usepackage{url}
\usepackage{amsmath} 
\usepackage{booktabs}
\usepackage{comment}
\usepackage{afterpage}
\usepackage{mathtools}
\usepackage{tabularx}
\usepackage{xspace}
\usepackage{relsize}
\usepackage[utf8]{inputenc}
\usepackage{newunicodechar}
\usepackage[most]{tcolorbox}
\usepackage{hyperref}

\newcommand{\msun}{\rm M_{\rm \odot}}

\newcommand\rhigh{$R_{\rm h}$\xspace}

\title[Circular Polarimetric Images of MADs]{Unraveling circular polarimetric images of magnetically arrested
  accretion flows near event horizon of a black hole}

\author[M. Mo\'scibrodzka et al.]{
M. Mo\'scibrodzka,$^{1}$\thanks{E-mail: m.moscibrodzka@astro.ru.nl}
A. Janiuk,$^{2}$
and M. De Laurentis$^{3,4}$
\\
$^{1}$Department of Astrophysics/IMAPP, Radboud University, P.O. Box 9010, 6500 GL
Nijmegen, The Netherlands\\
$^{2}$Center for Theoretical Physics PAS, Al. Lotnikow 32/46, 02-668 Warsaw, Poland\\
$^{3}$Dipartimento di Fisica ``E. Pancini'', Universit\'a di Napoli `Federico II'',
Via Cinthia, I-80126, Napoli, Italy\\
$^{4}$ INFN Sezione  di Napoli, Via Cinthia, I-80126, Napoli, Italy
}

\date{Accepted September 24 2021. Received February 26 2021; in original form
  February 26 2021}

\pubyear{2021}

\begin{document}
\label{firstpage}
\pagerange{\pageref{firstpage}--\pageref{lastpage}}
\maketitle

\begin{abstract}
Magnetically arrested accretion flows are thought to fuel some of the supermassive black holes and to power their relativistic jets. We calculate and study a time sequence of linear and
circular polarimetric images of numerical, high resolution and long duration simulations of magnetically dominated flows to investigate observational signatures of strong magnetic fields near the event horizon of a non-rotating black hole. We find that the magnitude of resolved linear and circular polarizations is rather sensitive to the assumption of the coupling of electron and ions in the accretion flow. Models with cooler electrons have higher Faraday rotation and conversion depths which results in scrambled linear polarization and enhanced circular polarization. In those high Faraday thickness cases the circular polarization is particularly sensitive to dynamics of toroidal-radial magnetic fields in the accretion flows. The models with high Faraday thickness are characterized by nearly constant handedness of circular polarization, consistent with observations of some accreting black holes. We also find that the
emission region produced by light which is lensed around the black hole shows inversion of circular polarization handedness with respect to the handedness of the circular polarization of the entire emission region. Such polarity inversions are unique to near horizon emission. \end{abstract}


\section{Introduction}

Low Luminosity Active Galactic Nulcei  (LLAGN) are believed to be powered by
magnetized radiatively inefficient accretion flows (MRIAFs) onto a supermassive black hole (e.g., \citealt{yuan:2014}).
One of the main uncertainties in these type of accretion flows is their magnetic field strength, geometry and dynamics which span from turbulent, organized to reconnecting (e.g., \citealt{beckwith:2008,ressler:2020}). Magnetic fields determine the accretion flow dynamics and likely play a leading role in the jet
formation process (\citealt{BZ:1977,BP:1982}) hence understanding their observational signatures close to the central black hole is a key to find connections between jet, accretion flow, and the black hole itself.

The geometry of magnetic fields around black holes is best studied in terms of linear and circular polarization of the emitted synchrotron light, the primary cooling process of MRIAFs. However, extracting information about the underlying field geometry from light polarization is not straightforward.
Firstly, the polarization signal from near black hole event horizon zone is distorted by gravitational lensing (\citealt{connors:1980,ishihara:1988,himwich:2020}) and Doppler boosting effects.
Secondly, the polarization signal can be affected by plasma effects changing polarization when light propagates through it. In \citet{moscibrodzka:2017} (see also \citealt{rosales:2018}) we have shown the Faraday effects present
internally in
MRIAFs are able to severely modify the linear
polarization of synchrotron light. While these internal Faraday effects strongly depend on the assumed model for plasma and field configuration in the LLAGN core, linear polarization can be also subject to additional modifications by Faraday screens which are physically displaced from the LLAGN core (in fact such external Faraday screen effect is often used to infer mass accretion rate towards LLAGNs; see e.g., \citealt{marrone:2007,kuo:2014}).
While linear polarization can be sensitive to the details of magnetic field and plasma configuration on large span of distances from the central black hole, the circular polarization is sensitive to near-horizon processes only. Although usually weaker (as shown in this work), circular polarization of synchrotron light may become an important probe of the near horizon plasma and magnetic field configuration and dynamics. Hence, the focus of this work is mostly on circular polarization.

The circular polarization is detected in radio observations of many AGN jets
(e.g., \citealt{wardle:1998}, \citealt{homan:2006}). It's been reported that
circular polarization magnitude in those brighter sources might correlate with
their spectral indices \citep{rayner:2000}. The circular polarization is also
found in couple of LLAGN sources (the notable examples are: Sgr~A*,
\citealt{bower:1999}, \citealt{munoz:2012}, M~81*, \citealt{brunthaler:2006},
\citealt{bower:2002} and recently also in M87,
\citealt{goddi:2021}). Interestingly, the handedness of circular polarization
in many of these accreting supermassive black holes
remains constant over the years and even decades (\citealt{homan_wardle:1999},
\citealt{bower:2002}, \citealt{bower:2018}). It has been proposed that such
long-term stability of circular polarization sign could be a results of: i)
the presence of stable global magnetic field component in the accretion flow
system (e.g., \citealt{beckert:2002});
ii) direction of rotation in an accreting system (e.g. \citealt{enslin:2003}).

\begin{figure}
\centering
\includegraphics[width=0.45\textwidth]{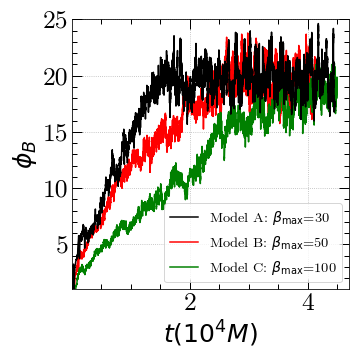}
\caption{Normalized magnetic field flux, $\phi_B=\Phi_B/\sqrt{\dot{M}}$, at the horizon as a function of time in three realizations of MAD model.}\label{fig:mdot}
\end{figure}

In the present article we analyse polarimetric properties of realistic MRIAFs
models obtained by combining high resolution, large domain, three dimensional
general relativistic magnetohydrodynamics (3-D GRMHD) simulations with fully polarized
radiative transfer model. Our 3-D GRMHD runs are evolved for relatively long times (up to about $t_f\sim$50000M) which enables us to study long term stability of the magnetic fields and associated polarization of the models. 
Our goal is to find the physical origin of the polarization signal and its characteristics. We introduce a diagnostic method which tracks, along the light paths, the components of the radiative transfer equation. The diagnostics allows us to identify the location and processes responsible for the observed model linear and circular polarization and, hence, helps to associate the emission polarization with, e.g., the underlying magnetic field geometry.
The models chosen for the study are those that are magnetically arrested
(hereafter MAD, e.g., \citealt{pb:2003,narayan:2003,mckinney:2012}). We
study three slightly different numerical realizations of MADs.
Mock observations are created at observer's frequency where the emission from black hole event horizon scale, where jet base and disk are connected, is visible. Our theoretical study is limited to near face-on observations corresponding to the viewing angle of LLAGN in M87 galaxy. M87 galaxy core is currently one of the most promising astrophysical sources where the details of the black hole-jet-disk connection theories can be quantitatively tested \citep{broderick:2009,dexter:2012,moscibrodzka:2016,moscibrodzka:2017,chael:2019,ehtIV:2019,ehtV:2019}. Recently, near-horizon linear polarimetric images of M87 have been published (\citealt{paper7:2021}, \citealt{paper8:2021}). The new images allow us to put much tighter constraints on our model free parameters.  

The article is structured as follows. In Section~\ref{sec:model} we present our methodology for modeling and investigating polarimetric signal of numerical models of MRIAFs. In Section~\ref{sec:results} we
present polarimetric maps of MAD flows and investigate the origin of the light linear and circular polarizations. In Section~\ref{sec:summary} we briefly discuss the new results and conclude. 

\section{Methodology}\label{sec:model}

\subsection{Magnetically Arrested Accretion Flow Models}

We use the general relativistic magneto-hydrodynamic code, \texttt{HARM}
\citep{Gammie_2003, Noble_et_all_2006, Sap2019ApJ}, with a fixed background
Kerr metric, i.e. neglecting the effects of self gravity and the BH spin
changes (cf. \citealt{Janiuk2018}).  The numerical scheme is based on GRMHD
equations where the energy-momentum tensor is contributed by the gas and
electromagnetic fields.  The code solves the fluid evolution based on
continuity and momentum conservation equations. 

\begin{figure*}
\centering
\includegraphics[width=0.45\textwidth,clip]{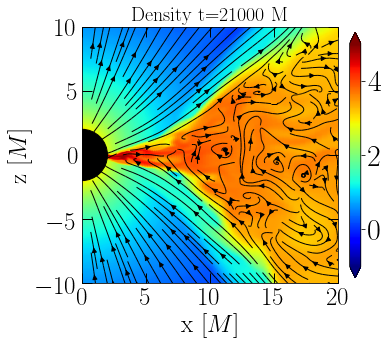}
\includegraphics[width=0.45\textwidth,clip]{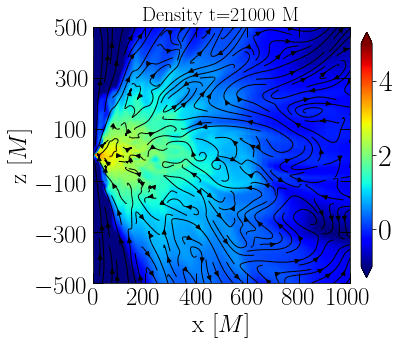}\\
\includegraphics[width=0.45\textwidth,clip]{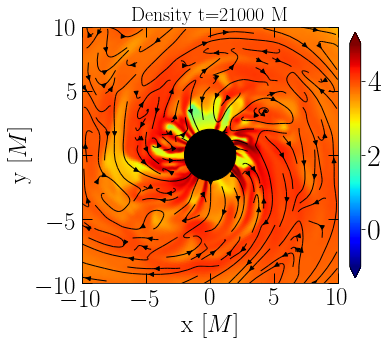}
\includegraphics[width=0.45\textwidth,clip]{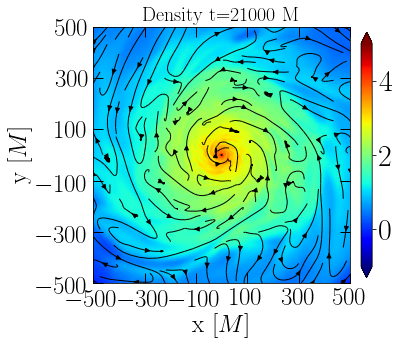}
\caption{Maps of the density (in arbitrary units), in the  polar plane XZ (top row) and in the equatorial plane XY (bottom row) of model A. Left and right panels show the same model on different scales. The color maps are in logarithmic scale and are taken at times denoted in the plots ($t \sim 21000 M$). Black contours follow magnetic field lines projected onto x-y and x-z Cartesian coordinates.
}\label{fig:models_beta30}
\end{figure*}

The \texttt{HARM} code performs the integration in Modified Kerr-Schild coordinates:
$\left(t,x^{(1)},x^{(2)},\phi\right)$. The assumed grid is exponential in radius,
while in the polar direction the cells are concentrated more on the mid-plane.
Moreover, cylindrified coordinates are used inside the radius $R_{\rm
  cyl}=3.5$ and inside the angle $\theta_{\rm cyl}=-1+1/N2$, to resolve better
the regions closest to the black hole horizon and the polar axis.  The inner
and outer radius are set to $R_{\rm in} = 0.87(1 + \sqrt{1 - a^{2}})$,
and $R_{\rm out} = 10^{5}$M, where $M\equiv GM/c^2$ is a length unit. The progressively sparser grid starts at $R_{\rm
  br}=400$, with $x^{(1)}_{i}=x^{(1)}_{i-1}+0.5 \times \delta x$, where
$\delta x = \log(R_{out} -2)^{1/4} + \log(R_{\rm br})$. Our grid resolution ($N_1,N_2,N_3$) in the ($r,\theta,\phi$) directions depends on the model (see paragraphs below). 

The initial configuration of the flow is based on the analytic solution of a
constant specific angular momentum, in a steady state configuration of a
pressure-supported ideal fluid in the Kerr black hole potential
\citep{Fishbone_Moncrief_1976_ApJ}.  
The initial torus embeds a poloidal magnetic field, prescribed with the vector potential of 
\begin{equation}
A_{\varphi}=({\bar{\rho} \over \rho_{\rm max}} - \rho_{0})  r^{5}
\end{equation}
where $\bar{\rho}$ is the density in the torus averaged over 2 neighboring
cells, and $\rho_{\rm max}$ is the density maximum, and we use offset of
$\rho_{0}=0.2$.  The factor of $r^{5}$ ensures that higher magnetic flux will
be brought onto the black hole horizon from larger distances, as the evolution
proceeds.  Other spatial components of the initial vector potential are zero,
$A_{r}=A_{\theta}=0$, hence the initial magnetic field vector will have only
$B_{r}$ and $B_{\theta}$ components.  The plasma $\beta-$parameter is defined
as the ratio of the fluid's thermal to the magnetic pressure, $\beta \equiv
p_{\rm g} / p_{\rm mag}$.  We normalize the magnetic field in the torus to
have $\beta_{\rm max}=(\gamma - 1) u_{max}/( b^{2}_{max}/2) $, where $u_{max}$ is the
internal energy at the torus pressure maximum radius.  In our models, the
adiabatic index of $\gamma=4/3$ is used.

The magnetically arrested flows are supposed to accrete onto black hole when
overpassing the magnetic barriers. Therefore, in order to introduce a
non-axisymmetric perturbation and allow the generation of azimuthal modes, we
impose an initial perturbation of internal energy. It is given by $u = u_{0}
(0.95+0.1C)$, where $C$ is a random number generated in the range $(0,1)$. This
perturbation on the order of less than 5\% is typically used in other 3-D
simulations of weakly magnetized flows (see \citealt{Mizuta2018}).

We carry out three high resolution simulations of MAD accretion flow onto
Schwarzschild black hole. Starting conditions differ by $\beta_{\rm max}$
parameter. Model A, B, and C starts with $\beta_{\rm max}=30,50,100$,
respectively. The resolutions ($N_1,N_2,N_3$) are (288,256,256),
(576,512,256), and (576,256,256) in model A, B, and C, respectively. The GRMHD
simulations A/C are evolved for $\sim45000$M and simulation B for $41000$M,
where $M \equiv GM/c^3$ is a time unit. In Figure~\ref{fig:mdot} we show time
evolution of the magnetic field flux through the horizon in our three
realizations of MAD model.
Note that our initial normalisation
of magnetic fields affected the time at which the MAD state is reached. For
smallest $\beta_{\rm max}$, in model A, the normalized magnetic flux on the
black hole horizon as time $t=21000$ M was equal to $\sim20.9$, while for
models B and C it was about $\sim16.0$ and $\sim10.7$, respectively. At the
end of these simulations, already all models resulted in $\phi_{B} \approx 16-21$,
which we consider as the magnetically arrested mode (largest instantaneous
value was reached in model B).
On the other hand, the accretion rate at the
black hole, measured at time $t=21000$ M, was largest for model C, which is
least magnetized. The time-averaged accretion rate, though, was maximal for
model A, and it was 0.981 (in code units), while the time averaged accretion
rates of models B and C were equal to 0.843 and 0.587, respectively.
In model A, the MAD state is reached at $t \approx 15000$M at which the magnetic field flux at the horizon saturates. In
Figure~\ref{fig:models_beta30} we show model A density maps overplotted with
magnetic field lines contours at $t=21000$M. Within inner-most stable orbit
the accretion proceeds in form of thin streams (fingers) of nearly radially
falling matter that is able to overpass the pressure of the magnetosphere.

\subsection{Radiative Transfer and Electron Heating Models}

\begin{figure*}
\def\arraystretch{0.0}
\centering
\setlength{\tabcolsep}{0pt}
\begin{tabular}{ccccc}
\rhigh=1 & \rhigh=10 & \rhigh=100 & \rhigh=200 &\\
 \includegraphics[width=0.23\linewidth]{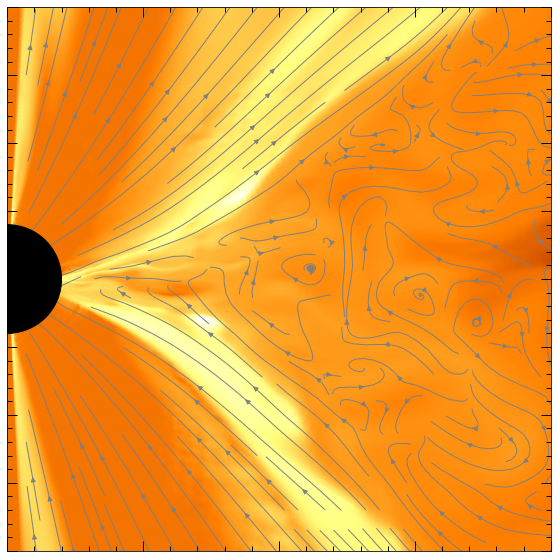}
&\includegraphics[width=0.23\linewidth]{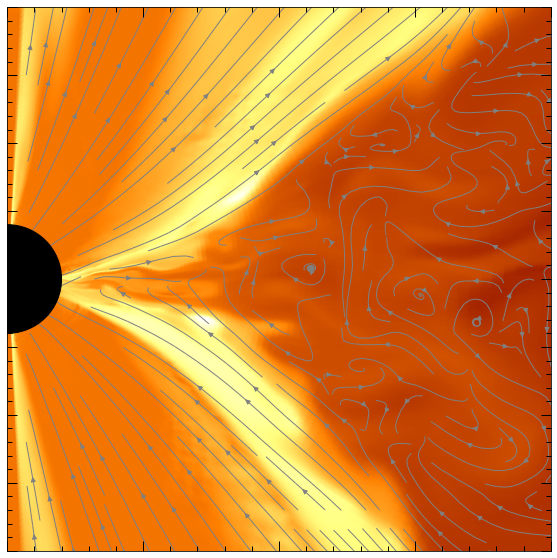}
& \includegraphics[width=0.23\linewidth]{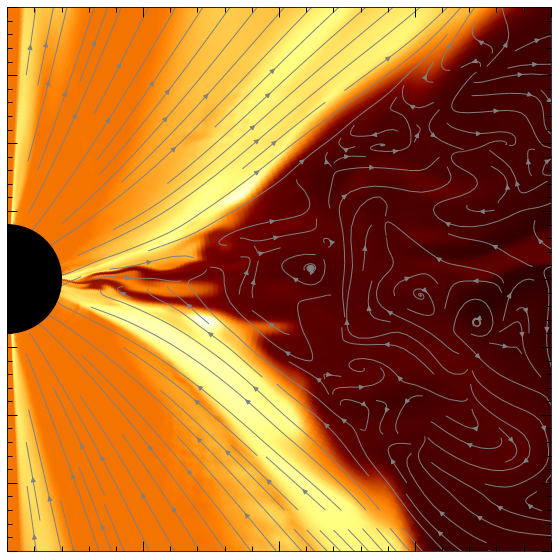}
& \includegraphics[width=0.23\linewidth]{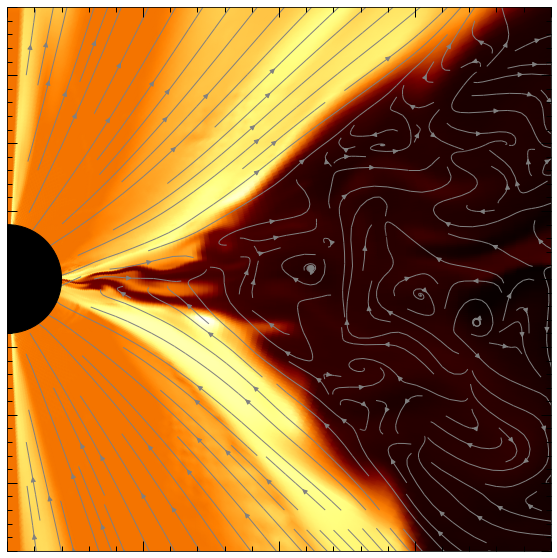}\\
\includegraphics[width=0.2\linewidth,trim=0 28cm 0 0,clip]{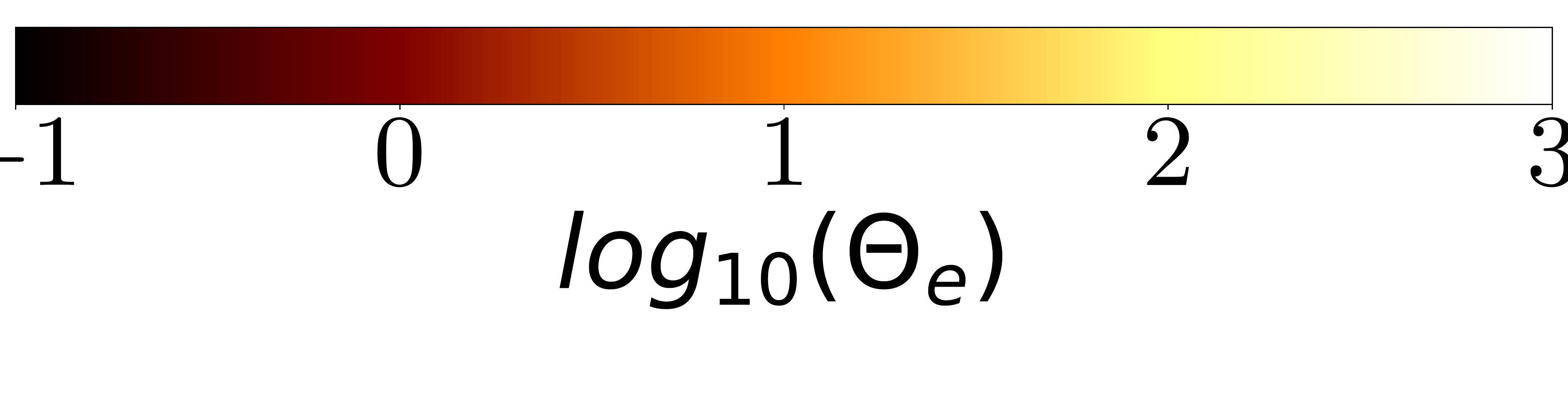} &
\includegraphics[width=0.2\linewidth,trim=0 28cm 0 0,clip]{plots/Model1a_beta30_a0/cbar_thetae.pdf} &
\includegraphics[width=0.2\linewidth,trim=0 28cm 0 0,clip]{plots/Model1a_beta30_a0/cbar_thetae.pdf} &
\includegraphics[width=0.2\linewidth,trim=0 28cm 0 0,clip]{plots/Model1a_beta30_a0/cbar_thetae.pdf} 
\end{tabular}
\caption{Impact of \rhigh parameter on electron temperature in MAD simulation
  A. Model snapshot at $t=21150$M within 20x20M is shown and the black hole
  is marked with the black circle.
  The gray contours mark magnetic fields lines projected from coordinate basis
  (modified Kerr-Schield coordinates) to x-z plane in Cartesian coordinates.}\label{fig:electrons}
\end{figure*}

Polarimetric images of the GRMHD models are computed by post-processing the simulations fluid variables with the polarized relativistic radiative transfer code \texttt{ipole~\footnote{www.github.com/moscibrodzka/ipole}} \citep{moscibrodzka:2018}. 
We scale the GRMHD simulations to black hole in M87 galaxy center, which we
use here as an example of astrophysical source, and we calculate polarimetric
images at frequency $f$=230\,GHz to focus the analysis on emission properties
on the near horizon scales. All images are computed for two viewing angles
$i=20, 160\deg$ corresponding to the putative viewing angles of M87
core.

The GRMHD simulations themselves do not follow the
evolution of radiating electrons temperatures which are
needed to calculate synchrotron emission. For the considered plasma number densities
($n_e \approx 10^2-10^5 {\rm cm^{-3}}$) we can assume that electrons
can be thermally
de-coupled from ions and electron distribution function is controlled by
other collisionless plasma processes. Following the approach introduced by
\citet{moscibrodzka:2016}, the collective  plasma effects heating electrons are
described by parameter \rhigh which sets the ratio of ion to
electron temperatures in the high-$\beta$ plasma regions (where plasma $\beta \equiv p_{\rm g}/p_{\rm mag}$). The electron temperature is calculated from expression:
\begin{equation}
T_{\rm e} = T_{\rm i} \left(\frac{R_{\rm h}\beta^2}{1+\beta^2} + \frac{1}{1+\beta^2} \right)^{-1},
\end{equation}
where the ion temperature, $T_{\rm i}$, is calculated from simulation gas pressure assuming ideal gas law.
As shown in Figure~\ref{fig:electrons}, by increasing \rhigh the decoupling of electron and ions in the
weakly magnetized plasma (higher $\beta$ regions) is stronger and electrons may become
sub-relativistic in these regions. In these models the emission from the places where the magnetic fields are relatively stronger and well organized
is expected to be more prominent. In \citet{moscibrodzka:2017} we have investigated linear polarization of the same emission models for M87 core but using non-MAD simulations. Here we extend our studies to MAD models and circular polarization.

In our emission modeling we use so called fast-light
  approximation where we post-process each simulation dump
  file separately as if light had infinite speed. In
  Appendix~\ref{app:fastlight} we explicitly show that this approximation
  holds very well for simulations of unpolarized/polarized radiative transfer
  through an accreting torus.

\subsection{Polarization Diagnostics}~\label{sec:probe}

In relativistic plasma, the circular polarization can be produced in two ways:
via intrinsic emission\footnote{The magnitude of the fractional circular
  polarization is expected to be smaller than fractional linear
  polarization. For a fixed frequency and $\Theta_e>1$, the ratio of both
  polarization emissivities $j_V/j_Q \sim 2 \Theta_e^{-1}$; hence if $|m|\sim
  10\%$ and $\Theta_e \approx 100$, $|v| \sim 0.1\%$.} and via Faraday
conversion. In our simulations both effects are contributing to final circular
polarimetric maps and Faraday conversion becomes increasingly important with
increasing \rhigh parameter. Because of the Faraday conversion (of linear to
circular polarization) circular polarization has to be analyzed simultaneously
with linear polarization. Apart of total intensity ${\mathcal I}$ and
fractional circular polarization ($v\equiv V/I$), we also examine maps of
linear polarization angles ($EVPA\equiv 0.5 \arctan(Q+iU)$) and fractional
linear polarization ($|m|\equiv \sqrt{Q^2+U^2}/I$). We also
define image net linear and circular polarizations as
$|m|_{\rm net} \equiv \sqrt{ (\sum_{i,j} Q_{i,j})^2 + (\sum_{i,j} U_{i,j})^2}  / (\sum_{i,j} I_{i,j})$
and
$v_{\rm net} \equiv (\sum_{i,j} V_{i,j})/(\sum_{i,j} I_{i,j})$, where sums
are over all image pixels.

The origin of linear and circular polarization in the polarization maps can be studied using a simplified but more formal approach than quantitative and qualitative description of polarization maps. 
The presented approach is most useful for investigating time variability of the polarimetric images. 
In the fluid rest-frame, the equations for relativistic polarized radiative transfer (see e.g, Eq. 19 in \citealt{moscibrodzka:2018} and notice that our fully covariant polarized radiative transfer equations accounts for polarization vector parallel transport outside of the fluid frame) can be recast into a form in which they have simple geometrical interpretation.
In analogy to \citet{book:2004} (Chapter 5.6) the relativistic polarized radiative transfer equations can be written down separately for invariant total intensity $I$ ($I \equiv I_\nu/\nu^3$) and fractional polarization three-vector $\vv{p}\equiv(Q/I,U/I,V/I)$ as follows:
\begin{equation}\label{eq:delin1}
\frac{dI}{d\lambda}=- (\alpha_I + \vv{\alpha} \cdot \vv{p} - \epsilon_I ) I,
\end{equation}
\begin{equation}\label{eq:delin2}
\frac{d\vv{p}}{d\lambda} = - \vv{\alpha} + (\vv{\alpha} \cdot \vv{p})\vv{p} + \vv{\rho} \times \vv{p} + \vv{\epsilon} - \epsilon_I \vv{p}.
\end{equation}
Here $\vv{\alpha}=(\alpha_Q,\alpha_U,\alpha_V)$ vector components are the
invariant synchrotron absorptivities ($\alpha_{QUV}\equiv \nu
\alpha_{\nu,QUV}$), $\vv{\rho}=(\rho_Q,\rho_U,\rho_V)$ are the invariant
Faraday rotativities ($\rho_{QUV}\equiv \nu \rho_{\nu,QUV}$),
$\vv{\epsilon}=(j_Q/I,j_U/I,j_V/I)$ are normalized invariant synchrotron
polarized emissivities ($j_{QUV}\equiv j_{\nu,QUV}/\nu^2$), $\epsilon_I=j_I/I$
is a scalar quantity and $\lambda$ is the affine parameter. Equations are
written down in the rest-frame comoving with the
fluid. Equation~\ref{eq:delin1} means that the polarization of light \text{color}{impacts}
the absorption of total intensity via term $(\vv{\alpha}\cdot\vv{p})$.
Notice that Faraday effects modify total intensity only via polarization
vector $\vv{p}$.

Here we are mostly interested in 
Equation~\ref{eq:delin2}, which describes the motion of a polarization-representing point inside of the Poincar{\'e} sphere. 
Following \citet{book:2004}, the first term (i) $-\vv{\alpha}$ tends to align
$\vv{p}$ with $-\vv{\alpha}$, n.b. components of polarization vector
$\vv{p}$ can be negative or positive.
This has a consequence for Equation~\ref{eq:delin1}, $(\vv{\alpha} \cdot \vv{p})$ may
  act as negative absorptivity for Stokes ${\mathcal I}$.
With (constant) term (i) alone, the fractional polarization
grows with increasing $\lambda$. However other terms may prevent this growth.
The second term (ii) $(\vv{\alpha} \cdot \vv{p}) \vv{p}$ is so called saturation term which prevents $\vv{p}$ from crossing the Poincar{\'e} sphere. The importance of term (ii) increases as $\vv{p}$ grows and results in polarization saturation. 
The third term (iii) in Equation~\ref{eq:delin2} describes precession of $\vv{p}$ about $\vv{\rho}$ or, in other words, Faraday rotation and conversion. Because Faraday rotation and conversion remove alignment between $\vv{p}$ and -$\vv{\alpha}$, the presence of Faraday effects reduce polarization saturation and helps polarization to increase. The fourth and fifth terms in Equation~\ref{eq:delin2} act in the same way as terms (i) and (ii), $\vv{\epsilon}$ aligns $\vv{p}$ with itself and $-\epsilon_I \vv{p}$ prevents saturation of fractional polarization. 

Our full radiative transfer calculation are done using the numerical code
\texttt{ipole} which provides us with $\vv{p}, \vv{\alpha}$ and
$\vv{\epsilon}$ along geodesics paths. We then examine the magnitude of all
terms in the right-hand-side of Equation~\ref{eq:delin2} and plasma and
magnetic field properties along chosen light rays. By measuring the strength
of each term which controls polarization we can associate it with location in
the flow and specific components of the magnetic field or plasma
conditions. Such polarization diagnostics is simplified because the full
relativistic transfer in \texttt{ipole} includes parallel transport of
polarization vector, hence the procedure allows us for discussion of
fractional polarizations but not EVPA, which is also modified 
due to parallel transport of polarization vector in strong gravity and Lorentz transformations from coordinate frame to frame comoving with the plasma. Notice that we set plasma co-moving frame in a way so that one of the tetrad vectors is always aligned with magnetic field line in the plasma frame; hence in plasma frame $(j_U,\alpha_U,\rho_U)=0$ by definition. 

\begin{figure*}
\def\arraystretch{0.0}
\centering
\setlength{\tabcolsep}{0pt}
\begin{tabular}{ccccccc}
  \multicolumn{3}{c}{$i=20$ deg} & \multicolumn{3}{c}{$i=160$ deg}\\
   $\mathcal{I}$ & $\mathcal{I + LP}$ & $\mathcal{CP}$ & $\mathcal{I}$ &  $  \mathcal{I +LP}$ &  ${\mathcal CP}$ \\
\raisebox{0.07\linewidth}[0pt][0pt]{\rotatebox[origin=c]{90}{A1}}\phantom{.}
 \includegraphics[width=0.16\linewidth]{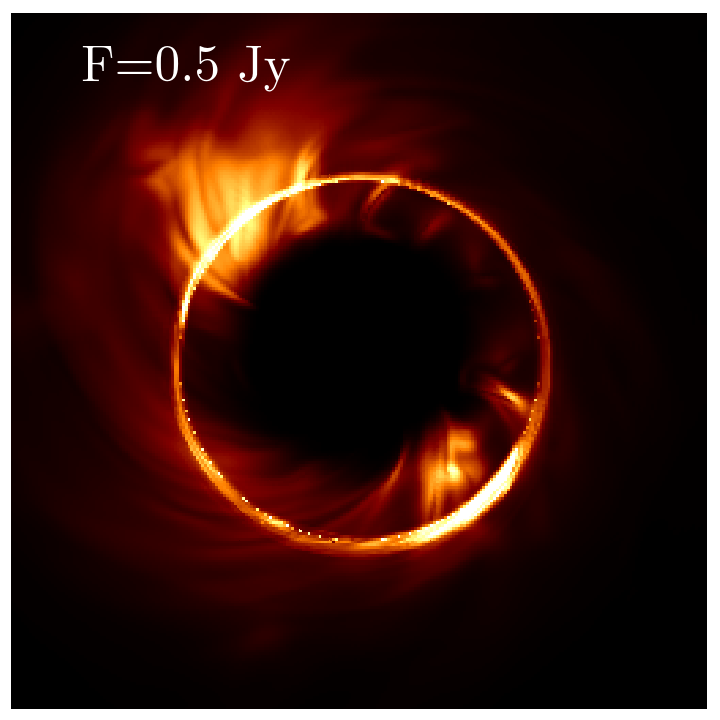} 
&\includegraphics[width=0.16\linewidth]{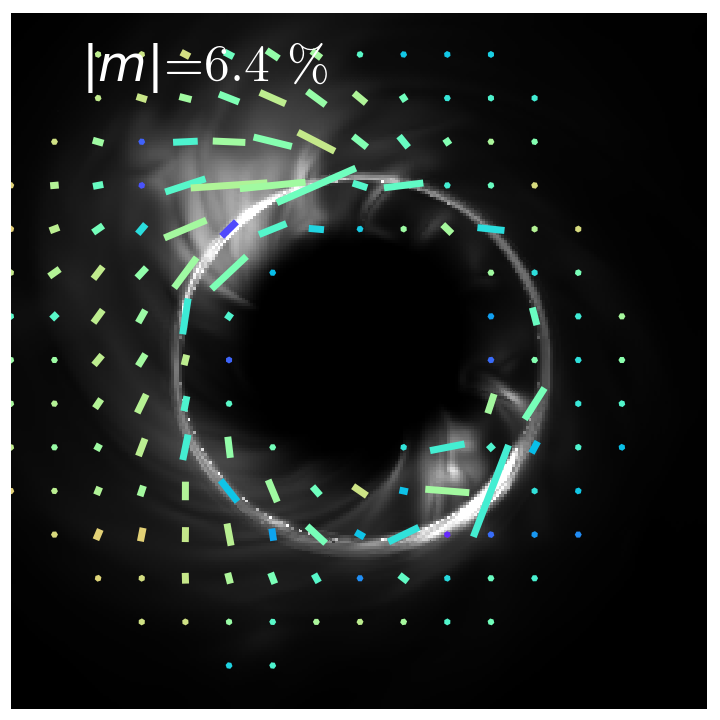}
& \includegraphics[width=0.16\linewidth]{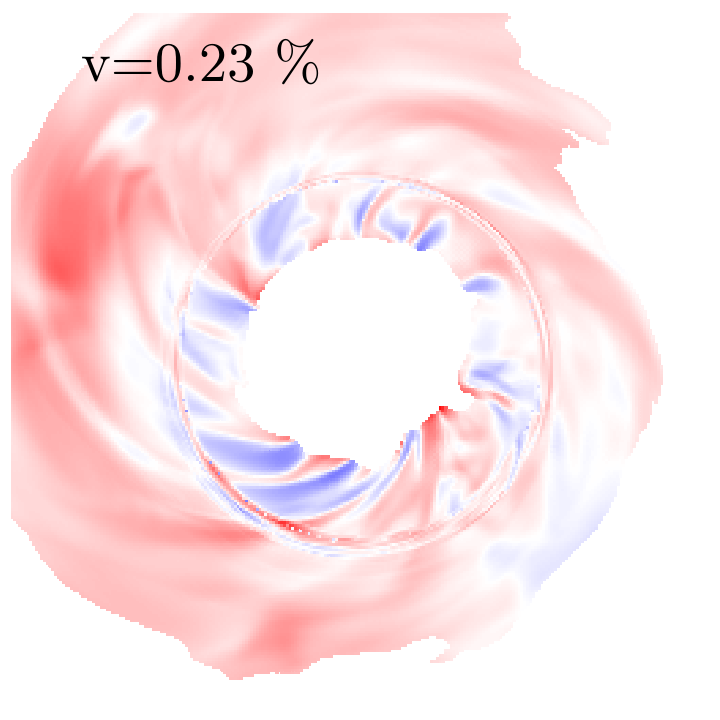}
& \includegraphics[width=0.16\linewidth]{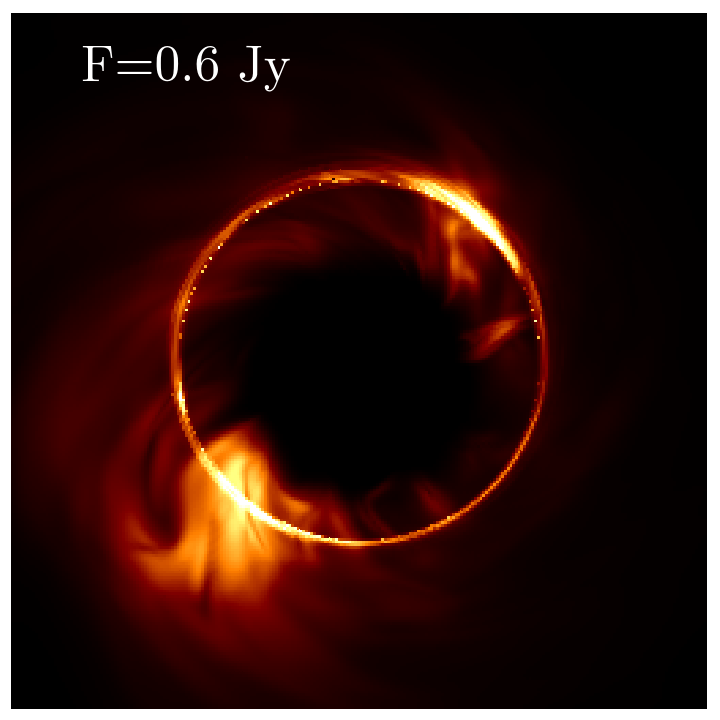}
& \includegraphics[width=0.16\linewidth]{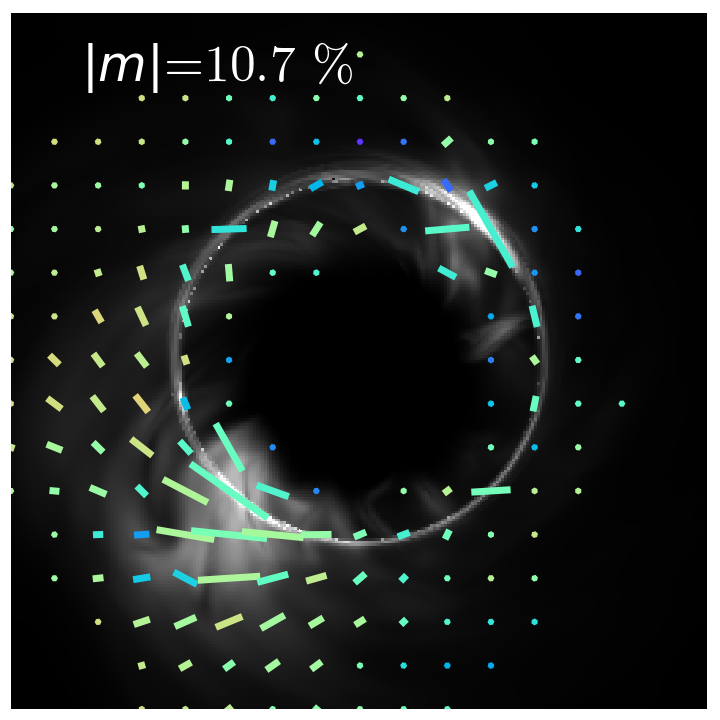}
& \includegraphics[width=0.16\linewidth]{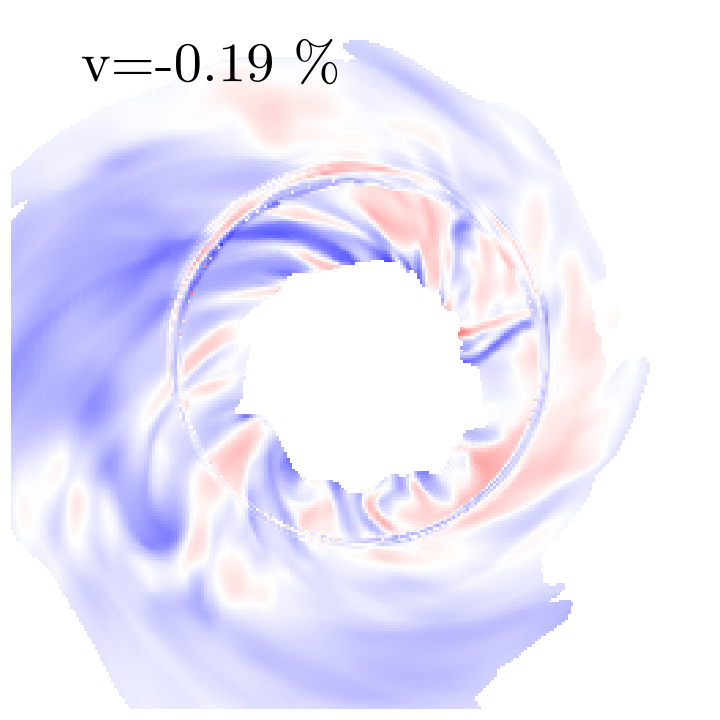}\\
\raisebox{0.07\linewidth}[0pt][0pt]{\rotatebox[origin=c]{90}{A10}}\phantom{.} 
 \includegraphics[width=0.16\linewidth]{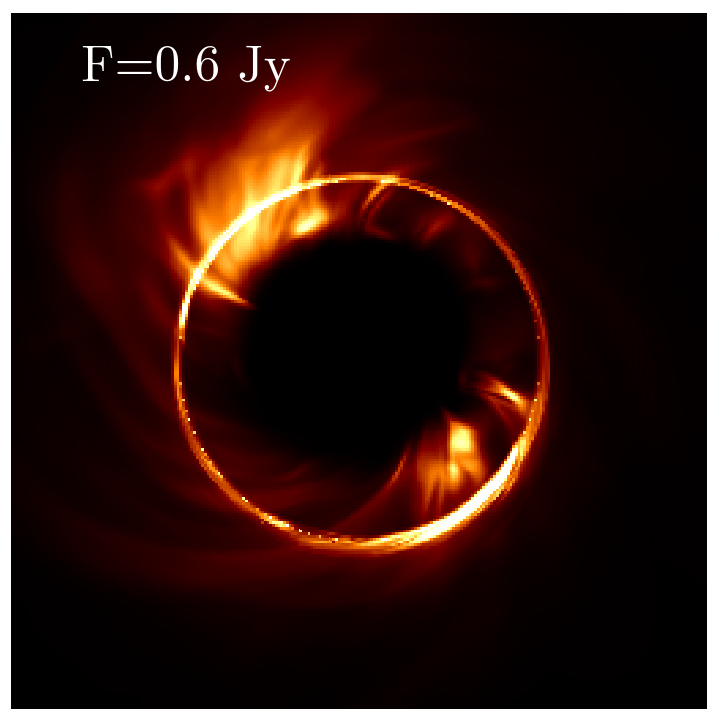}
&  \includegraphics[width=0.16\linewidth]{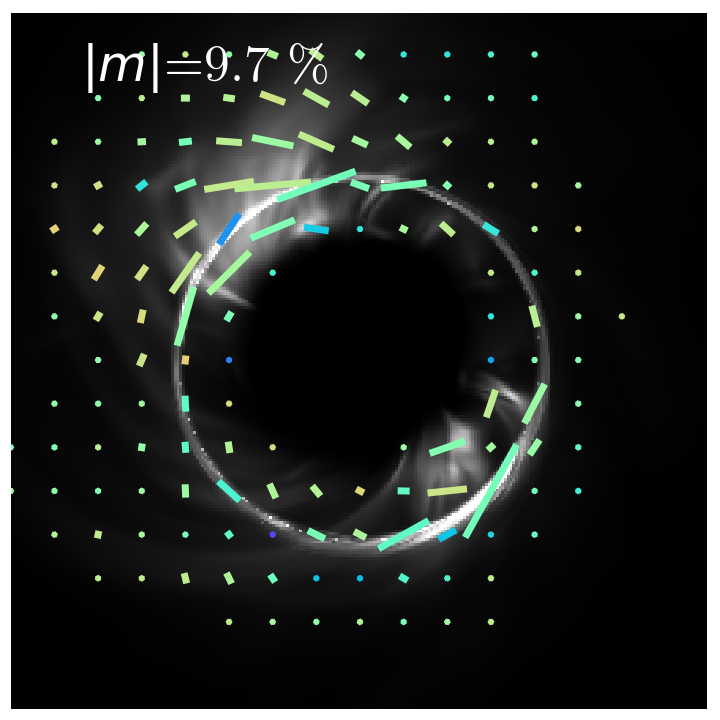} 
& \includegraphics[width=0.16\linewidth]{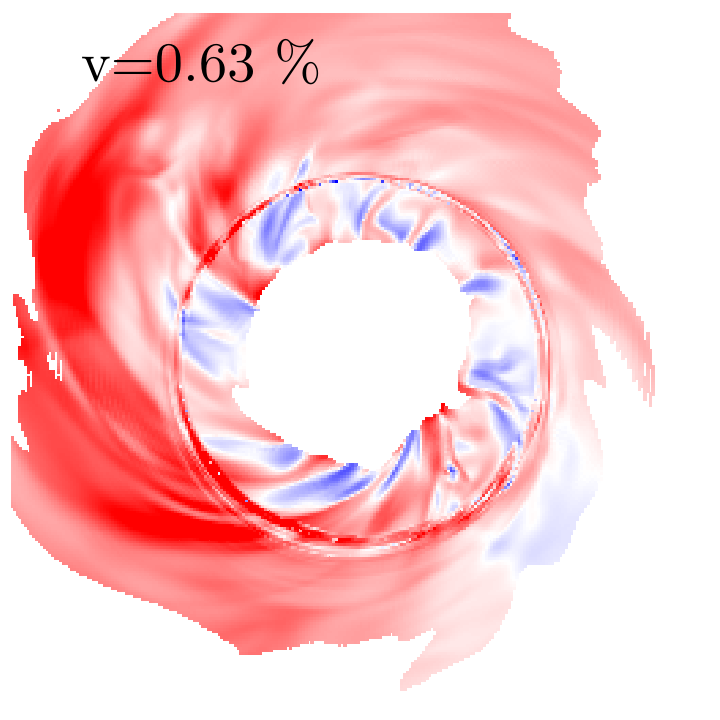}
& \includegraphics[width=0.16\linewidth]{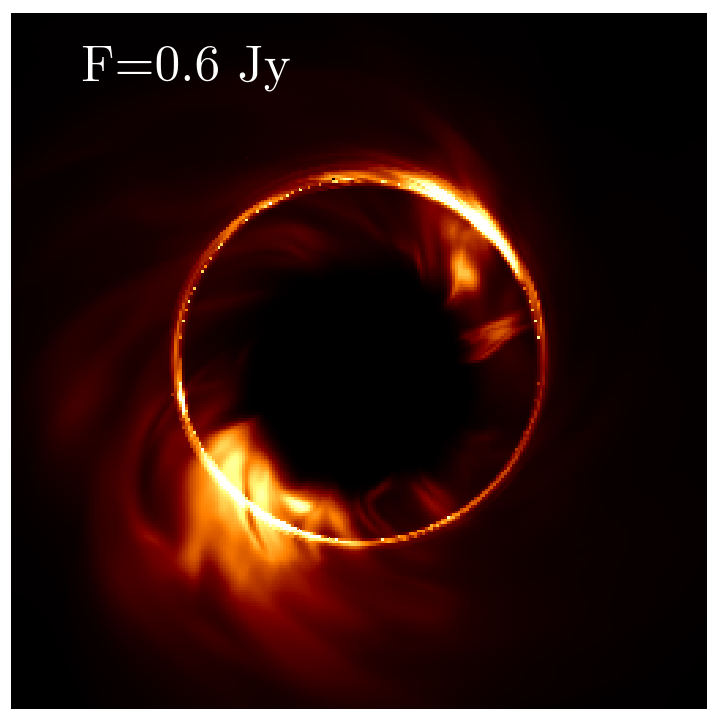}
& \includegraphics[width=0.16\linewidth]{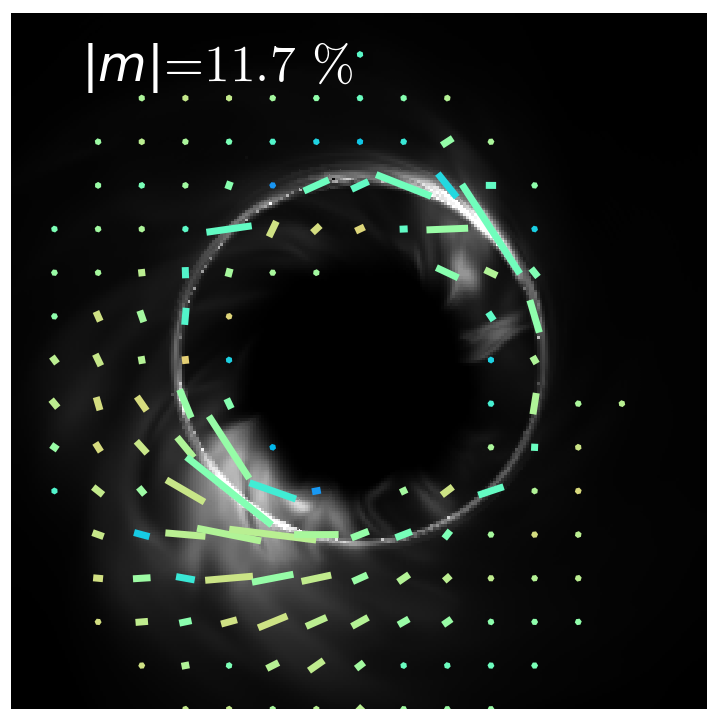}
& \includegraphics[width=0.16\linewidth]{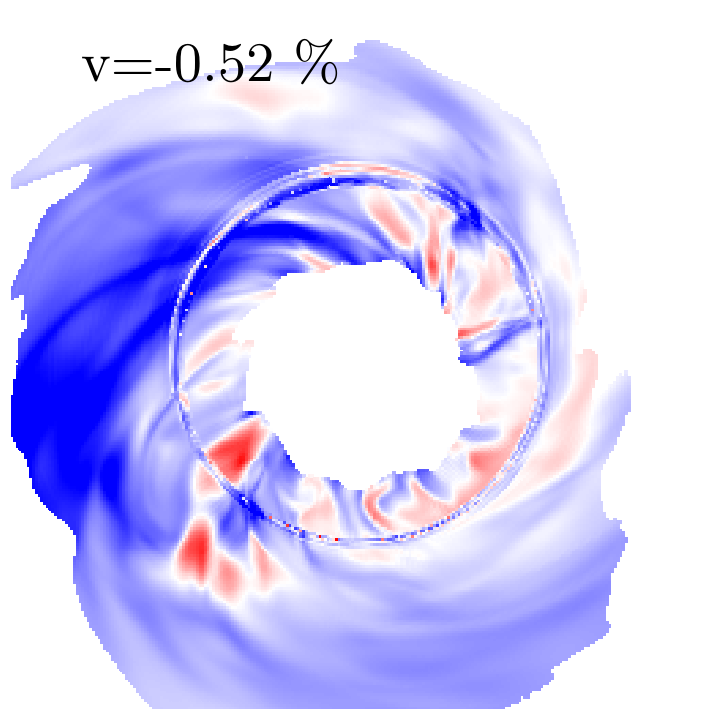}\\
\raisebox{0.07\linewidth}[0pt][0pt]{\rotatebox[origin=c]{90}{A100}}\phantom{.}
 \includegraphics[width=0.16\linewidth]{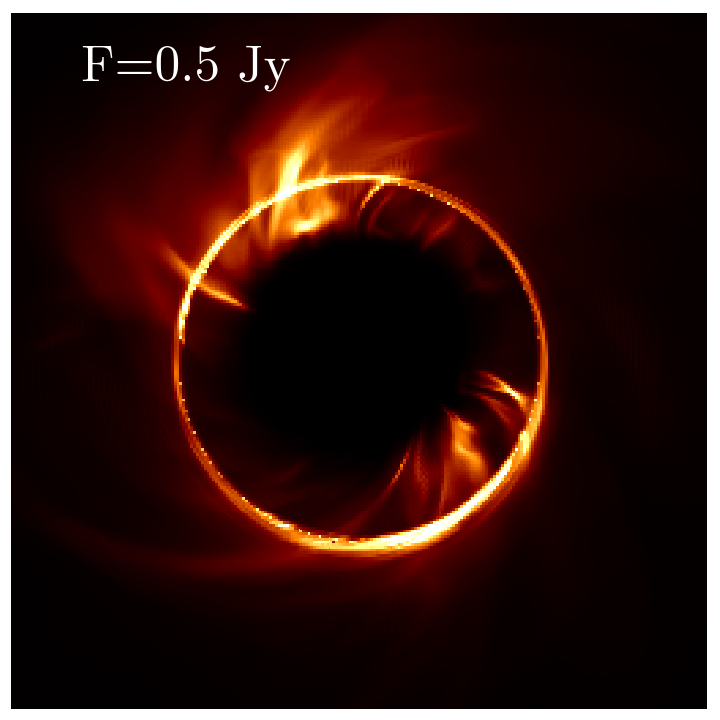}
& \includegraphics[width=0.16\linewidth]{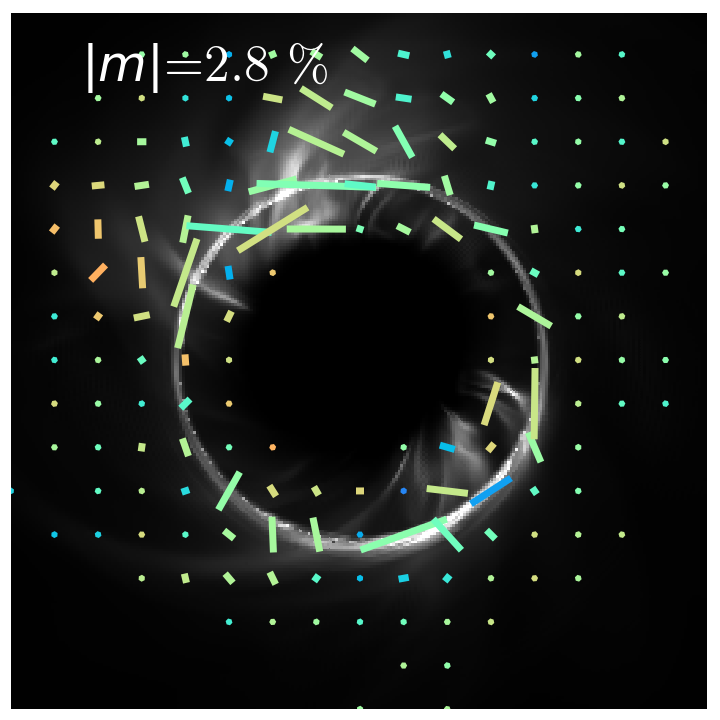}
& \includegraphics[width=0.16\linewidth]{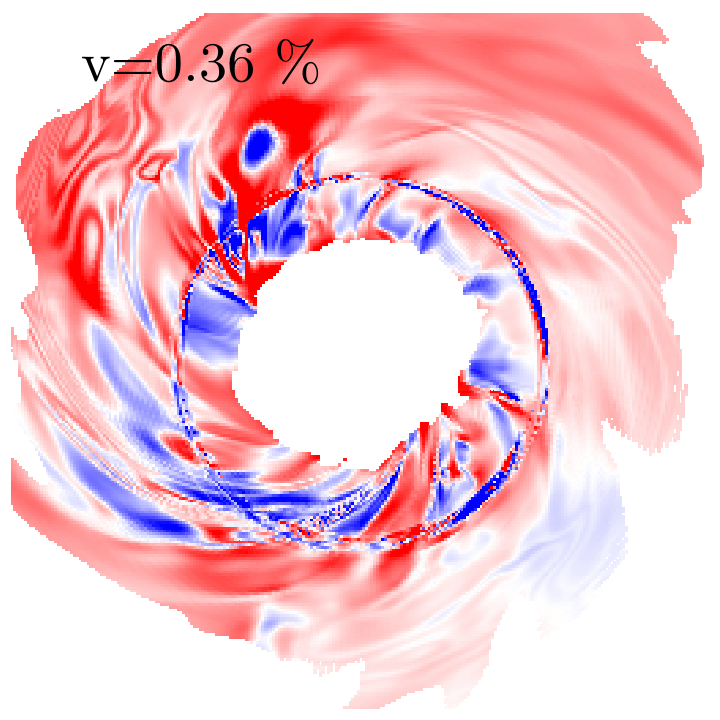}
& \includegraphics[width=0.16\linewidth]{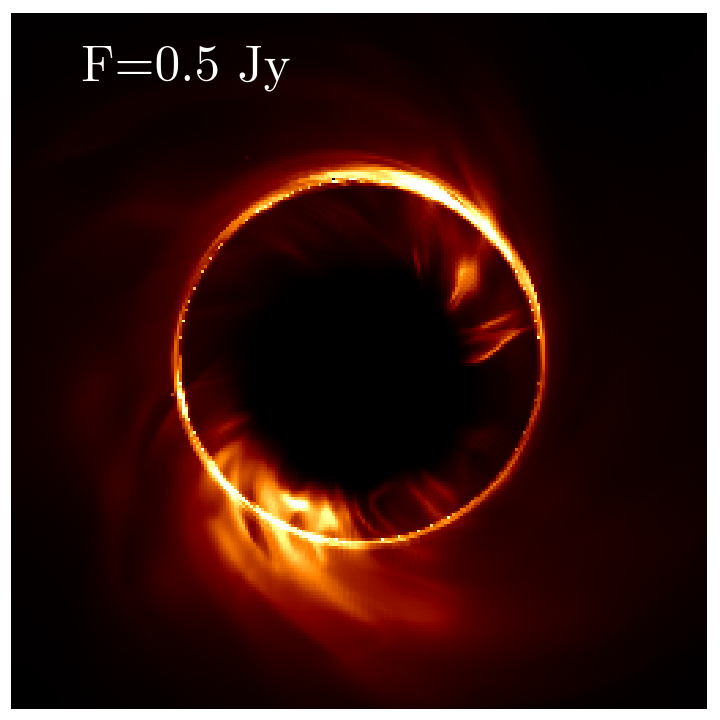}
& \includegraphics[width=0.16\linewidth]{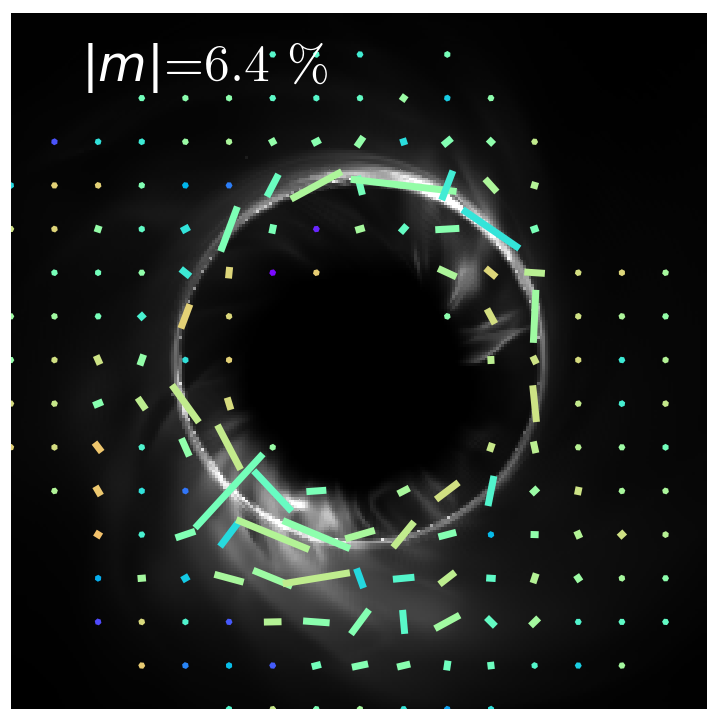}
& \includegraphics[width=0.16\linewidth]{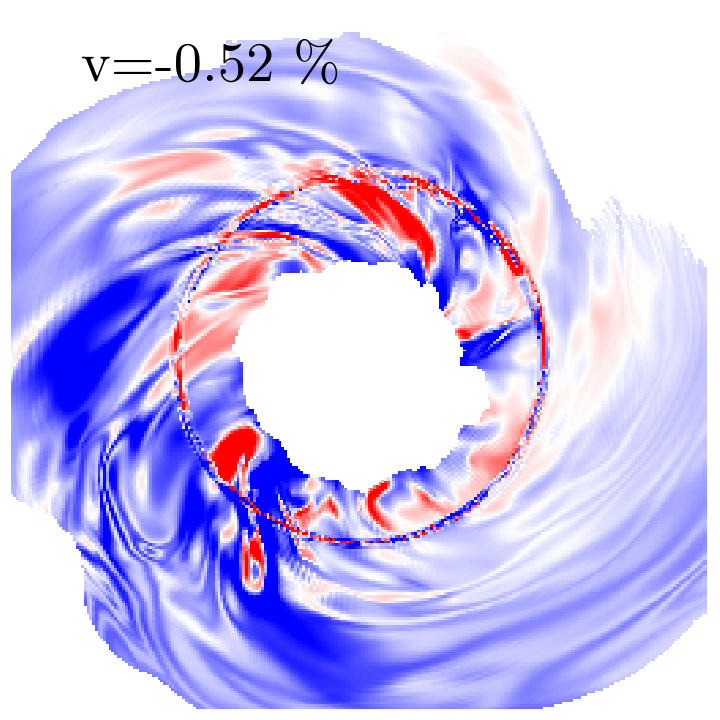}\\
\raisebox{0.07\linewidth}[0pt][0pt]{\rotatebox[origin=c]{90}{A200}}\phantom{.}
\includegraphics[width=0.16\linewidth]{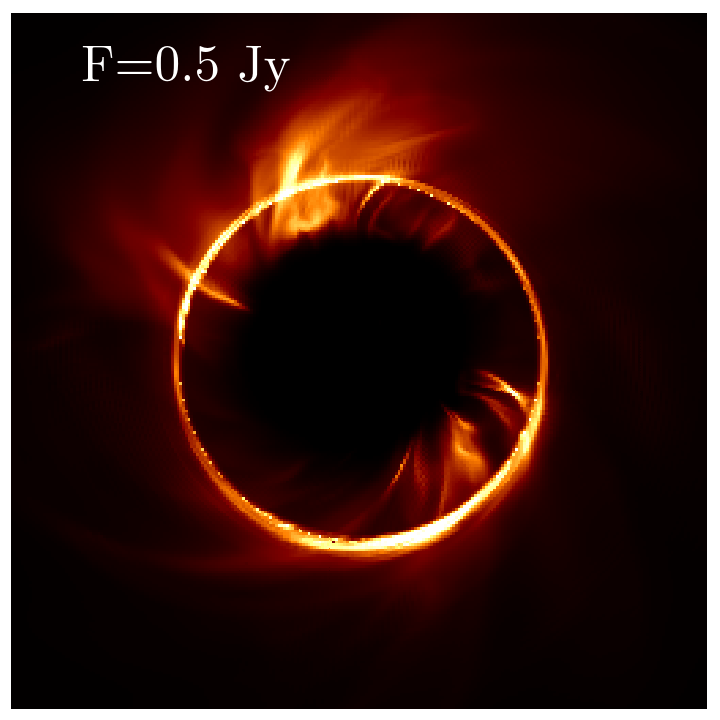} & \includegraphics[width=0.16\linewidth]{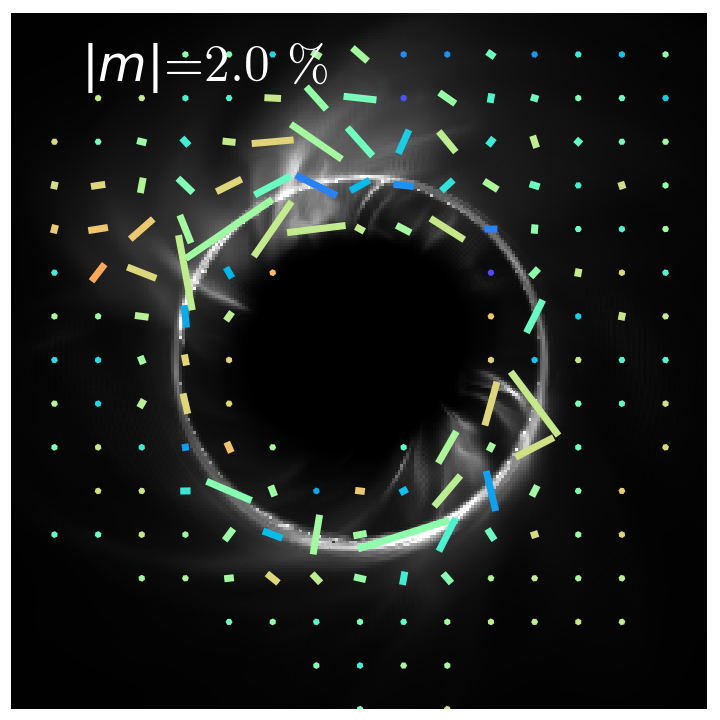}
 & \includegraphics[width=0.16\linewidth]{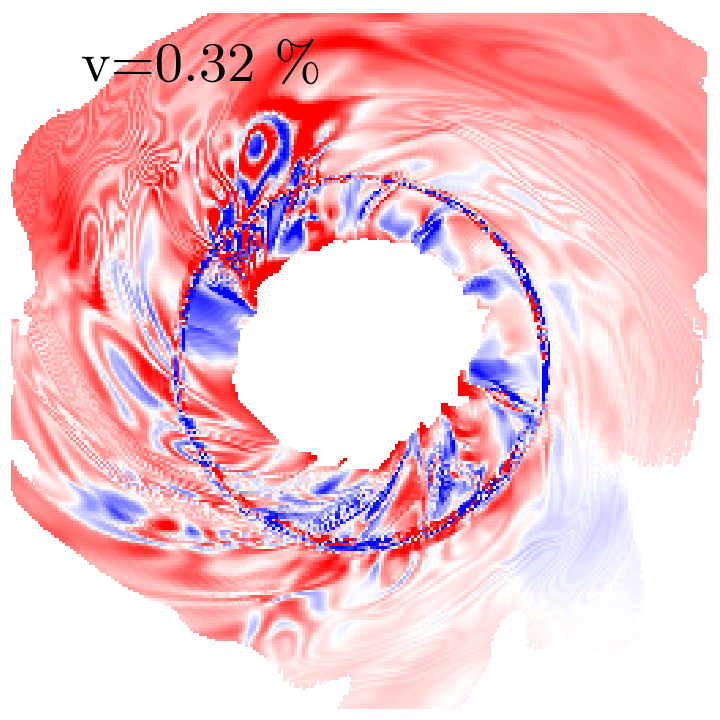}&
 \includegraphics[width=0.16\linewidth]{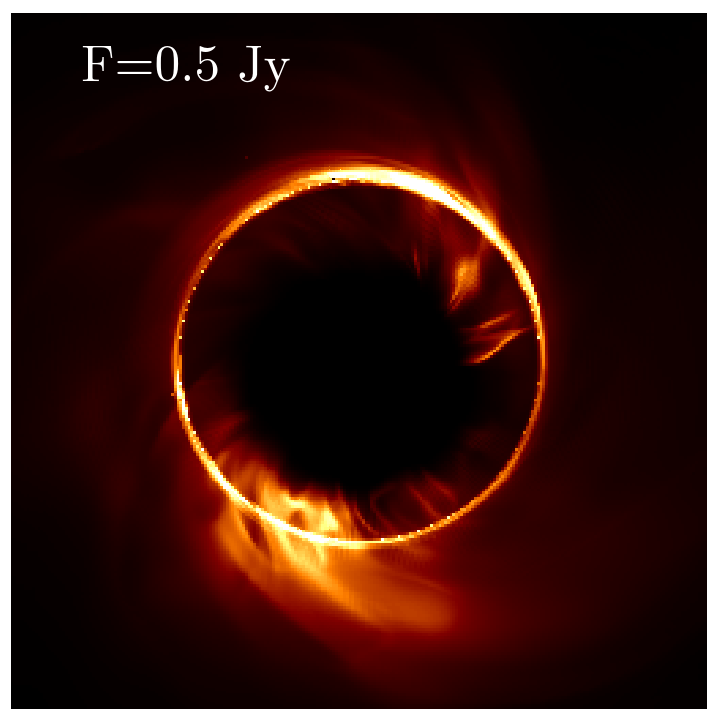} & \includegraphics[width=0.16\linewidth]{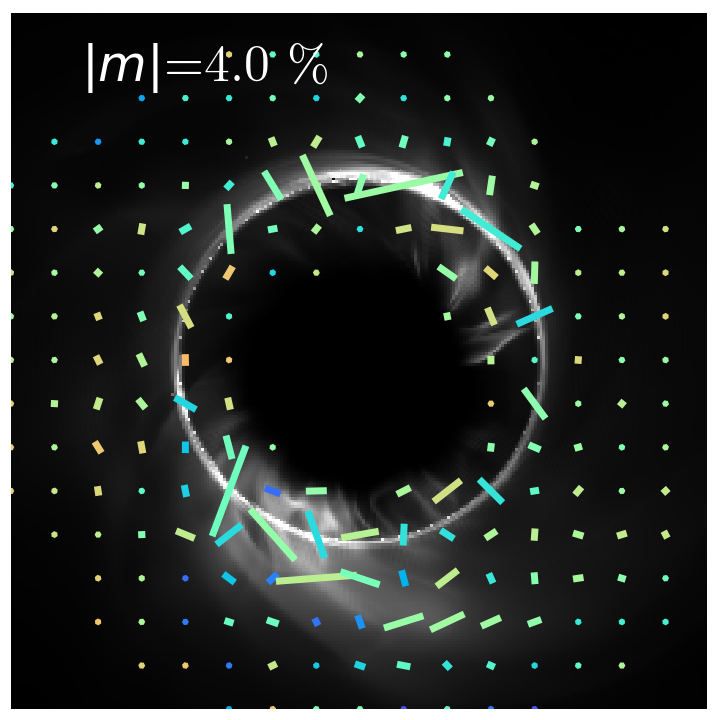}
& \includegraphics[width=0.16\linewidth]{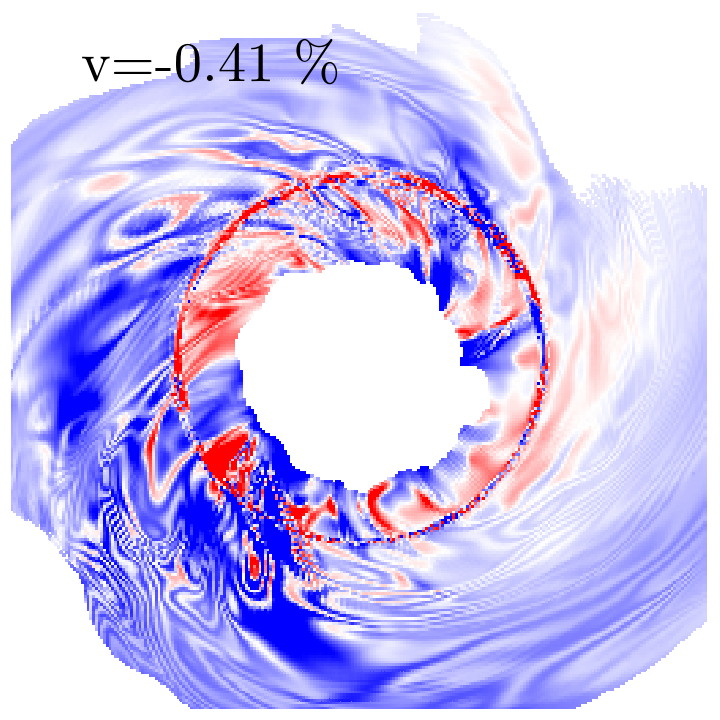}\\
\hspace{+0.02\linewidth}\includegraphics[width=0.14\linewidth,trim=0 18cm 0 0,clip]{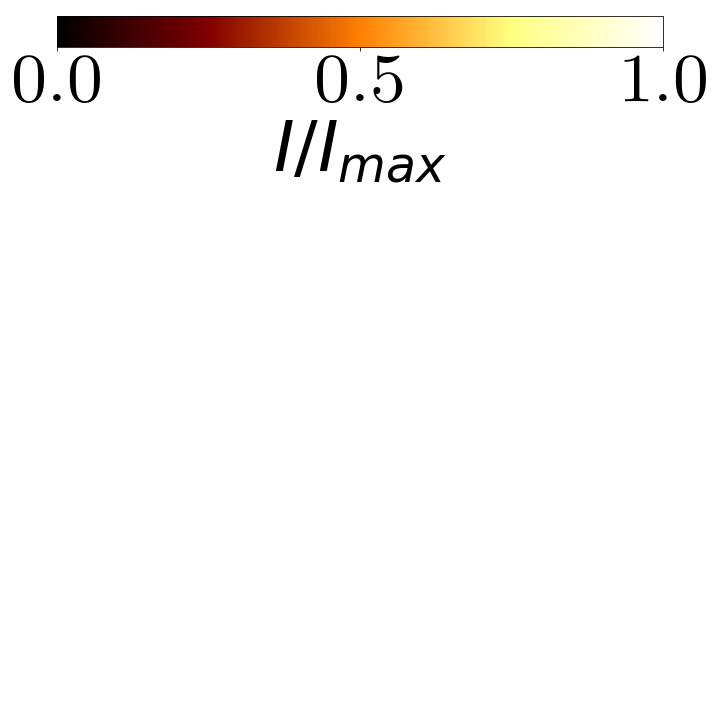} &
\includegraphics[width=0.14\linewidth,trim=0 18cm 0 0,clip]{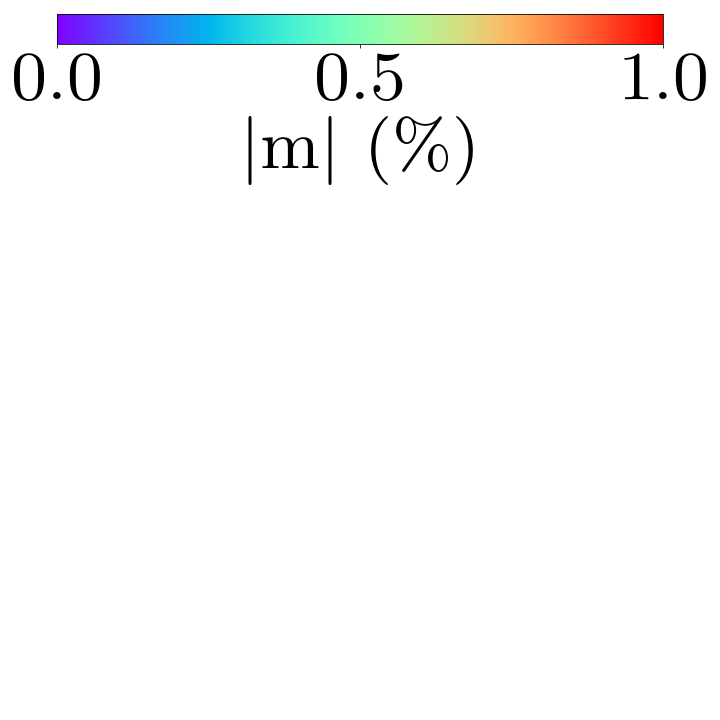} &
\includegraphics[width=0.14\linewidth,trim=0 18cm 0 0, clip]{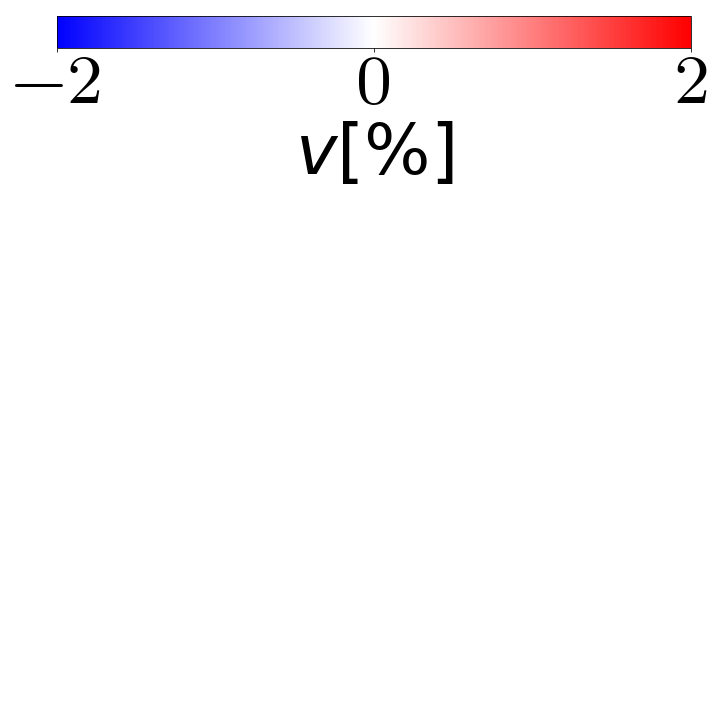} &
\includegraphics[width=0.14\linewidth,trim=0 18cm 0 0,clip]{plots/Model1a_beta30_a0/cbar_I_hor_2.png} &
\includegraphics[width=0.14\linewidth,trim=0 18cm 0 0,clip]{plots/Model1a_beta30_a0/cbar_LP_hor.png} &
\includegraphics[width=0.14\linewidth,trim=0 18cm 0 0, clip]{plots/Model1a_beta30_a0/cbar_CP.png} \\
\end{tabular}
\caption{Polarimetric images of model A snapshot for different electron heating ratios (from top to bottom) and two close to face-on viewing angles (left and right panels). Panels from left to right display images of in total intensity (Stokes $\mathcal{I}$), linear polarimetric images (where total intensity is shown in gray scale in the background and polarization is shown in form of polarization ticks which color indicates the fractional polarization, length - total flux in polarization and polarization plane position angle - the EVPA), and circular polarization maps (using red-white-blue color scale). In both linear and circular polarization images cuts has been made to omit displaying polarization regions where $I<1\% I_{max}$. Each panels field of view is 20x20M.}\label{fig:Rh}
\end{figure*}

\section{Results}\label{sec:results}

\subsection{Effects of Faraday thickness}

\begin{table}
\caption{Electron heating parameters and polarimetric properties of simulation A.}\label{tab:param}
\begin{center}
\begin{tabular}{cccccccc}
\hline 
\hline 
ID & $R_{\rm h}$ & $\frac{\left<\dot{M}\right>}{10^{-4}}$  & i &
\multicolumn{2}{c}{$|m|_{\rm net}$ (\%)} &\multicolumn{2}{c}{$v_{\rm net}$ (\%)} \\
& & [$ \msun/yr$]  & [$\deg$] & Min & Max & Min& Max\\
\hline 
\multirow{2}{*}{A1} & \multirow{2}{*}{1} &  \multirow{2}{*}{$0.7 $} & 160 & 0.4 &16 &  -0.2 & 0.1  \\
&&&20&0.1&12.7&0&0.4\\\\
\multirow{2}{*}{A10}& \multirow{2}{*}{10} & \multirow{2}{*}{$1.13 $} & 160 &1.3 & 12 & -0.75& -0.12\\
&&&20&0.74&16.5&0.21&0.96\\\\
\multirow{2}{*}{A100} & \multirow{2}{*}{100} & \multirow{2}{*}{$1.9 $} & 160 &0.8& 9.9&-0.8& -0.15\\
&&&20&0.7&11.2&0.08&0.77\\\\
\multirow{2}{*}{A200} & \multirow{2}{*}{200} & \multirow{2}{*}{$2.24$} & 160 & 0.35 &  7.6 &-0.6& 0.11\\
&&&20&0.5&6.4&0&0.5\\
\hline
\end{tabular}
\end{center}
Note -- Four electron temperature parameters were used to simulate various modes of electron heating in collisionless plasma. Model characteristics for different electron models are shown in rows. Columns from left to right show: (1) model ID; (2) electron heating parameter (see text for description); (3) average mass accretion rate in physical units; (4,5) range of fractional linear polarization of near horizon emission measured using a time sequence of model images; (6,7) range of fractional circular polarization of near horizon emission measured using the same sequence of models.
\end{table}

Our radiative transfer simulations assume four values of \rhigh parameter ranging between 1 and 200. Models with different electron temperature parameter have different mass scaling factors to produce total flux similar to that observed in astrophysical source at $f=$230\,GHz (in M87 that is about 0.5 Jy, \citealt{ehtIV:2019}). Differences in the mass scaling results in a different accretion rates in physical units but the time dependence of accretion rate remains fixed. We report the averaged mass accretion rates obtained for each \rhigh parameter in Table~\ref{tab:param}. By changing overall mass and temperature of the simulations we change the Faraday thickness of the model which affects the polarimetric images. 
The Faraday rotation thickness is
defined as $\tau_{\rm FR}=\rho_V d\lambda$ where $\rho_V$ is the Faraday
rotativity, is expected to increase with \rhigh and accretion rate as
$\tau_{\rm FR}=\dot{M}^{3/2} R_{\rm h}^2$ (see \citealt{moscibrodzka:2017} for
more details and \citealt{tsunetoe:2020} for recent discussion of
axi-symmetric GRMHD simulations). Faraday conversion thickness is also expected to increase with mass accretion rate, $\tau_{\rm FC} \sim n_e B^2 \sim \dot{M}^{2}$. Notice that while Faraday rotation is sensitive to parallel field component, the conversion is larger in places where field lines are perpendicular to the line of sight.

In Figure~\ref{fig:Rh} we display total intensity and polarimetric images of the chosen snapshot ($t=21150$M) of model A for different values of \rhigh and two opposite, low viewing angles to present the appearance of the system from both sides. Rows in Figure~\ref{fig:Rh} correspond to the models with different electron temperature parameter, whose maps are shown Figure~\ref{fig:electrons}. 
The polarimetric images of MADs are sensitive to the assumption of \rhigh parameter. 
In models with hot electrons in the accretion disk (\rhigh=1,10) the EVPA pattern is organized into a {\it vortex-like} structure. In models with sub-relativistic electrons in the accretion disk (\rhigh=100, 200) the EVPA pattern becomes disorganized. The scrambling of the EVPA ticks is expected as a result of increasing Faraday depth in colder plasma as pointed out by \citet{moscibrodzka:2017}.  Indeed we have measured that the Faraday rotation depth in our models is increasing with \rhigh and $\dot{M}$ but the dependency is less steep function of these parameters (intensity weighted averaged $\left<\tau_{\rm FR}\right>_I$=8,22,87,135 for \rhigh=1,10,100,200 models, respectively). The enhanced Faraday rotations result in the significant decrease of the image net linear polarization via beam depolarization. 

In Figure~\ref{fig:Rh}, the maps of circular polarization highlight fractional
circular polarization in the regions of relatively high total intensity. In
contrast to linear polarization, the net fractional circular polarization
increases with \rhigh. Note that although the resolved fractional circular
polarization can be higher than $1\%$, the $v_{\rm net}$ is less than unit for the given snapshot and all parameters. In models A1-A200 the total intensity weighted Faraday conversion depth range is $\left<\tau_{FC}\right>_I=0.02-0.1$ (Faraday conversion depth is proportional to the fraction of $|m|$ converted to $v$, \citealt{homan:2009}.).

It is also evident that models observed at opposite viewing angles show circular polarization of opposite signs. 
Noteworthy, the sign of the circular polarization in the lensing ring around the black hole shadow can be inverted with respect the net image circular polarization. The sign inversion is more visible in models with \rhigh=100 and 200 (high Faraday thickness models), for both inclination angles.

\subsection{Sign and strength of circular polarization in direct and lensed emission}\label{sec:3.2}

\begin{figure*}
\centering
\def\arraystretch{0.0}
\centering
\setlength{\tabcolsep}{0pt}
\begin{tabular}{ccccc}
$\mathcal{I}$& $\mathcal{I + LP}$ & $\mathcal{CP}$,all & $\mathcal{CP},\rho_Q=0$ & $\mathcal{CP},\rho_V=0$ \\
\raisebox{0.07\linewidth}[0pt][0pt]{\rotatebox[origin=c]{90}{Full}}\phantom{.}
\includegraphics[width=0.18\linewidth]{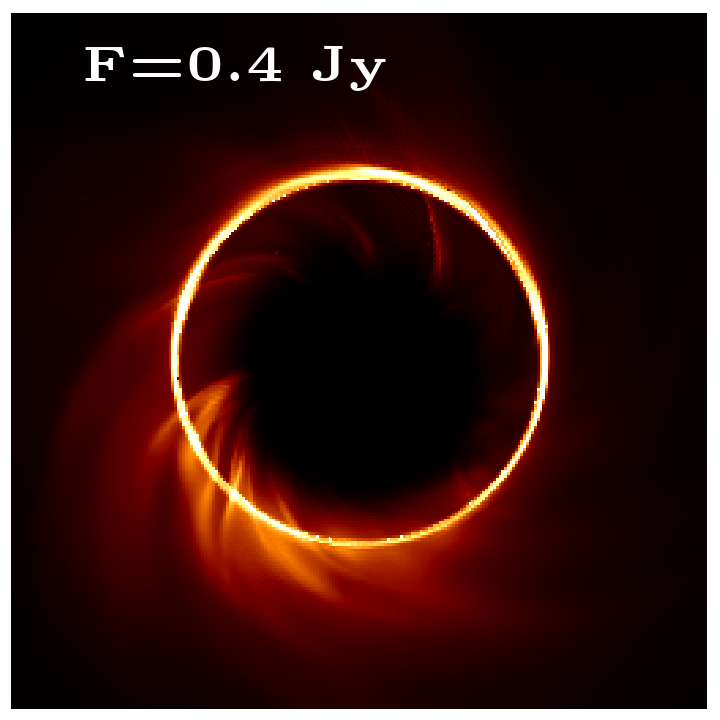} & 
\includegraphics[width=0.18\linewidth]{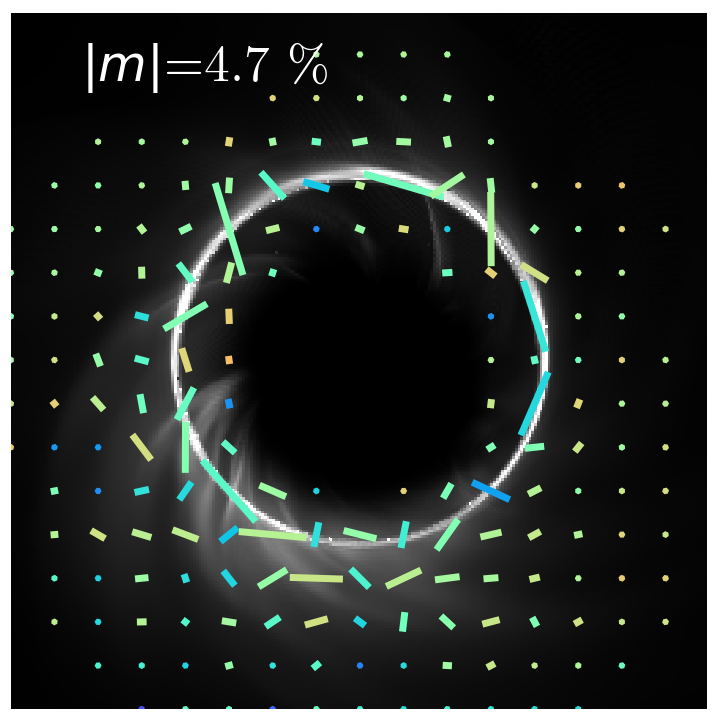} & 
\includegraphics[width=0.18\linewidth]{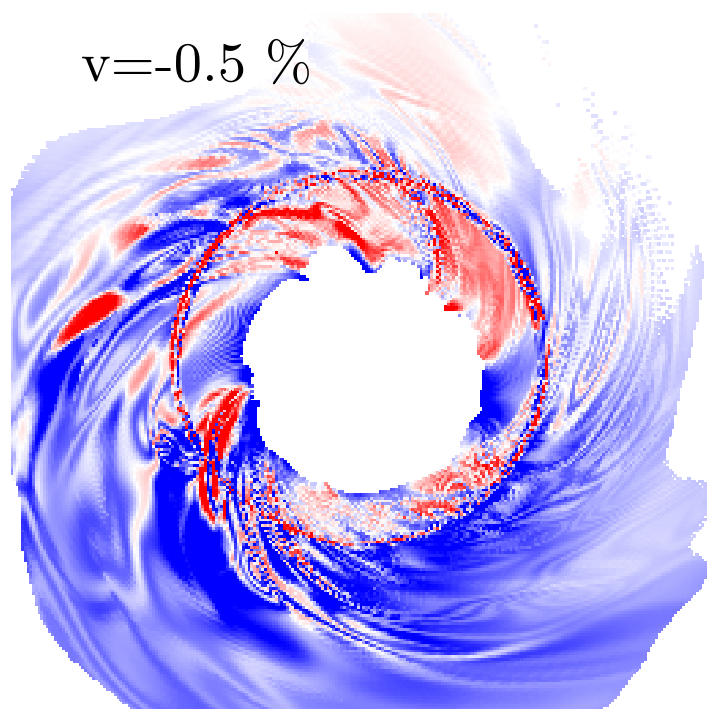} &
\includegraphics[width=0.18\linewidth]{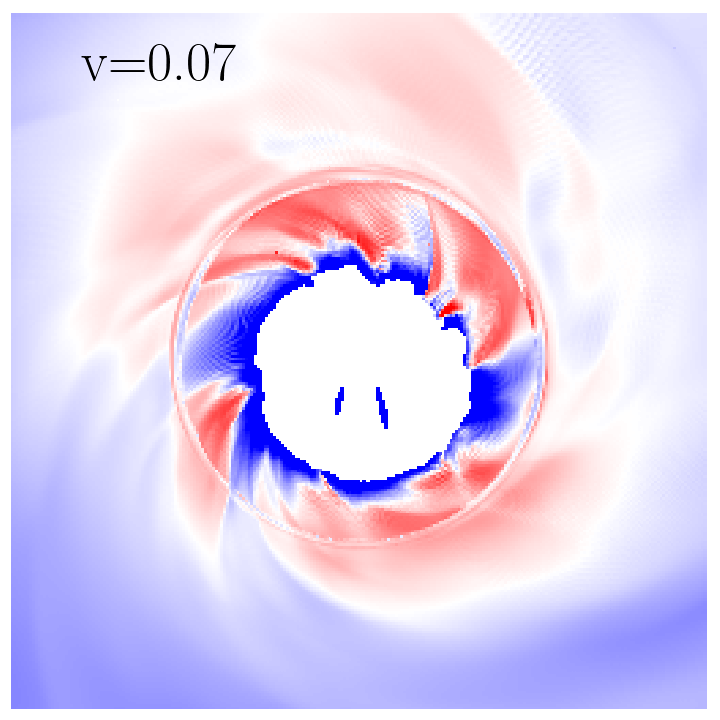} &
\includegraphics[width=0.18\linewidth]{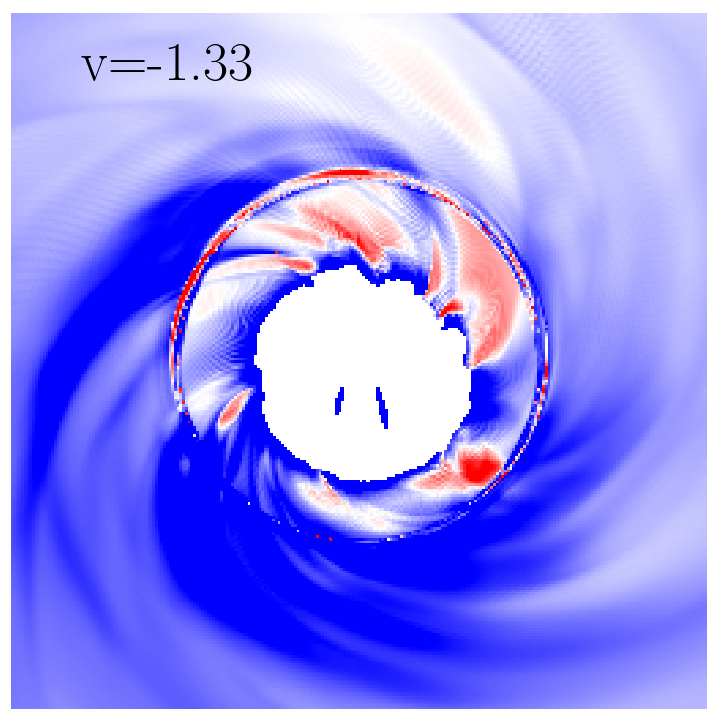}\\
\raisebox{0.07\linewidth}[0pt][0pt]{\rotatebox[origin=c]{90}{Farside}}\phantom{.}
\includegraphics[width=0.18\linewidth]{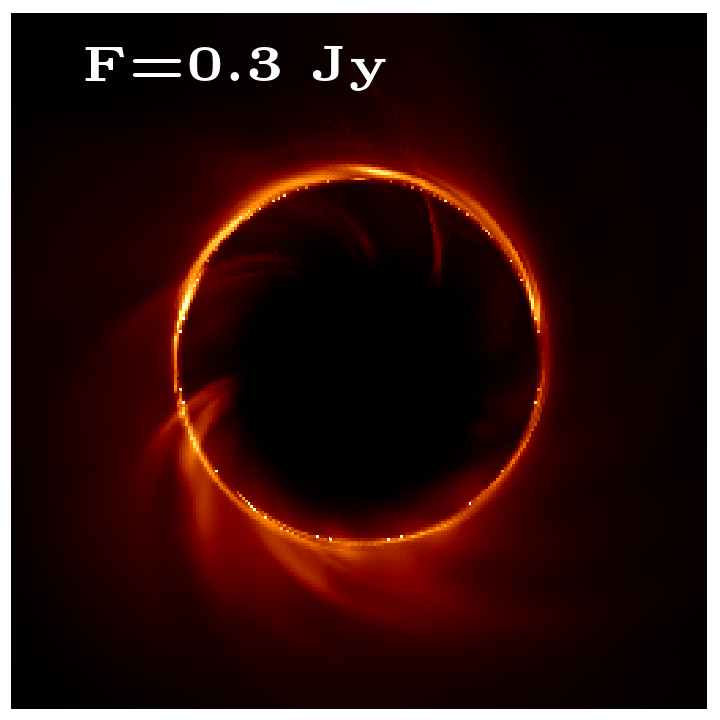} & 
\includegraphics[width=0.18\linewidth]{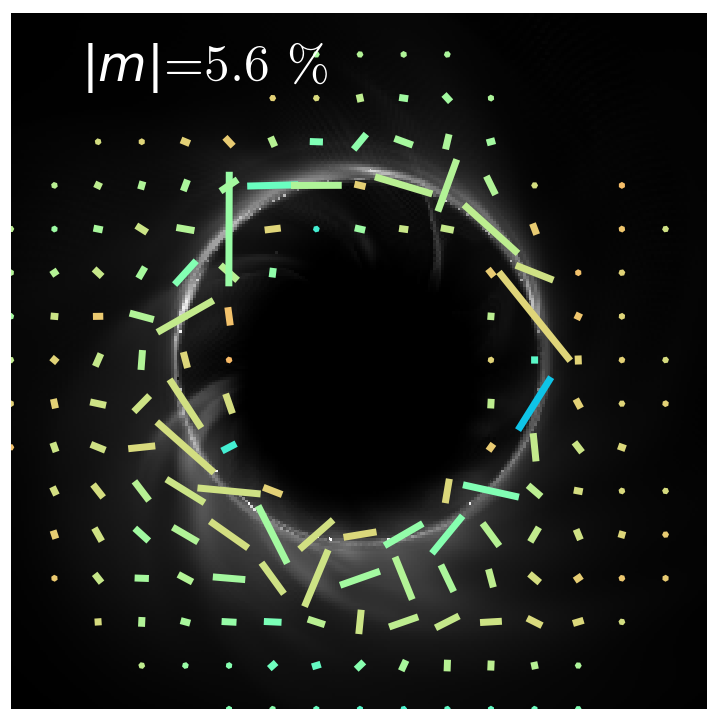} & 
\includegraphics[width=0.18\linewidth]{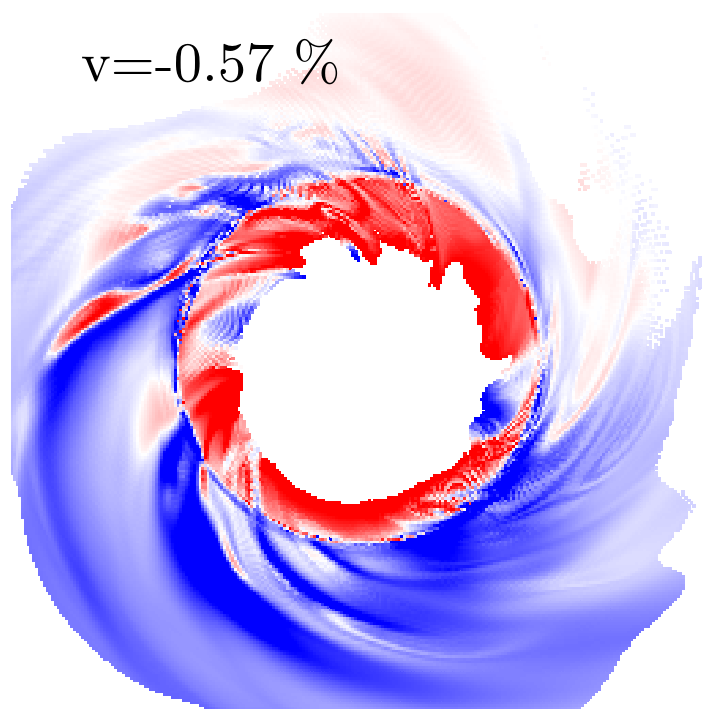} &
\includegraphics[width=0.18\linewidth]{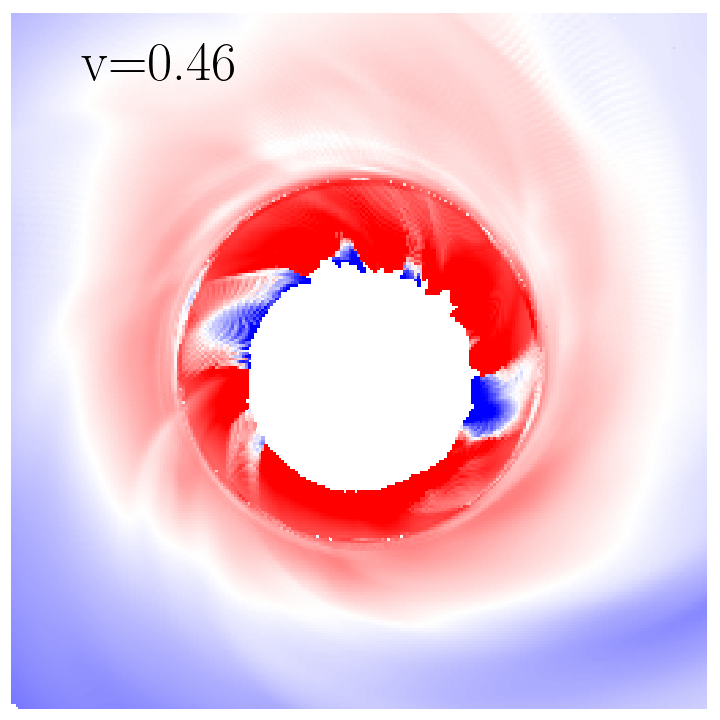} &
\includegraphics[width=0.18\linewidth]{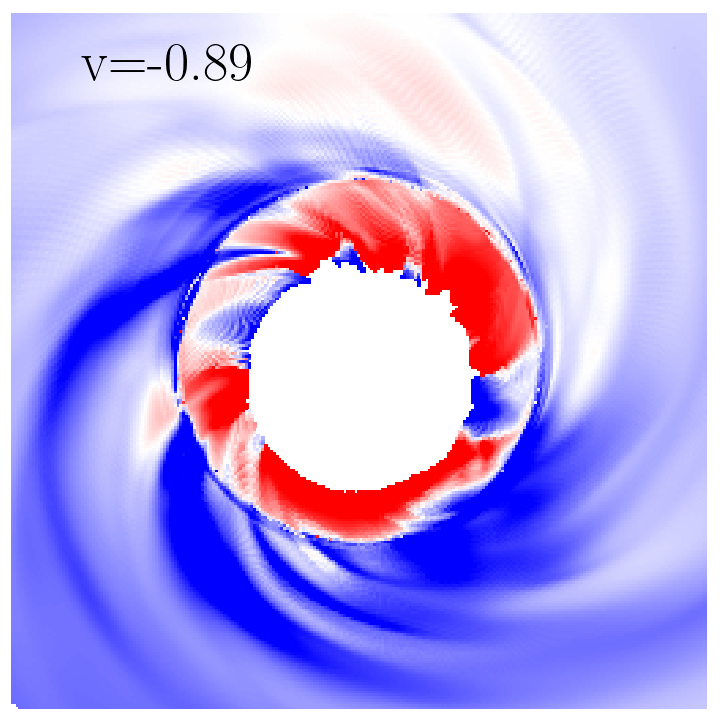}\\
\raisebox{0.07\linewidth}[0pt][0pt]{\rotatebox[origin=c]{90}{Nearside}}\phantom{.}
\includegraphics[width=0.18\linewidth]{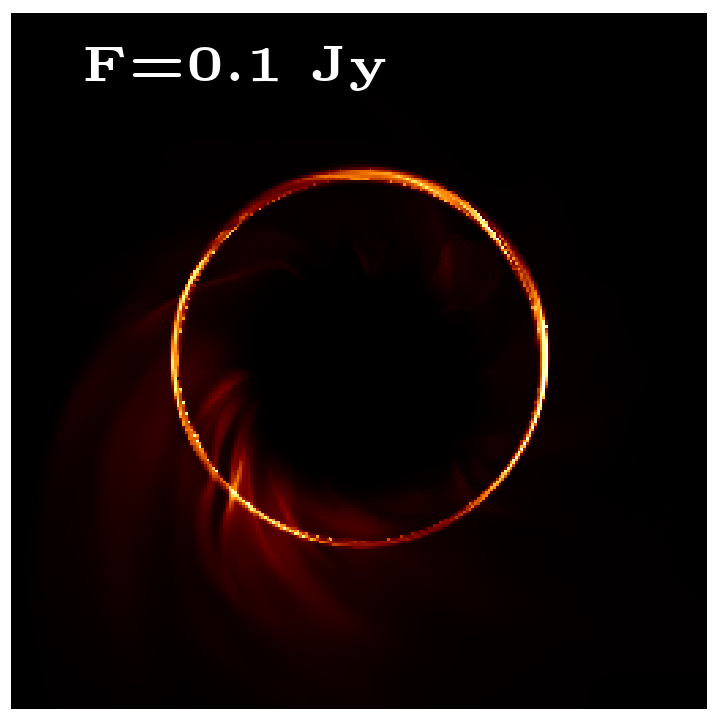} & 
\includegraphics[width=0.18\linewidth]{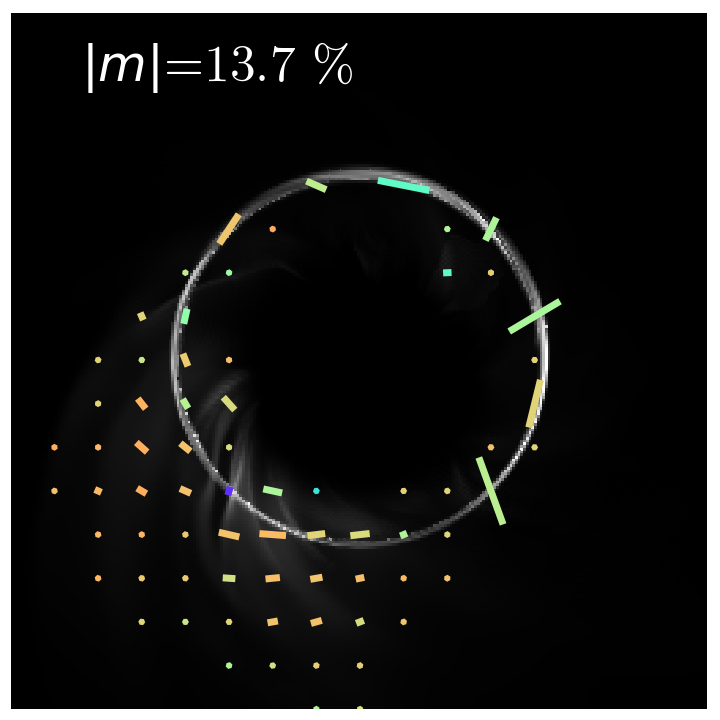} & 
\includegraphics[width=0.18\linewidth]{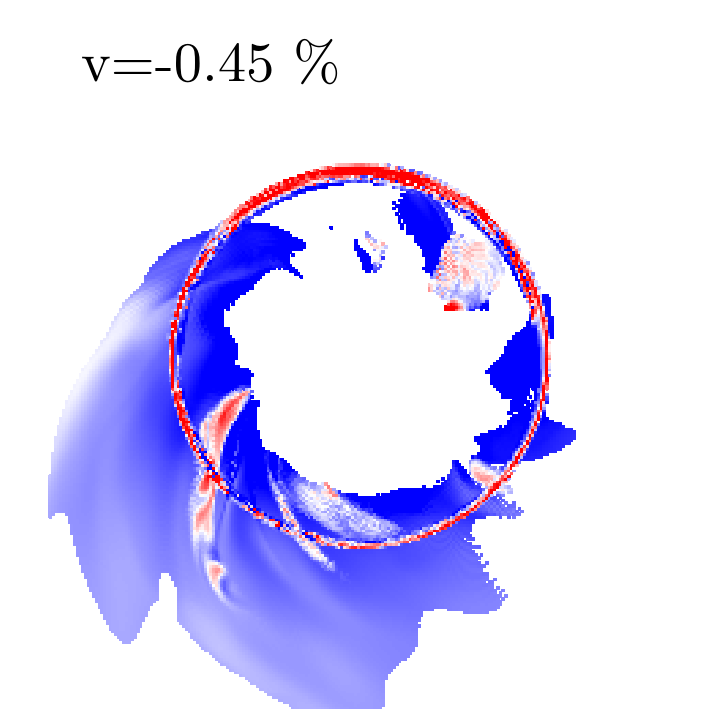} &
\includegraphics[width=0.18\linewidth]{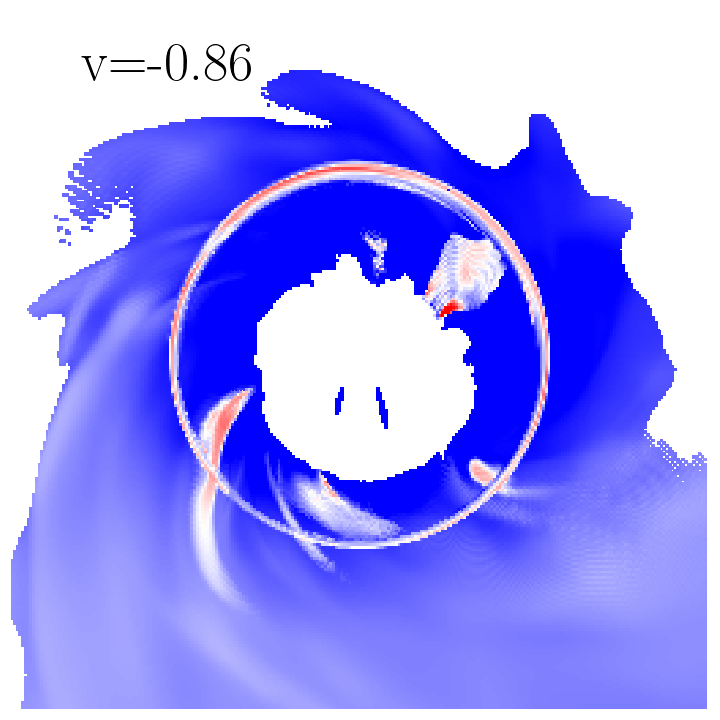} &
\includegraphics[width=0.18\linewidth]{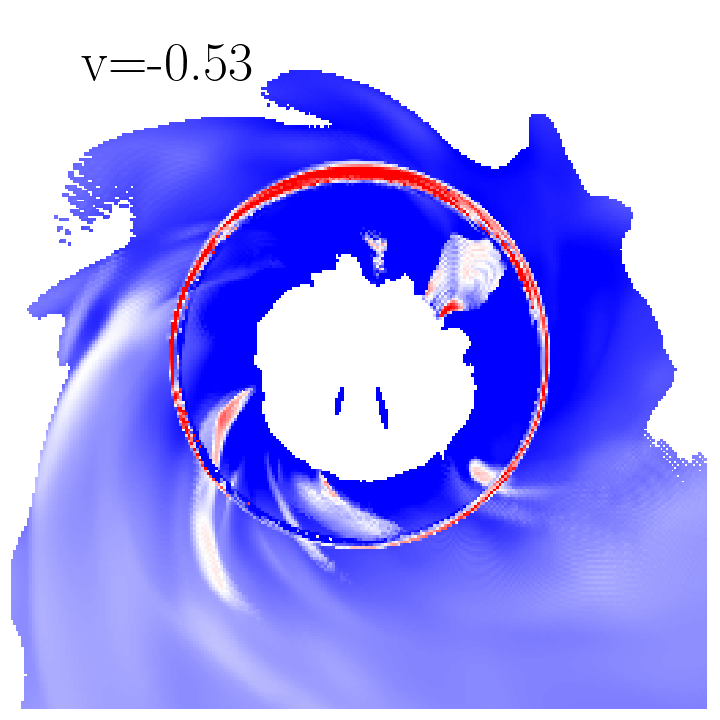}\\
\hspace{+0.02\linewidth}
\includegraphics[width=0.17\linewidth,trim=0 18cm 0 0,clip]{plots/Model1a_beta30_a0/cbar_I_hor_2.png} &
\includegraphics[width=0.17\linewidth,trim=0 18cm 0 0,clip]{plots/Model1a_beta30_a0/cbar_LP_hor.png} &
\includegraphics[width=0.17\linewidth,trim=0 18cm 0 0, clip]{plots/Model1a_beta30_a0/cbar_CP.png} &
\includegraphics[width=0.17\linewidth,trim=0 18cm 0 0, clip]{plots/Model1a_beta30_a0/cbar_CP.png} &
\includegraphics[width=0.17\linewidth,trim=0 18cm 0 0, clip]{plots/Model1a_beta30_a0/cbar_CP.png}\\
\end{tabular}
\caption{Decomposition of model A200 snapshot total intensity and polarimetric
  images (shown in top row) into farside (middle row) and nearside
  contributions (bottom row). In farside images, the radiative transfer
  equations are integrated until equatorial plane of the coordinate system,
  i.e. on the farside of the black hole with respect to the observer. In the
  nearside images (bottom panels) the integration is done on the nearside of
  the black hole. Columns show total intensity, linear polarimetric images and
  circular polarimetric images from left to right. The two rightmost circular
  polarimetric images assume no Faraday conversion, $\rho_Q=0$ (total intensity and linear polarization remain unaffected by Faraday conversion) or no Faraday rotation $\rho_V=0$ (for which only total intensity remains unaffected). All images are displayed in the same manner as in Figure~\ref{fig:Rh}.
}\label{fig:invert}
\end{figure*}

GRMHD simulations and their polarization signal are complex. We first investigate the strength and sign of the circular polarization in the direct and lensed emission by decomposing the polarimetric images into farside and near side components\footnote{Notice that fractional polarizations are not additive quantities so the plot is only indicative.}. In the farside/nearside images the radiative transfer equations are integrated only below/above the mid-plane of the disk with respect to the observer. 

We decompose different snapshot of model A200 into
far and nearside images, the same model with Faraday conversion set to zero,
$\rho_Q=0$, and Faraday rotation set to zero, $\rho_V=0$ are also shown (see the
rightmost panels). Top panels in Figure~\ref{fig:invert} show
that Faraday conversion plays a key role in circular polarization formation
even when Faraday rotation is not present.
Without both Faraday effects the intrinsic circular polarization
is significantly reduced. In contrast, without Faraday rotation alone, the circular
polarization is significantly enhanced and has almost single handedness in the
most of image regions. The latter suggest that the sign of the
circular polarization is more sensitive to the direction of the rotation of
the flow \citep{enslin:2003} rather then polarity of the magnetic fields and
Faraday rotation process \citep{beckert:2002}. In Appendix~\ref{app:polarityB} we discuss the impact of
changing the global polarity of B-fields in the GRMHD simulation on
polarimetric images. Interestingly, the results presented in Appendix~\ref{app:polarityB} further support the idea that
the Stokes ${\mathcal QU}$ rotation and production of Stokes ${\mathcal V}$
via conversion process mainly arises from rotation of plasma.
Nevertheless, by comparing panels with full, nearside and
farside images and models with and without Faraday effects, we find that in
the Faraday thicker flows, such as model A200, th details of
  the circular polarization depend on both Faraday effects: rotation and
conversion.
The nearside images of model with $\rho_V=0$ and $\rho_V\neq0$
suggest that Faraday rotation on the nearside is counteracting
the increase of circular polarization. As a result only negative circular polarization emission regions below the equator visibly contribute to the final circular polarization map. 
These results can be understood as follows. The Faraday conversion changes
Stokes ${\mathcal U}$ together with its sign into ${\mathcal V}$ because $\rho_Q
\sim n_e B^2 \sin^2 \theta$  (where $\theta$ is an angle between the light
wavevector and magnetic field vector measured in the fluid frame) is always
positive independently of the B-field sign.
However, Faraday rotation does depend on the polarity of magnetic field along the line of sight, since $\rho_V \sim n_e B \cos\theta$~\footnote{See \citealt{dexter:2016} for all Stokes synchrotron emissivities and rotativities used in our \texttt{ipole} code.}. Hence when rays cross the nearside region the Faraday rotation changes the sign of ${\mathcal U}$ because of overall opposite polarity of magnetic fields on the nearside compared to the farside, hence the conversion process starts to subtract from the total ${\mathcal V}$ rather than contribute to it.
Finally, in Figure~\ref{fig:invert}, in all cases the Stokes ${\mathcal V}$ polarity inversion in the ring is present and it seems to be associated with the geometry of the light propagation. 

\begin{figure*}
\def\arraystretch{0.0}
\centering
\setlength{\tabcolsep}{0pt}
\begin{tabular}{ccccccc}
&\multicolumn{3}{c}{\rhigh=1} & \multicolumn{3}{c}{\rhigh=200} \\ 
Density & ${\bf \mathcal{I}}$ & ${\bf \mathcal{I+LP}}$ & ${\bf \mathcal{CP}}$ & ${\bf \mathcal{I}}$ & ${\bf \mathcal{I+LP}}$ & ${\bf \mathcal{CP}}$\\
\raisebox{0.06\linewidth}[0pt][0pt]{\rotatebox[origin=c]{90}{ Model A}}\phantom{.}
\includegraphics[width=0.14\linewidth]{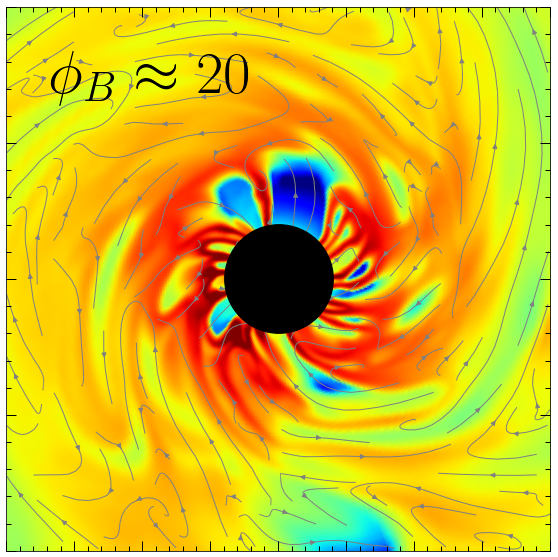} 
& \includegraphics[width=0.14\linewidth]{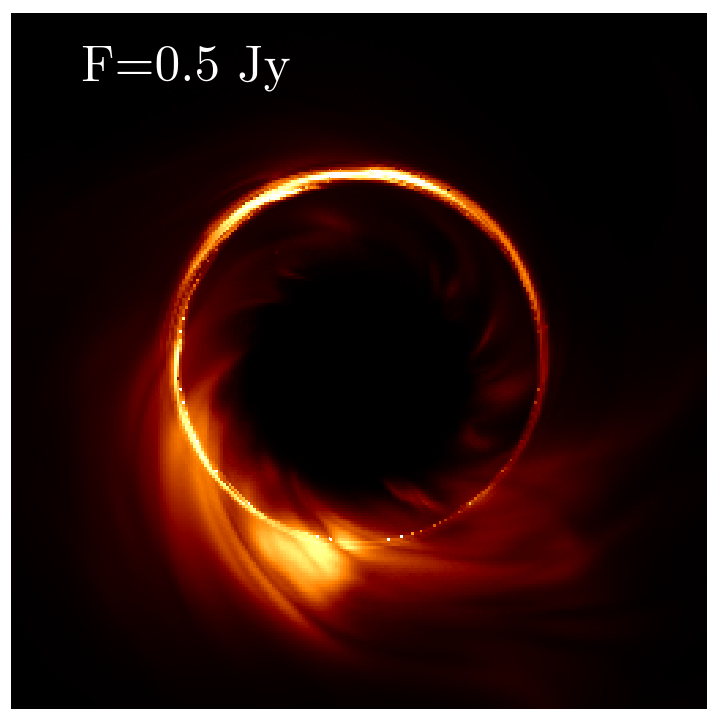}
& \includegraphics[width=0.14\linewidth]{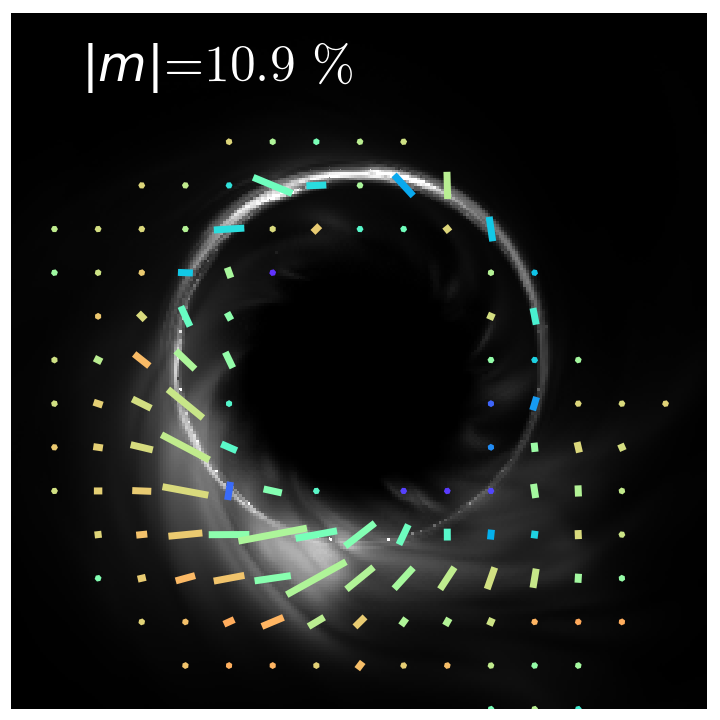}
& \includegraphics[width=0.14\linewidth]{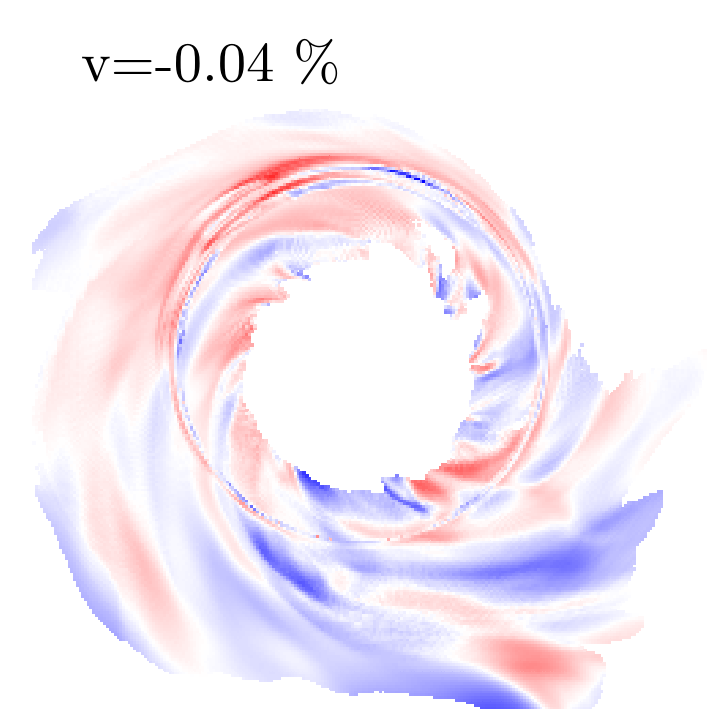}
& \includegraphics[width=0.14\linewidth]{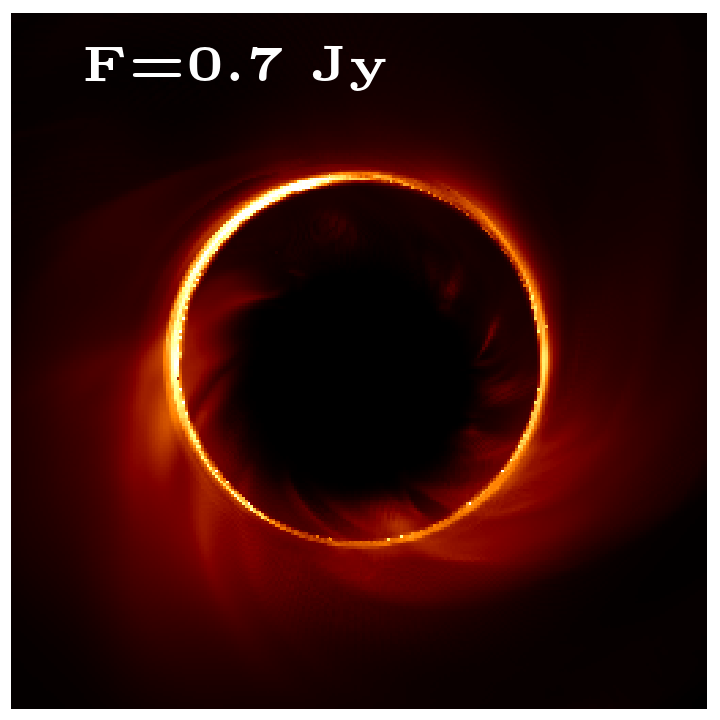}
& \includegraphics[width=0.14\linewidth]{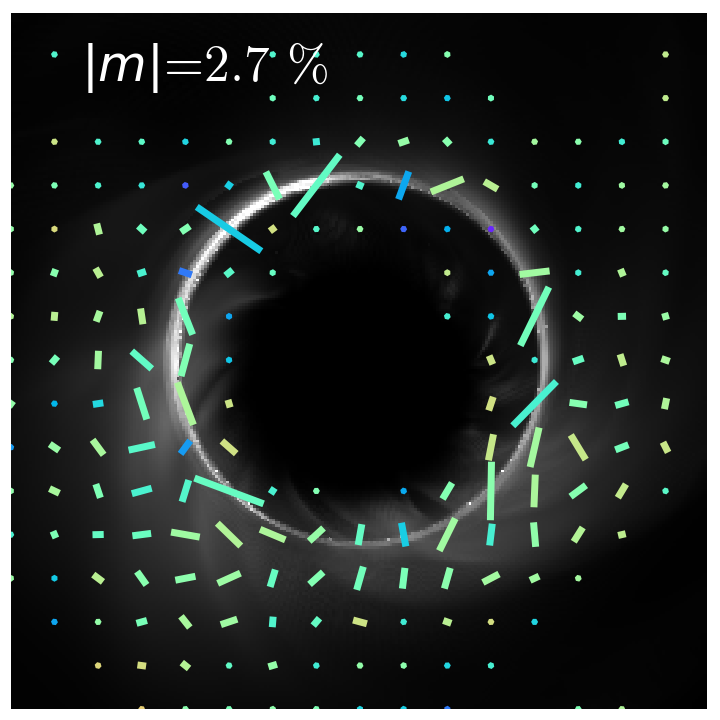}
& \includegraphics[width=0.14\linewidth]{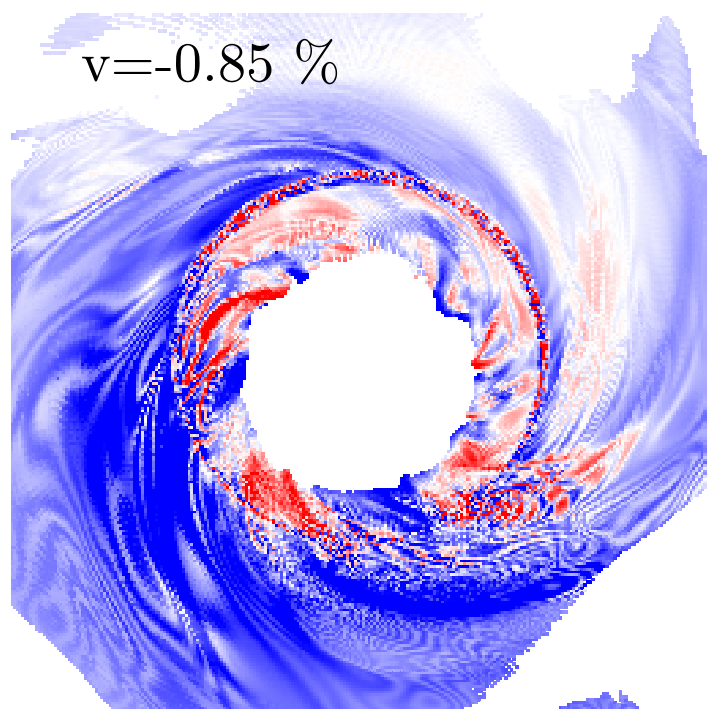}\\
\raisebox{0.06\linewidth}[0pt][0pt]{\rotatebox[origin=c]{90}{ Model B}}\phantom{.} 
\includegraphics[width=0.14\linewidth]{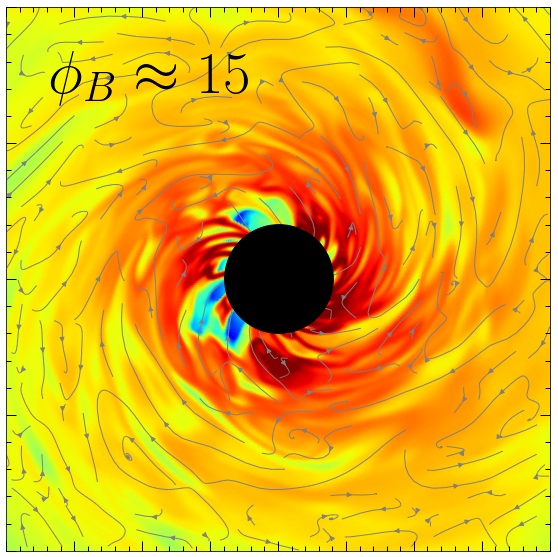}
& \includegraphics[width=0.14\linewidth]{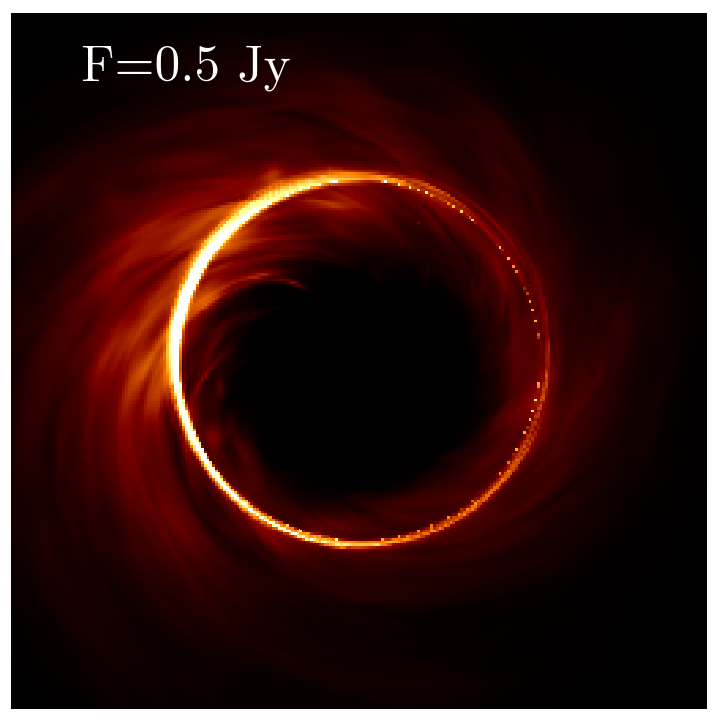}
& \includegraphics[width=0.14\linewidth]{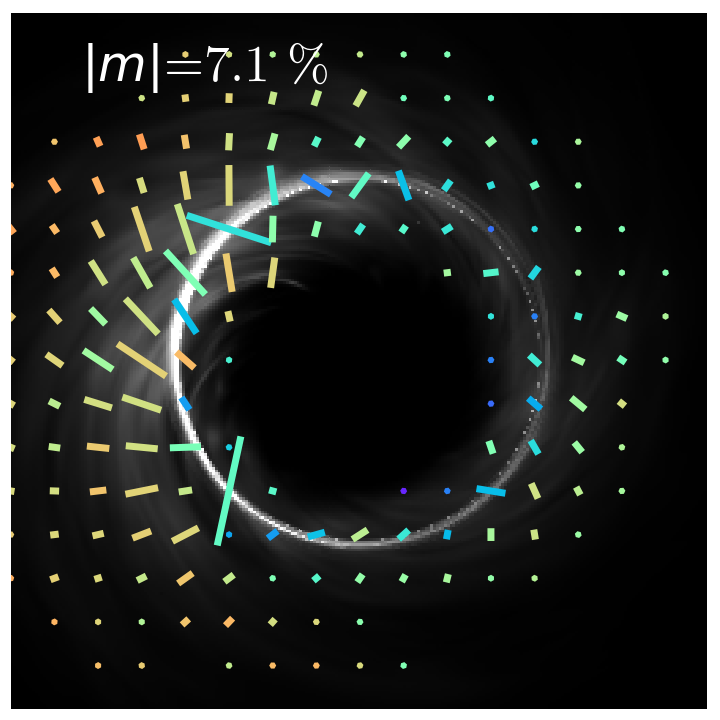}
& \includegraphics[width=0.14\linewidth]{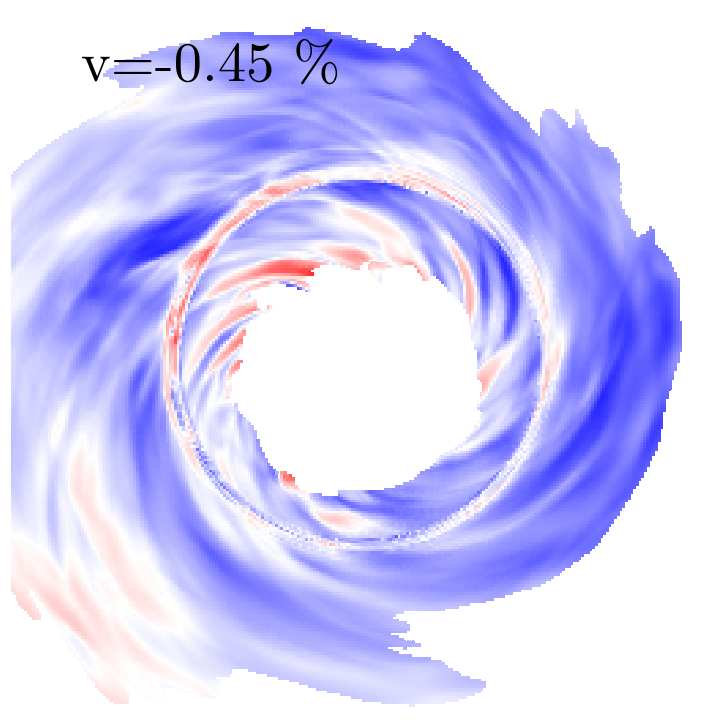} 
& \includegraphics[width=0.14\linewidth]{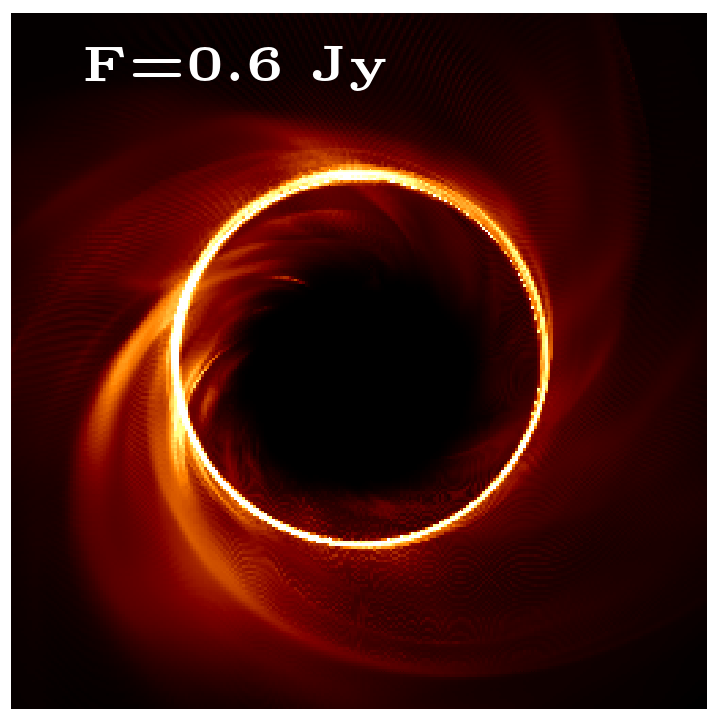}
& \includegraphics[width=0.14\linewidth]{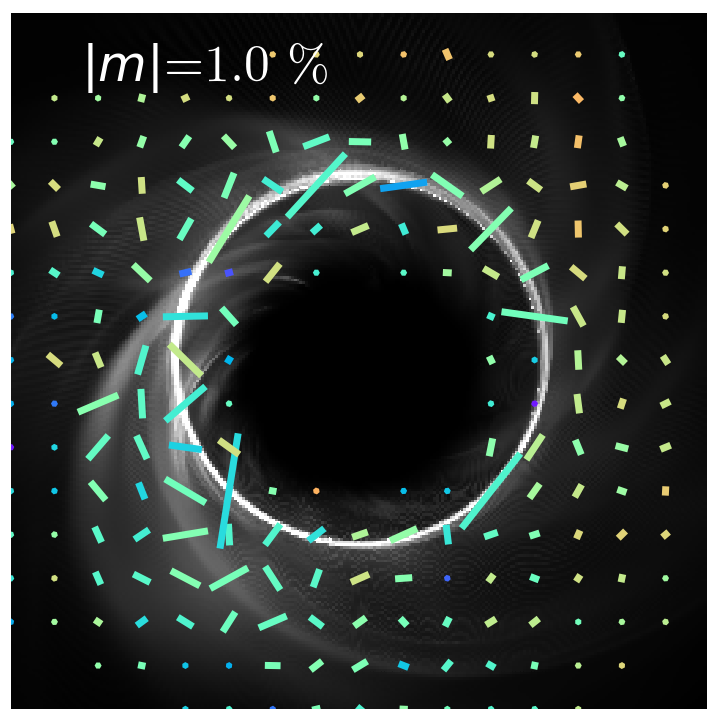}
& \includegraphics[width=0.14\linewidth]{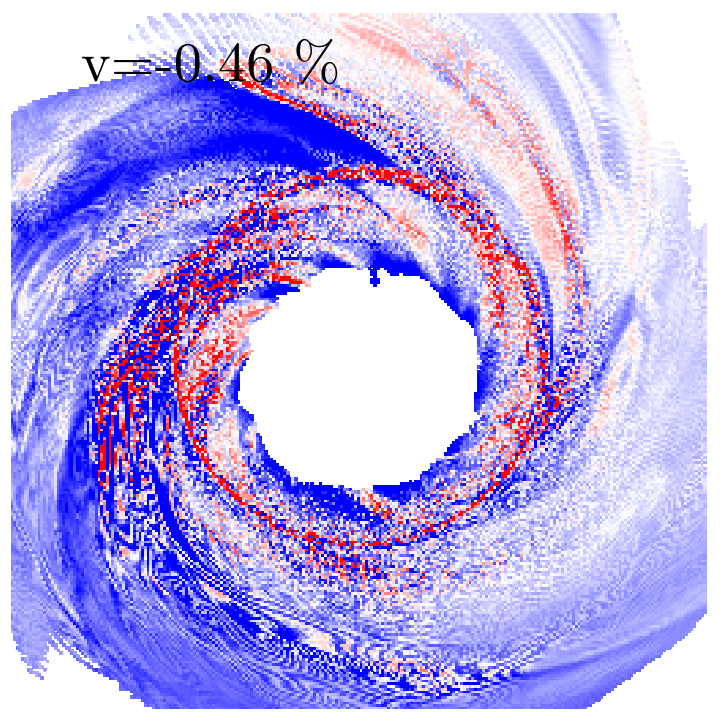}\\
\raisebox{0.06\linewidth}[0pt][0pt]{\rotatebox[origin=c]{90}{Model C}}\phantom{.}
\includegraphics[width=0.14\linewidth]{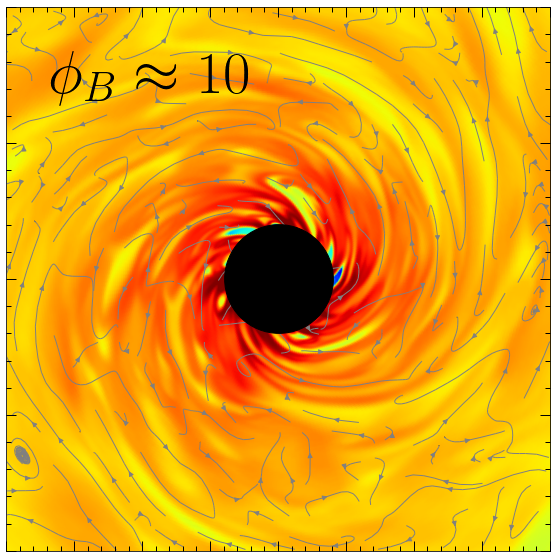} 
& \includegraphics[width=0.14\linewidth]{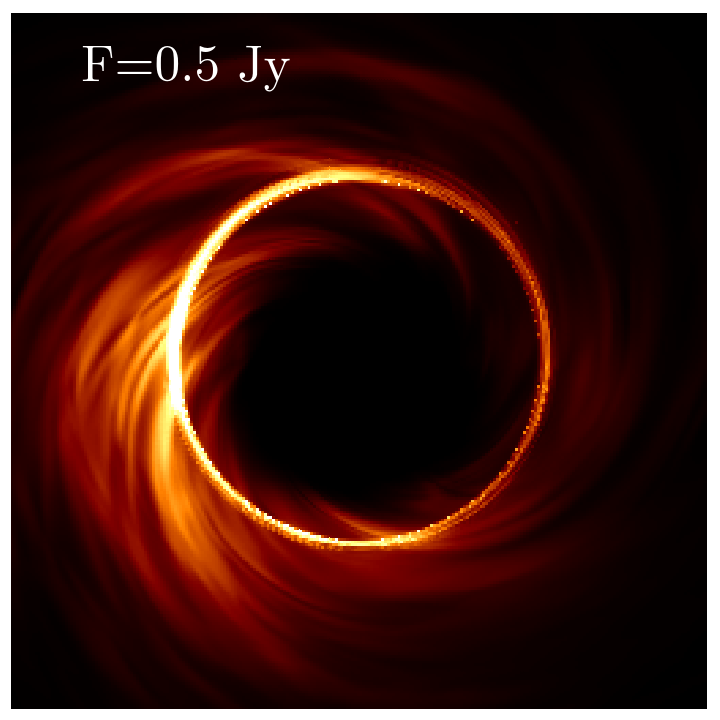}
& \includegraphics[width=0.14\linewidth]{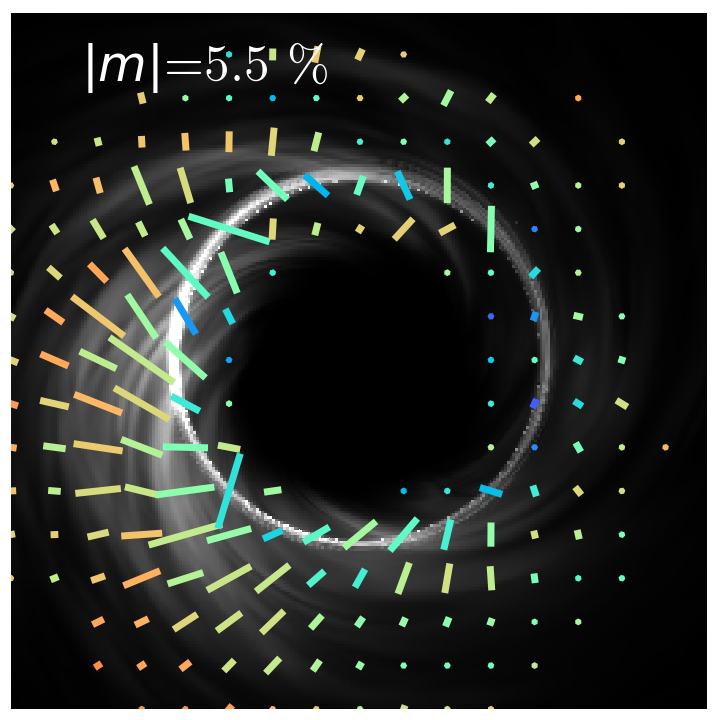}
& \includegraphics[width=0.14\linewidth]{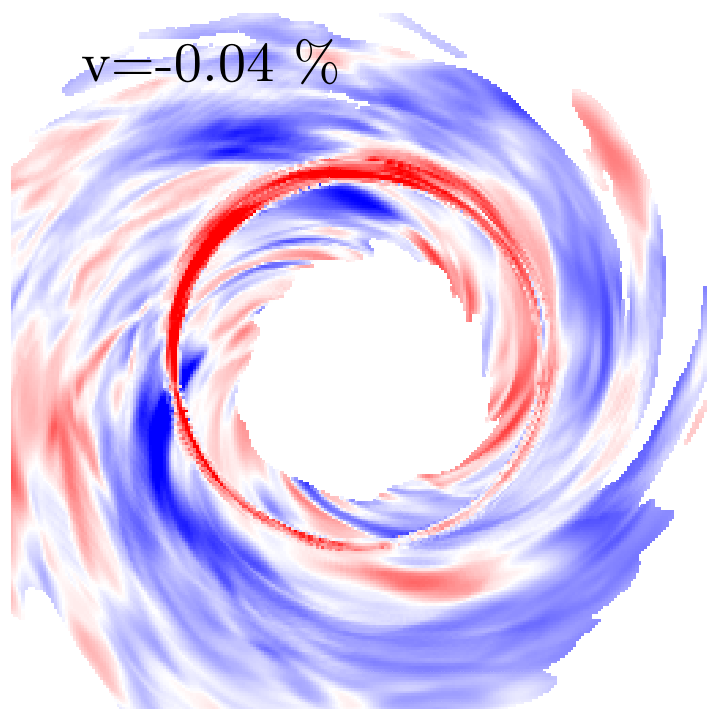} 
&\includegraphics[width=0.14\linewidth]{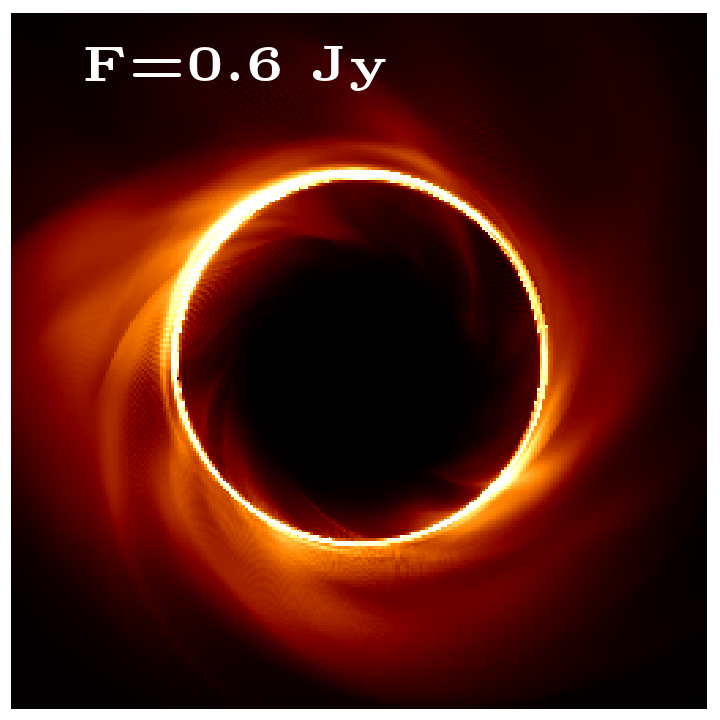}
& \includegraphics[width=0.14\linewidth]{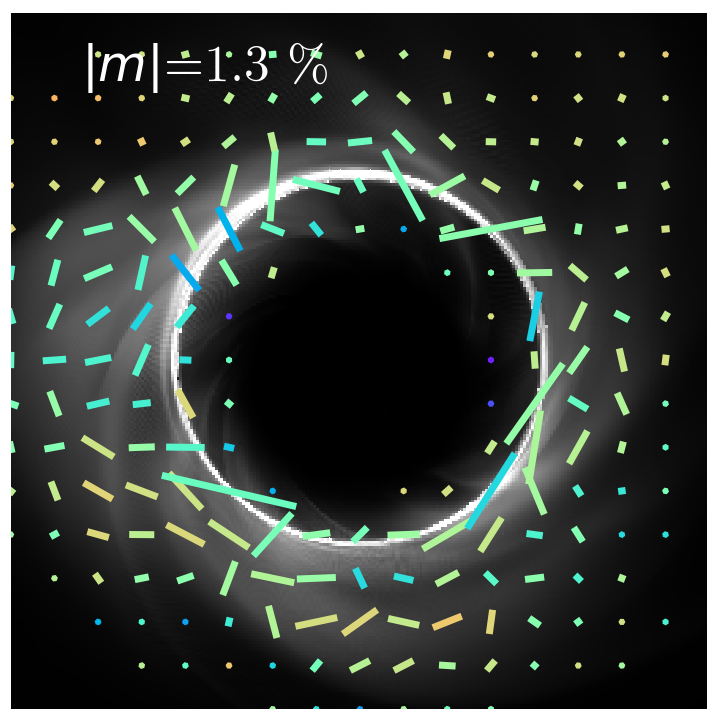}
& \includegraphics[width=0.14\linewidth]{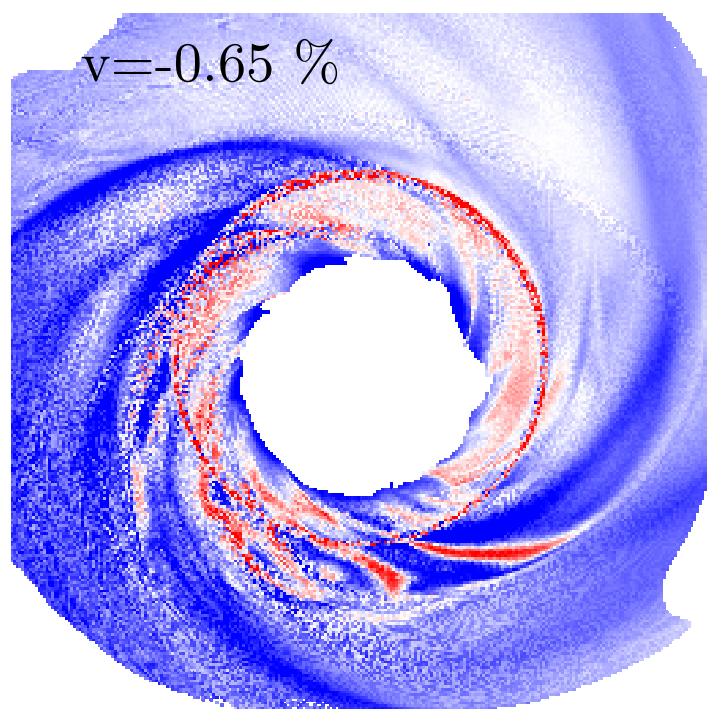}\\
\hspace{+0.02\linewidth}\includegraphics[width=0.13\linewidth,trim=0 28cm 0 0,clip]{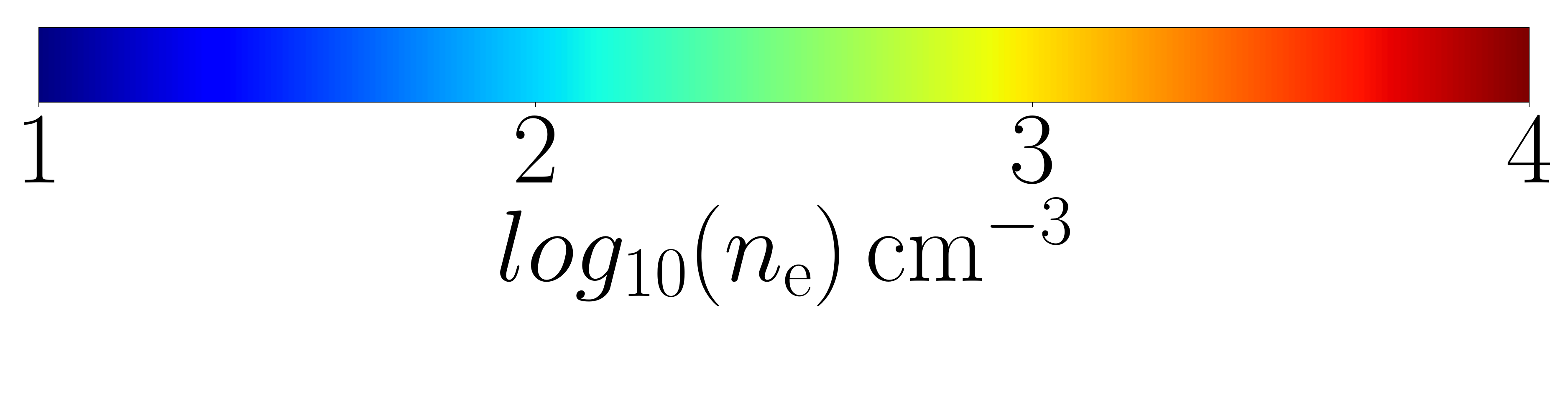} &
\includegraphics[width=0.13\linewidth,trim=0 18cm 0 0,clip]{plots/Model1a_beta30_a0/cbar_I_hor_2.png} &
\includegraphics[width=0.13\linewidth,trim=0 18cm 0 0,clip]{plots/Model1a_beta30_a0/cbar_LP_hor.png} &
\includegraphics[width=0.13\linewidth,trim=0 18cm 0 0, clip]{plots/Model1a_beta30_a0/cbar_CP.png} &
\includegraphics[width=0.13\linewidth,trim=0 18cm 0 0,clip]{plots/Model1a_beta30_a0/cbar_I_hor_2.png} &
\includegraphics[width=0.13\linewidth,trim=0 18cm 0 0,clip]{plots/Model1a_beta30_a0/cbar_LP_hor.png} &
\includegraphics[width=0.13\linewidth,trim=0 18cm 0 0, clip]{plots/Model1a_beta30_a0/cbar_CP.png} \\
\end{tabular}
\caption{Polarimetric images of models A, B and C (from top to bottom) for \rhigh=1 and 200 and nearly face-on viewing angle, $i=160 \deg$. The left-most panels show maps of plasma density distribution at the equatorial plane together with projected magnetic field lines. Second/fifth, third/sixth and fourth/seventh panels display the total intensity (Stokes $\mathcal{I}$), linear and circular polarimetric images. Images are displayed in the same manner as in Figure~\ref{fig:Rh}.}\label{fig:modelsABC}
\end{figure*}

\begin{figure}
\centering
\includegraphics[width=0.45\linewidth]{./plots/Model1_beta100_a0/ipole_datV.png}
\includegraphics[width=0.45\linewidth]{./plots/Model1_beta100_a0/ipole_test_datV.png}\\
\includegraphics[width=0.45\linewidth]{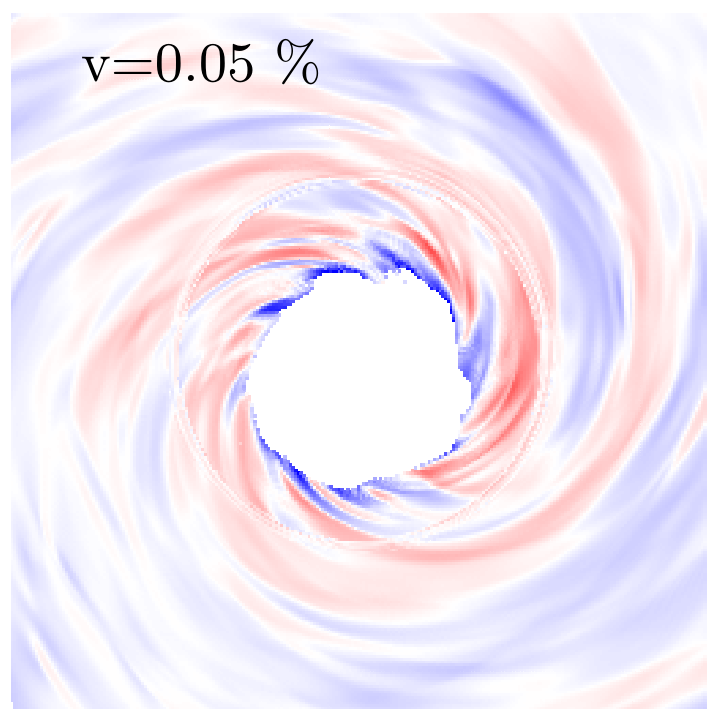}
\includegraphics[width=0.45\linewidth]{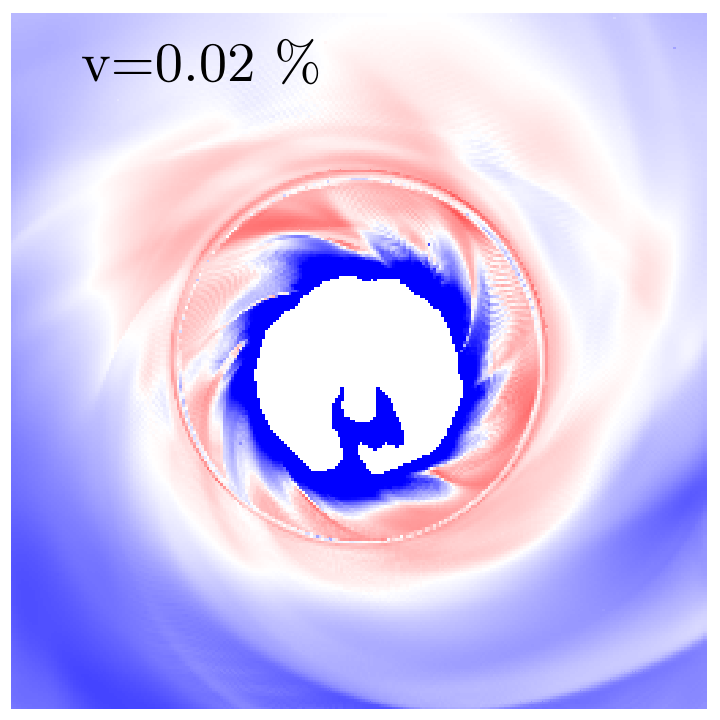}\\
 \includegraphics[width=0.44\linewidth,trim=0 18cm 0 0, clip]{plots/Model1a_beta30_a0/cbar_CP.png}
\includegraphics[width=0.44\linewidth,trim=0 18cm 0 0, clip]{plots/Model1a_beta30_a0/cbar_CP.png}
\caption{Circular polarization maps of model C1 (left panels)
  and C200 (right panels) by performing the full radiative transfer (top panels) and with $\rho_Q=0$ (i.e. no Faraday conversion, bottom panels). The inversion of the ring polarity disappears when Faraday conversion effects are not included.}\label{fig:C_cp}
\end{figure}

Circular polarization produced in the Faraday thicker models
has mostly one handedness and handedness inversion in the
  lensing ring. We check if similar effects are present in different realization of MAD models (B and C). In Figure~\ref{fig:modelsABC} we present examples of density maps, total intensity and polarimetric images of models A, B and C, all at $t\approx19000$M. At this time moment in each model, while accretion rate is similar, a different magnetic flux is accumulated on the horizon causing the dynamics of the flow near the horizon to be distinct. The stronger magnetic barrier, created by extended magnetosphere in the polar regions (visible in Fig~\ref{fig:models_beta30}, top left panel), makes the inner disk in equatorial plane more inhomogeneous in azimuthal direction. Evidently, in models A and B the streams of matter have to overpass stronger magnetic barrier compared to model C. While the streams in model A and B are not purely radial but slightly bend, the structure of density in C model indicates stronger toroidal component of magnetic fields. In Figure~\ref{fig:modelsABC}, we show the corresponding polarimetric maps of models A, B and C for \rhigh=1 and \rhigh=200. The model B and C behaviour as a function of \rhigh is similar to model A. Linear polarization maps become disorganized with increasing \rhigh parameter. 
Circular polarization maps of model B and C display similar features to their counterpart model A, as expected. The net circular polarization is negative and small but increases with \rhigh parameter. All models show polarity inversion in the lensing ring when \rhigh=200. Notice that in model C the ring inversion for \rhigh=1.  

In Figure~\ref{fig:C_cp} we show circular polarization maps of model C1 and C200 (model C with \rhigh=1 and 200 in analogy to models A1 and A200). These two snapshots have similar conversion thickness ($\left<\tau_{FC}\right>_I=$ 0.1 and 0.17 in models C1 and C200, respectively) but 
different rotation thickness ($\left<\tau_{FR}\right>_I=$ 0.49 and 322 in models C1 and C200, respectively).
Again, Faraday conversion play a key role in the production of circular polarization in the direct and in the lensed emission. 
The polarity inversion effect in the lensing ring seems to be robust in all three realizations of MADs as long as Faraday conversion thickness is significant.  

\subsection{Polarization variability vs underlying magnetic fields and plasma conditions}\label{sec:3.3}

\begin{figure}
\centering
\includegraphics[width=0.99\linewidth]{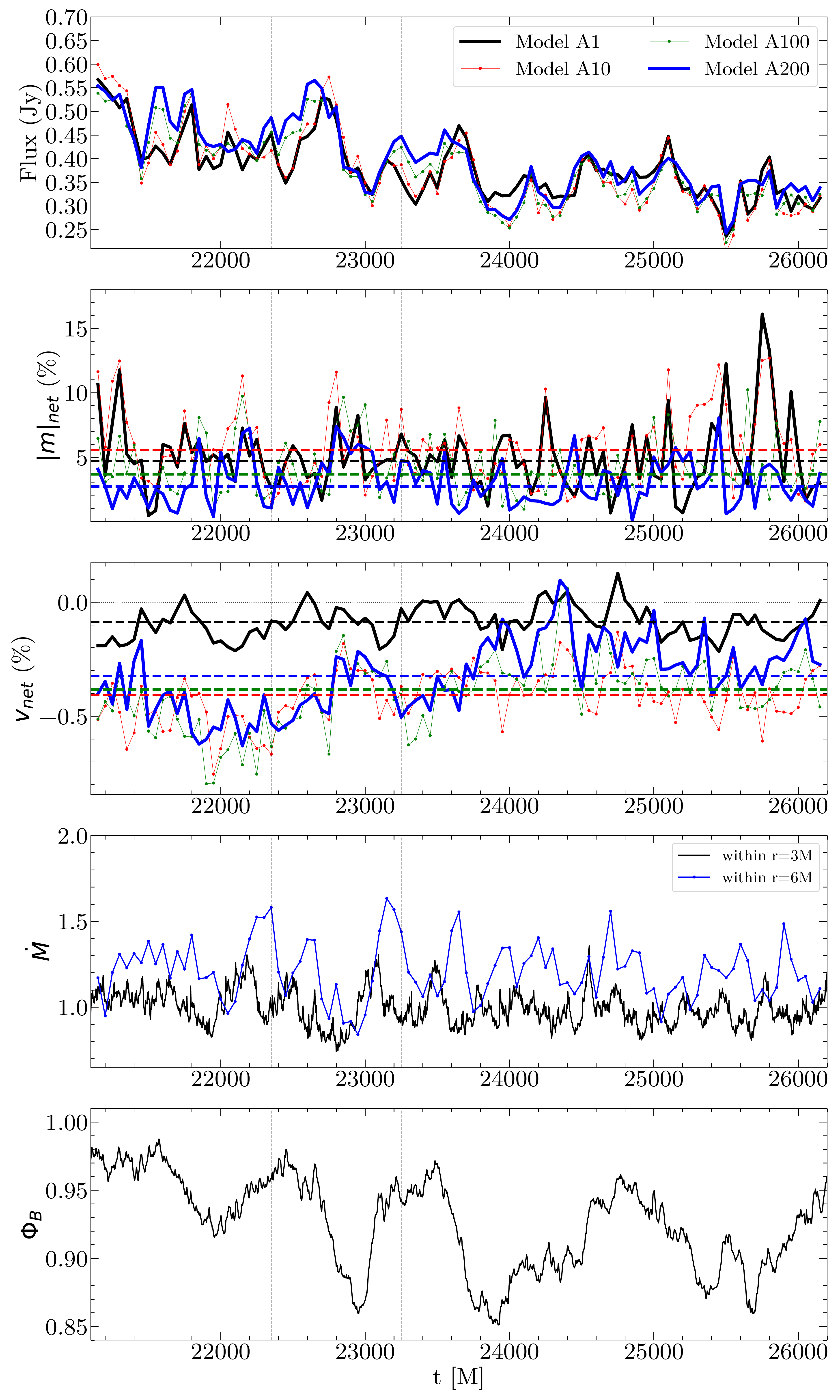}
\caption{Time variability in
models A1, A10, A100, A200 (models with different electron temperatures are
color coded). Three top panels display time variations in total flux, net
fractional linear polarization $|m|_{\rm net}$, and the net fractional
circular polarization $v_{\rm net}$ for observer at i=160 deg. In panels with
$|m|_{\rm net}$ and $v_{\rm net}$ the horizontal lines mark corresponding median polarizations. Two bottom panels display the normalized to unity mass accretion rate through the event horizon and the magnetic field flux accumulated at the event horizon (in code units).  Two mass accretion rates are averaged within 3M and within 6M. The thin vertical lines mark time moments shown in Figures~\ref{fig:poincare1} and ~\ref{fig:poincare2}.}\label{fig:lp_cp}
\end{figure}

\begin{figure}
\centering
\includegraphics[width=0.45\linewidth]{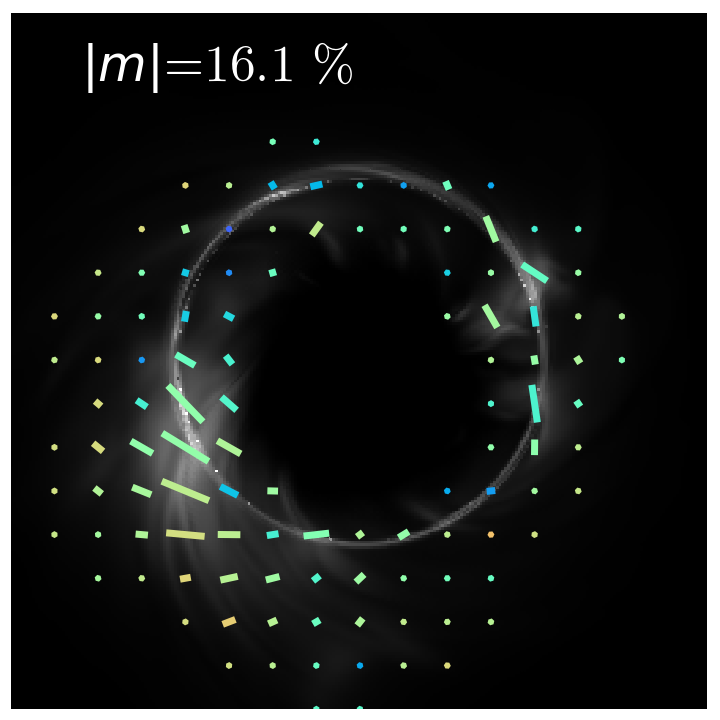}
\includegraphics[width=0.45\linewidth]{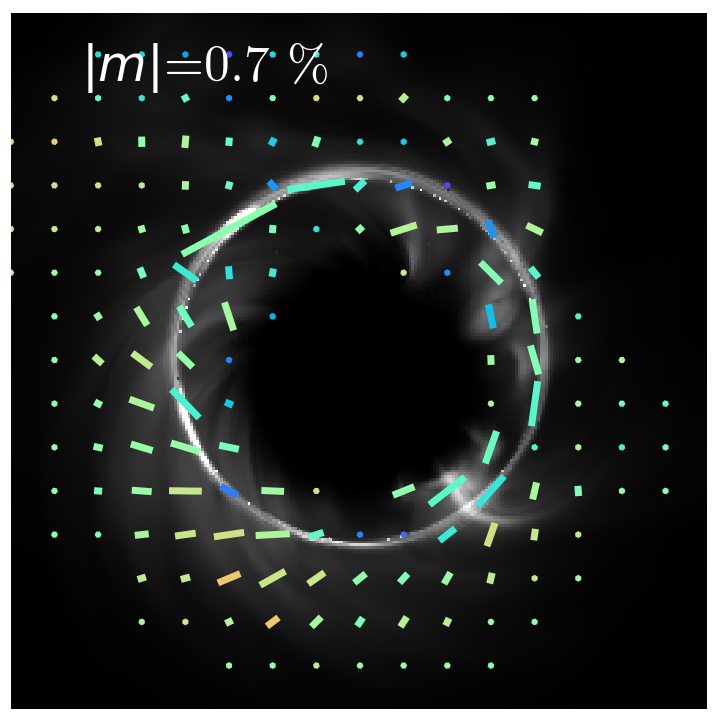}\\
\includegraphics[width=0.45\linewidth]{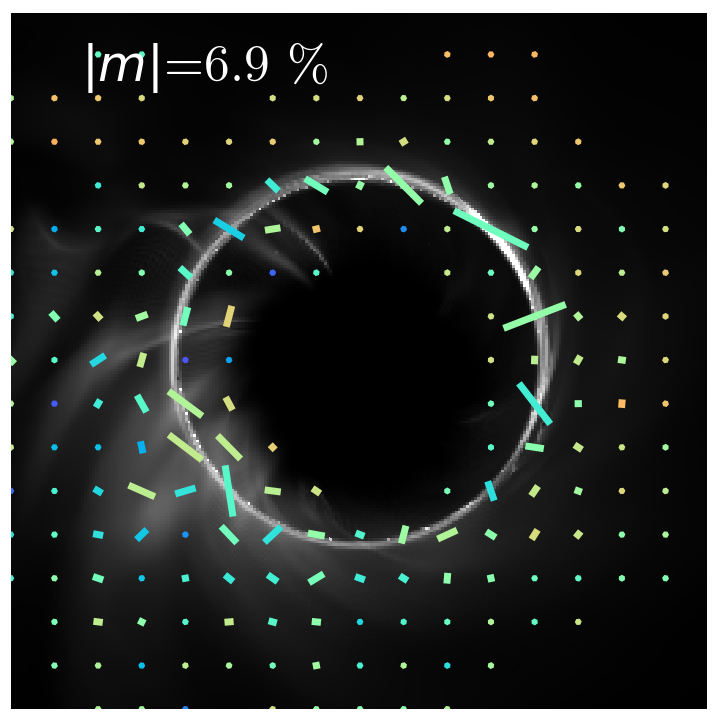}
\includegraphics[width=0.45\linewidth]{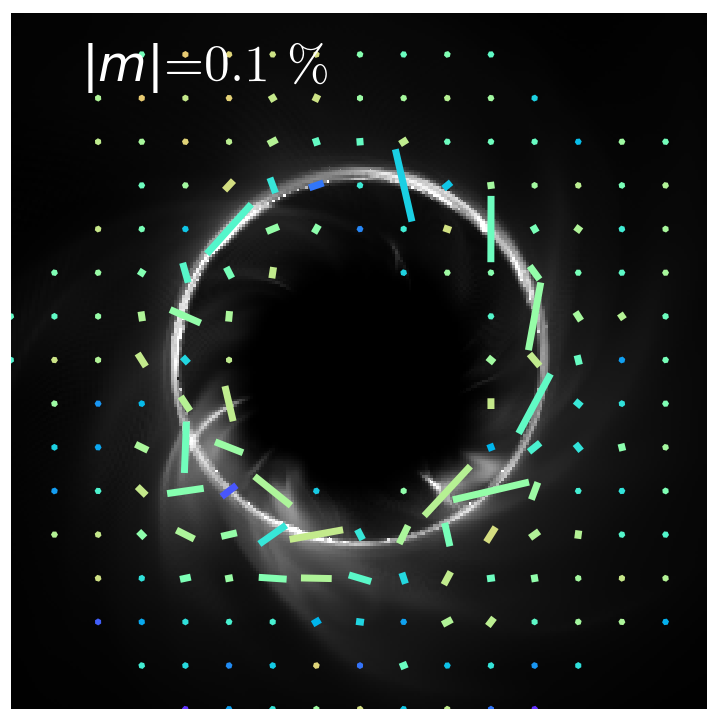}\\
\includegraphics[width=0.44\linewidth,trim=0 18cm 0 0,clip]{plots/Model1a_beta30_a0/cbar_LP_hor.png}
\includegraphics[width=0.44\linewidth,trim=0 18cm 0 0,clip]{plots/Model1a_beta30_a0/cbar_LP_hor.png}
\caption{Examples of linear polarimetric images of model A1 (top panels) and
  A200 (lower panels)
  which display relatively high (left panels) and low (right panels) net linear polarizations. 
  Total intensity and polarizations are displayed in the same manner as in
  Figure~\ref{fig:Rh}. Snapshots times are: $t=25750$ M (top left), $t=25200$ M (top right), $t=22150$ M (bottom left) and $t=24850$ M (bottom right).}\label{fig:high_low_lp_pol}
\end{figure}

\begin{figure}
\centering
\includegraphics[width=0.45\linewidth]{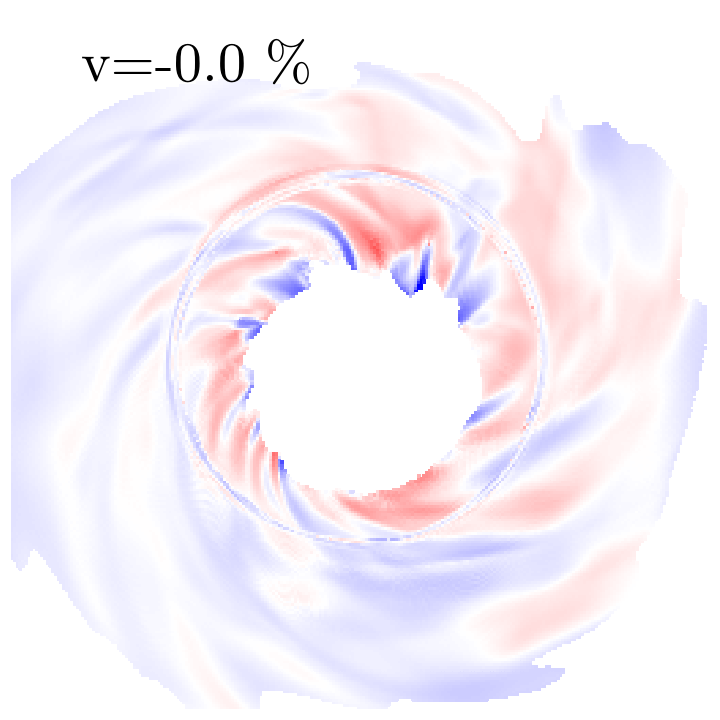}
\includegraphics[width=0.45\linewidth]{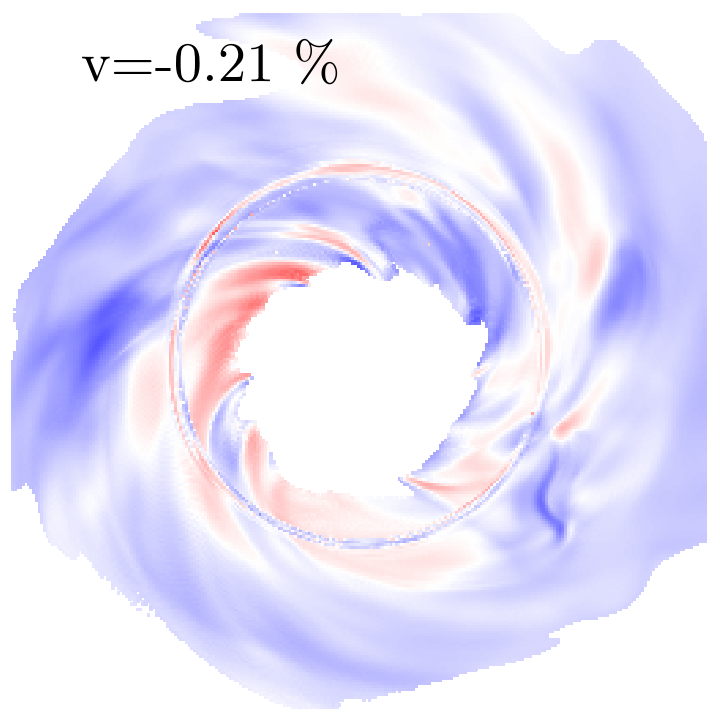}\\
\includegraphics[width=0.45\linewidth]{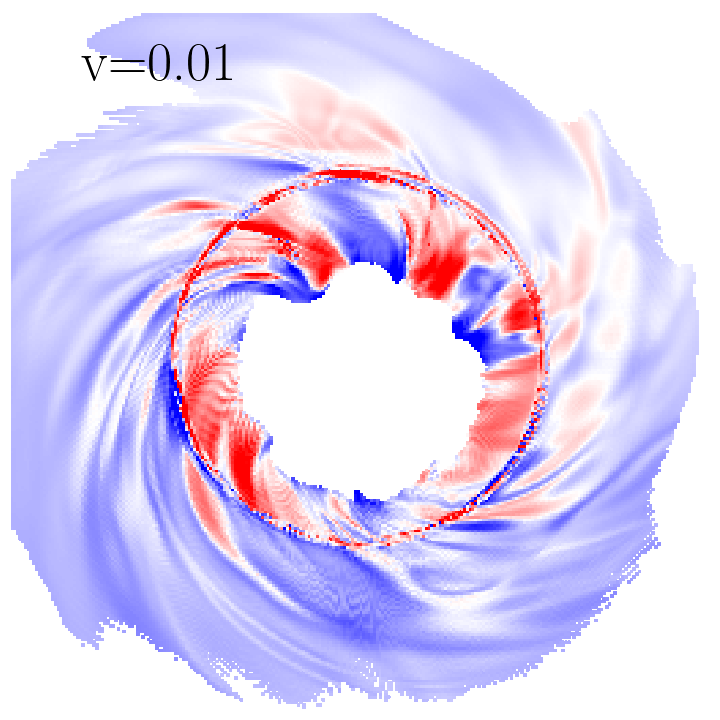}
\includegraphics[width=0.45\linewidth]{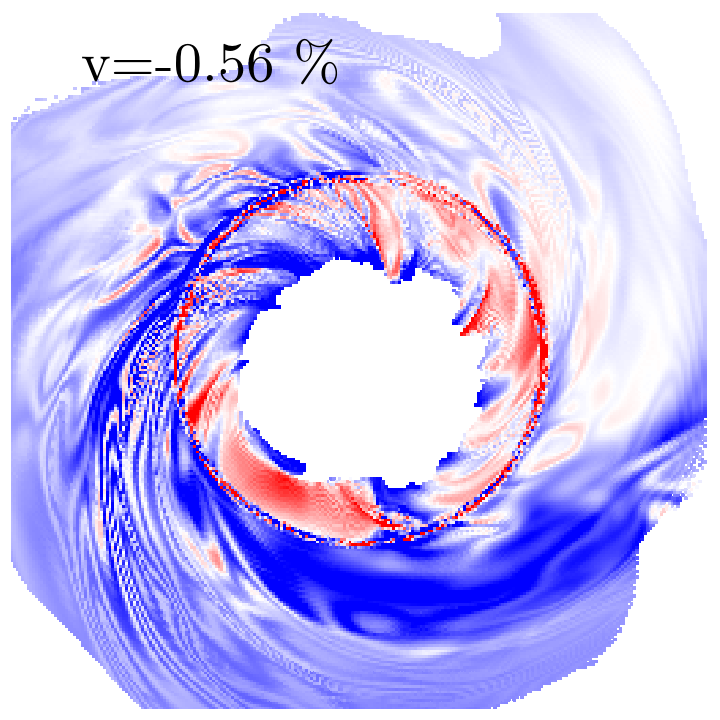}\\
\includegraphics[width=0.44\linewidth,trim=0 18cm 0 0,clip]{plots/Model1a_beta30_a0/cbar_CP.png}
\includegraphics[width=0.44\linewidth,trim=0 18cm 0 0,clip]{plots/Model1a_beta30_a0/cbar_CP.png}
\caption{Examples of circular polarimetric images of model A1 (top panels) and
  model A200 (bottom panels) which display relatively high (left
    panels) and low (right panels) net circular polarizations.
  The circular polarizations are displayed in the same manner as in
  Figure~\ref{fig:Rh}. Snapshots times are: $t=24300$ M (top left), $t=23100$
  M (top right), $t=24350$ M (bottom left) and $t=22400$ M (bottom right).
  Notice that snapshots chosen for display here are different from those shown in Figure~\ref{fig:high_low_lp_pol}.}\label{fig:high_low_cp_pol}
\end{figure}

Does the variability of the linear and circular polarization reflect variability in the underlying magnetic field or mass accretion rate?
To further investigate the physical origin of circular polarization and its variations as a function of time, we analyse a longer sequence of model A snapshots ($\Delta t$=5000M) corresponding to 5~yrs of M87 core evolution. 

In Figure~\ref{fig:lp_cp} we show models A1 through A200 total intensity
fluxes, net fractional linear ($|m|_{\rm net}$) and net circular ($v_{\rm
  net}$) polarization integrated over entire images, mass accretion rate and
magnetic flux threading horizon as a function of time. The latter two are the
global physical parameters of the model that one is interested to constrain
using observables so they are shown for a comparison. Horizontal lines in
panels with fractional polarizations mark median values of fractional
polarizations. Results are shown for runs with viewing angle of 160 deg but
the images at 20 deg display similar properties. In Table~\ref{tab:param} we
report ranges of net fractional linear and circular polarizations integrated
over all images within the considered time period for both viewing
angles. Models with higher \rhigh parameter tend to display lower net linear
polarization and higher (in the absolute sense) net circular polarization. In
our runs, the magnitude of net circular polarization remains rather small,
$v_{\rm net}<1\%$, with median value of $v_{\rm net}=-0.05 \%$ in model A1 and
median value of $v_{\rm net} \approx -0.4\%$ in A10, A100, and A200 runs. For
all displayed models the sign of the net circular polarization is most of the
time constant for the assumed viewing angle. The sign flips of $v_{\rm net}$
occur slightly more frequently in model A1 in comparison to other models
(where other models have higher Faraday thickness for rotations and conversions). 

In Figure~\ref{fig:high_low_lp_pol} we show examples of linear polarization images of two
models (with thoroughly different Faraday thicknesses) A1 and A200 ($i=160
\deg$) when $|m|_{\rm net}$ is relatively high (well above the median value)
or extremely low ($|m|_{\rm net}<1\%$). All images display complex
polarimetric structures due to interplay of different effects. In frames with
high net linear polarization larger patches of highly polarized emission
regions are present. Images with low net linear polarization either display a
few smaller regions of organized polarization in different directions or are
scrambled by Faraday rotation effect, both resulting in low $|m|_{\rm net}$. 
In models with higher Faraday thickness (A200) the patches with coherent high
polarization are also possible and so the high net polarization is. In
Figure~\ref{fig:high_low_cp_pol} we show a few examples of model A1 and A200
circular polarization maps with relatively high and low net circular
polarizations. Those with high absolute $v_{\rm net}$ often show stronger features.

The connections between the observed linear/circular polarization and physical parameters such as mass accretion rate and magnetic flux at the horizon are not immediately obvious (see Figure~\ref{fig:lp_cp}, bottom panels). 
We have examined all images in our time sequence and no characteristic linear polarization pattern associated with enhanced accretion rate or magnetic field flux near the event horizon was found. Instead we found examples of model A1 and A200 images with similar linear polarization patterns and net polarizations but accretion rates different by a factor of two. 

\begin{figure*}
\def\arraystretch{0.0}
\centering
\setlength{\tabcolsep}{0pt}
\begin{tabular}{cc}
 &\multirow{4}{*}{\includegraphics[width=0.51\linewidth]{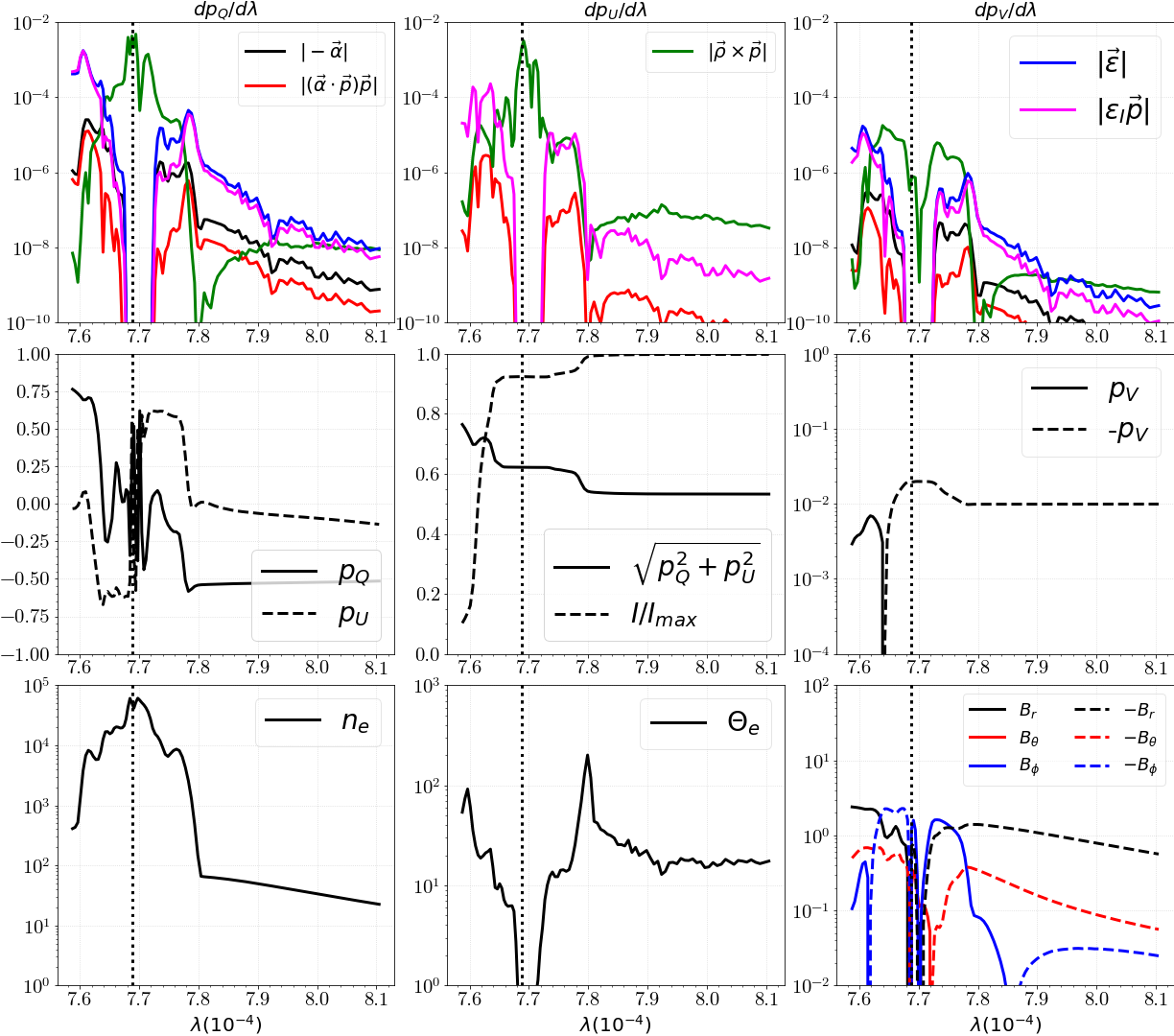}} \\
\includegraphics[width=0.19\linewidth]{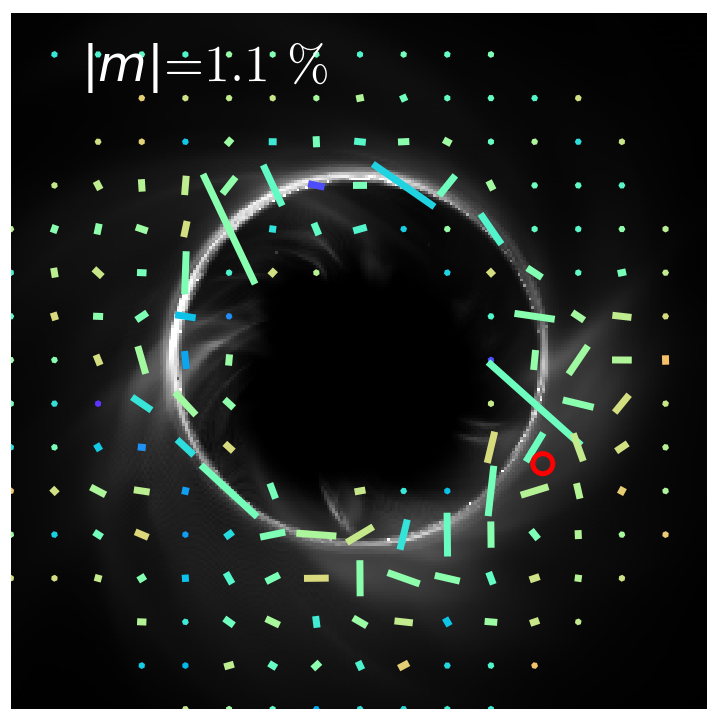} &\\
\includegraphics[width=0.18\linewidth,trim=0 18cm 0 0,clip]{plots/Model1a_beta30_a0/cbar_LP_hor.png} &\\
\includegraphics[width=0.19\linewidth]{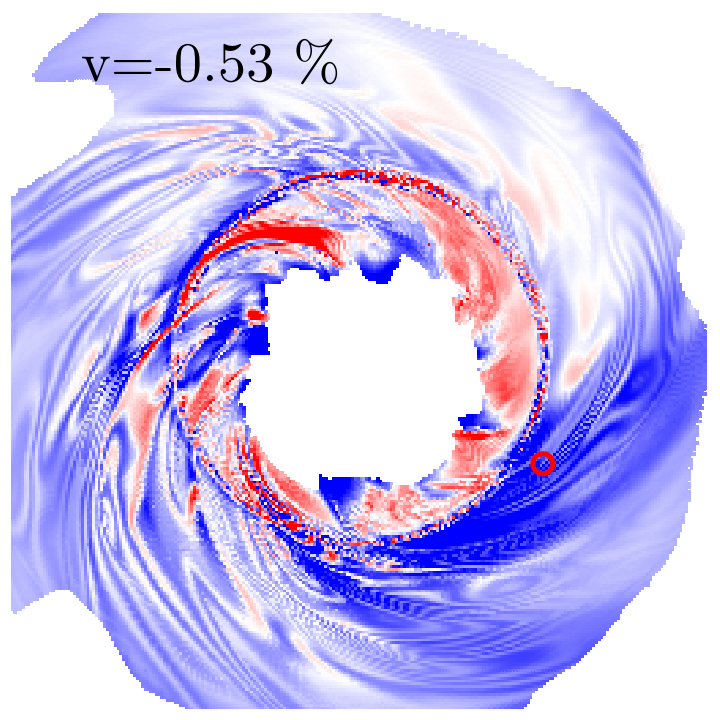} &\\
\includegraphics[width=0.18\linewidth,trim=0 18cm 0 0,clip]{plots/Model1a_beta30_a0/cbar_CP.png} &\\
\end{tabular}
\caption{
  Polarimetric images (left panels) and polarization diagnostic plot (right
  panels) showing profiles of various quantities along a single geodesic line
  marked as a red circle in the left panel. We display model A200 snapshot
  that has relatively high circular polarization. In the right panels, we
  follow integration details from the geodesics point where the total
  intensity ${\mathcal I}$ is equal 10 per cent of the ray final intensity until
  radius of 8M (within a region where most of the radiation is
  generated). Vertical dotted line marks passage of light ray through the
  equatorial plane. First row: Equation~\ref{eq:delin2} terms (i)
  through (v) for each of the Stokes parameters ${\mathcal Q}$, ${\mathcal
  U}$, and ${\mathcal V}$ (shown in columns from left to right; notice that
   labels displayed in each panel are valid for all other panels in this
   row). Second row: fractional polarizations in all Stokes parameters and
   normalized total intensity.
  Third row: plasma number density ($cm^{-3}$), dimensionless
  electron temperature ($\Theta_{\rm e}=k_{\rm B} T_{\rm e} / m_{\rm e} c^2$)
  and the three magnetic field components (measured in the coordinate frame
  in units of Gauss). Here one can notice that the mid-plane of the disk is
  dominated by Faraday effect. Linear polarization is mostly produced away
  from the midplane and scrambled by Faraday rotation. Majority of circular
  polarization is produced below the disk mid-plane where toroidal-radial
  component of magnetic fields dominates.}\label{fig:poincare1}
\end{figure*}

\begin{figure*}
\def\arraystretch{0.0}
\centering
\setlength{\tabcolsep}{0pt}
\begin{tabular}{cc}
 &\multirow{3}{*}{\includegraphics[width=0.51\linewidth]{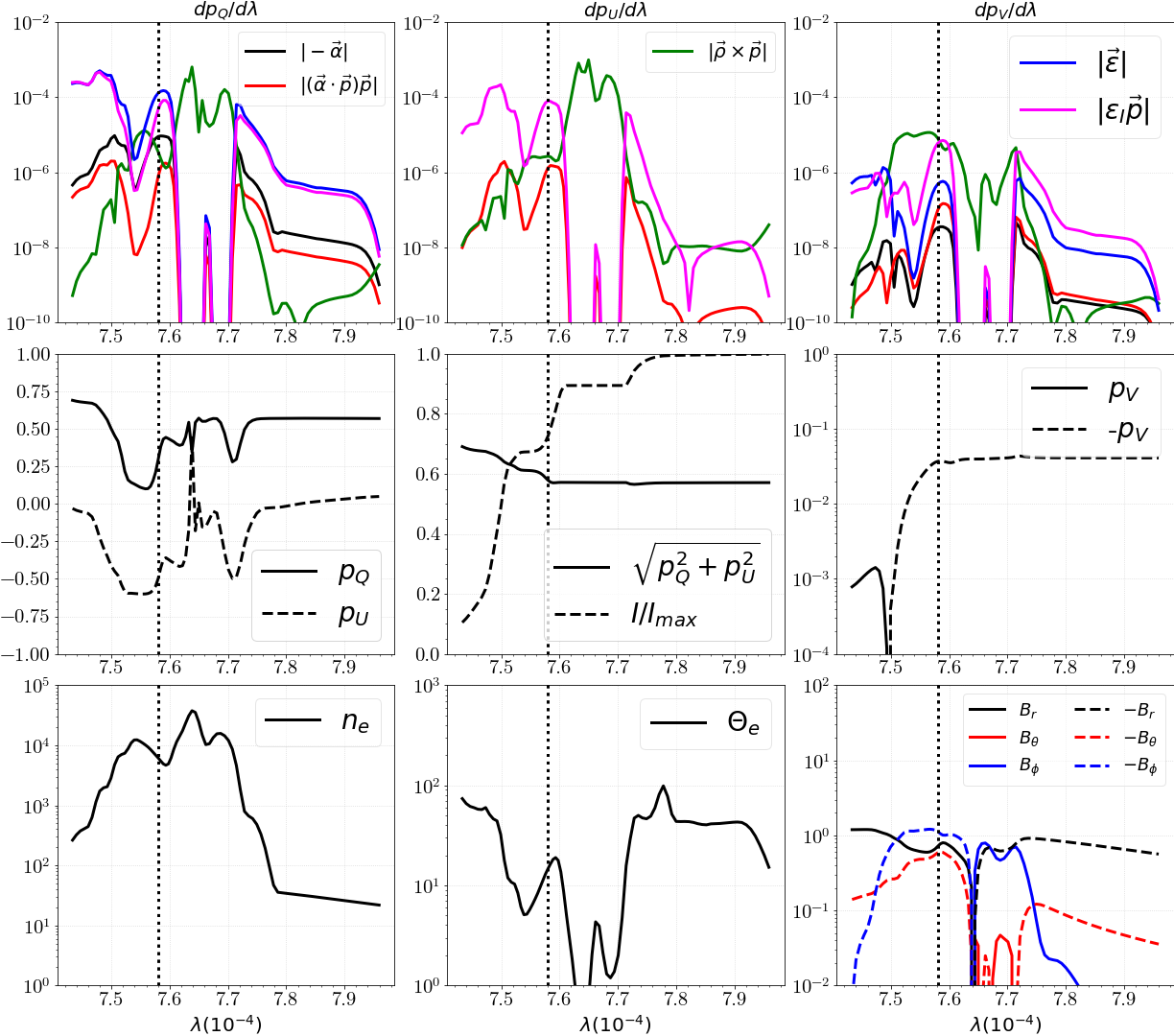}} \\
\includegraphics[width=0.18\linewidth]{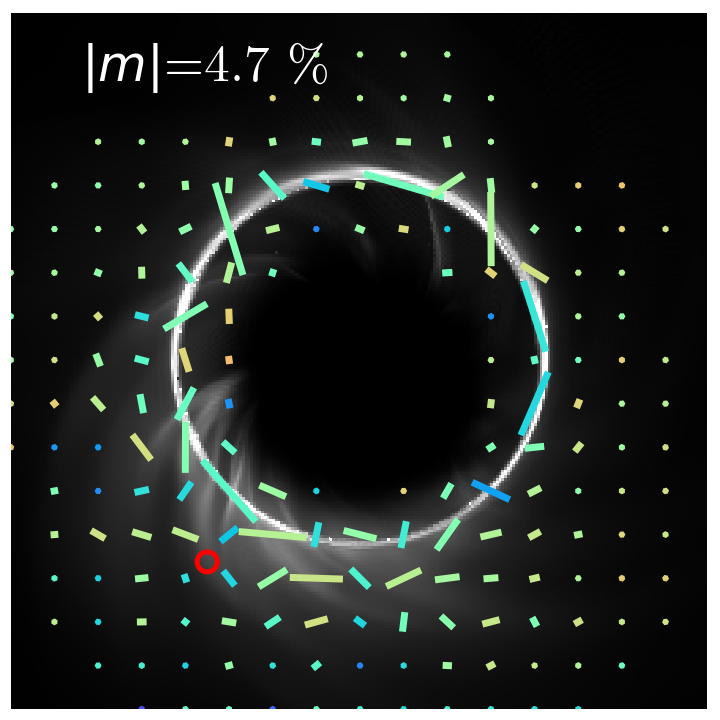} &\\
\includegraphics[width=0.18\linewidth,trim=0 18cm 0 0,clip]{plots/Model1a_beta30_a0/cbar_LP_hor.png} &\\
\includegraphics[width=0.19\linewidth]{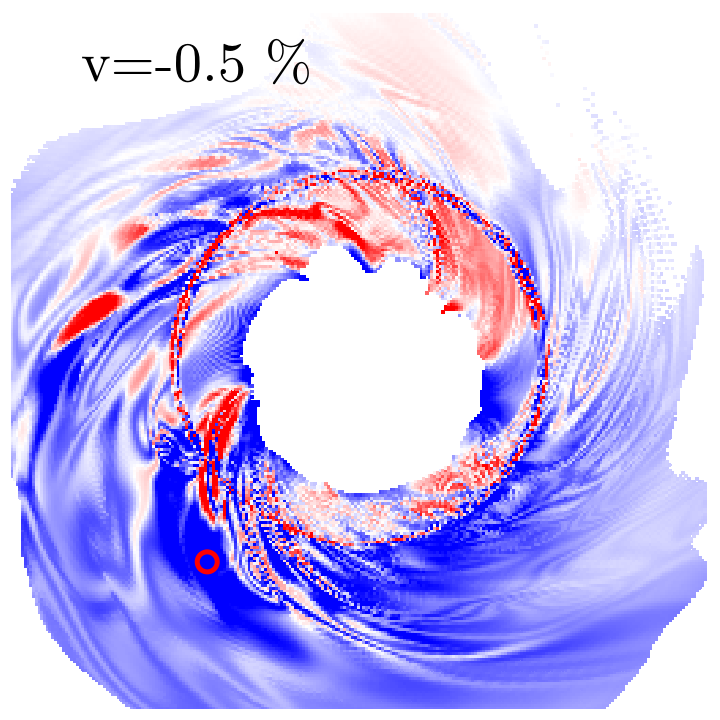} &\\
\includegraphics[width=0.18\linewidth,trim=0 18cm 0 0,clip]{plots/Model1a_beta30_a0/cbar_CP.png} &\\
\end{tabular}
\caption{Polarimetric images (left panels) and polarization diagnostic plot (right panels) for another highly circularly polarized snapshot of model A200. All panels display various quantities in the same fashion as in Figure~\ref{fig:poincare1}. Here the polarization is formed in a similar way as in case shown in Figure~\ref{fig:poincare1}.}\label{fig:poincare2}
\end{figure*}

Circular  polarization  maps, with fractional polarizations smaller compared to linear polarization, are likely more sensitive  to  changes  in  the  plasma  and  magnetic  field  conditions. Using method outlined in Section~\ref{sec:probe}, we examine the origin of circular (and  linear) polarization in two example snapshots of models A200 which show high absolute circular polarizations (these times are $t$=22350M and $t$=23250M marked in Figure~\ref{fig:lp_cp} with thin vertical lines). 

In Figures~\ref{fig:poincare1}-\ref{fig:poincare2} we show the chosen polarimetric images (left panels) and various quantities along a single light ray towards the highly circularly polarized emission region (marked with a circle in the right panels).
Figures~\ref{fig:poincare1}-\ref{fig:poincare2} (panels with $dp_Q / d\lambda$,$dp_U / d\lambda$, $dp_V / d\lambda$) show that the linear polarization along the chosen ray is produced by intrinsic emission away from the equatorial plane and strongly modified by Faraday rotation in the flow mid-plane (position of the equator is marked as vertical dotted line in each panel) where temperature of plasma is cooler (even sub-relativistic). The self-absorption effects for linear and circular polarization are negligible.
Circular polarization is predominantly produced in the conversion process in
the disk material where the toroidal-radial magnetic field components is
stronger (above 1 Gauss). The latter is observed in both examples. In
Figure~\ref{fig:poincare1} the sign of Stokes ${\mathcal U}$ changes sign at
the equator due to Faraday rotation, hence the Stokes ${\mathcal V}$ slightly
decreases above the equatorial plane.
This finding is consistent with the results presented in Section~\ref{sec:3.2}.

\section{Summary}\label{sec:summary}

We have examined simulated polarimetric images of accreting black hole produced by ray-tracing high resolution simulation of magnetically dominated accretion flow. We arrive at the following general conclusions:

\begin{itemize}
\item The magnitude of the resolved $|m|$ and $v$ in black hole images removes degeneracy between parameters of the models for electron temperatures. Models with lower/higher \rhigh parameter display 
stronger/weaker $|m|$ and weaker/stronger $v$. This finding is consistent with the idea of \citet{moscibrodzka:2017} where we pointed out that parameter \rhigh controls Faraday depth of the models which may cause linear depolarization and lead to enhanced circular polarization. 
\item We do not attempt to fit our models to M87 black hole observations, but rather use the source as an example to point out observational signatures of dynamical magnetic fields around this type of objects when accretion flow is in MAD state. Nevertheless, models with low net linear polarization ($|m|_{\rm net} \sim 1-3.7\%$), or rather high \rhigh parameter, are favored by M87 observations (\citealt{kuo:2014}, \citealt{paper7:2021}). 
Using model a scoring procedure introduced in \citealt{paper8:2021}, we have checked that our model A200 does contain snapshots which are consistent with EHT polarimetric observations of M87 core.
If indeed M87 black hole is surrounded by MAD with \rhigh=200 parameter then
the expected resolved circular polarization around the black hole could be as
large as a few per cent. However, our prediction is inconsistent
  with results of \citealt{tsunetoe:2020} who based on their best-bet model predicted circular polarization
  of the M87 core to be larger than 10 per cent. The difference may be due to
  different GRMHD setups (2D vs 3D models studied here or different model magnetization).
\item Our findings about the circular polarization handedness are consisitent with the idea of
  \citet{enslin:2003} who associates it with the
  direction of rotation. If this is indeed the case, M87 core should have
  negative/positive circular polarization in case if the
  source is rotating in the clock-wise/counter-clock-wise direction on the sky. 
\item We discover the polarity inversion in circular polarimetric images of black hole shadow. The polarity inversion in the lensing ring is also associated with the stronger Faraday conversion in regions on the nearside of the black hole. Detection of the polarity inversion in Stokes ${\mathcal V}$ in emission region in M87 could provide evidence that the ring visible in the EHT observations \citep{ehtIV:2019} is produced near event horizon rather than being a feature of the forward jet (see e.g. models in Appendix A in \citealt{davelaar:2019} or \citealt{kawashima:2020}). The frequency of $f=$230\,GHz may be an optimal frequency for such observation because Faraday conversion decreases with observing frequency as $f^{-3}$ which will decrease fractional circular polarization at higher frequencies (see example images in Appendix~\ref{app:freq}). At $f$<230\,GHz, on the other hand, the near horizon emission is rather obscured by the synchrotron photosphere so no such inversion effects would be visible. 
\item We find that variations in the circular polarization maps reflect the variations in the underlying magnetic field strength around it. These variations are best highlighted in models with high Faraday conversion thickness. 
The Faraday conversion strength is proportional to $B_\perp$ component along
the line of sight. Hence for our low viewing angles (nearly face-on) Faraday
rotation would be enhanced in regions where toroidal (or other component of
field that is perpendicular to the line of sight in a given setup) field is
significant. The Faraday conversion seems to be highlighting polarity of
magnetic fields on one side of the disk where polarity is more or less the
same so it results in increased  $v_{\rm net}$ and images with uniform
polarity. In contrast to resolved quantities, the $|m|_{\rm net}$ and $v_{\rm net}$ do not correlate with physical parameters of the accreting system such as accretion rate or magnetic field flux at the black hole event horizon.
\item Finally, the effects described in this paper may not be universal to all possible accretion models and viewing angles. Investigating circular polarization magnitude and variations in different accretion models (e.g. Standard and Normal Evolution instead of MADs) around black hole with different spin values or models with non-thermal electrons are all possible directions for the future research. 
\end{itemize}

\section*{Acknowledgements}
The authors thank Daniel Palumbo, Alejandra Jimenez-Rosales, 
Hector Olivares and Jesse Vos for useful comments and suggestions which significantly improved the quality of the manuscript.
MM acknowledges support by the NWO grant no. OCENW.KLEIN.113. AJ acknowledges support by the grant no. 2019/35/B/ST9/04000 from the Polish National
Science Center, and using computational resources of the Warsaw ICM through
grant Gb79-9, and the PL-Grid through the grant plggrb4.
The paper uses public code {\tt ipole} available at www.github.com/moscibrodzka/ipole \citep{moscibrodzka:2018}.

\section*{Data availability}
The data underlying this article will be shared on reasonable request to the corresponding author.


\appendix

\section{Slow-light vs. fast-light polarized radiative transfer model}\label{app:fastlight}

\begin{figure*}
\centering
\includegraphics[width=0.32\linewidth]{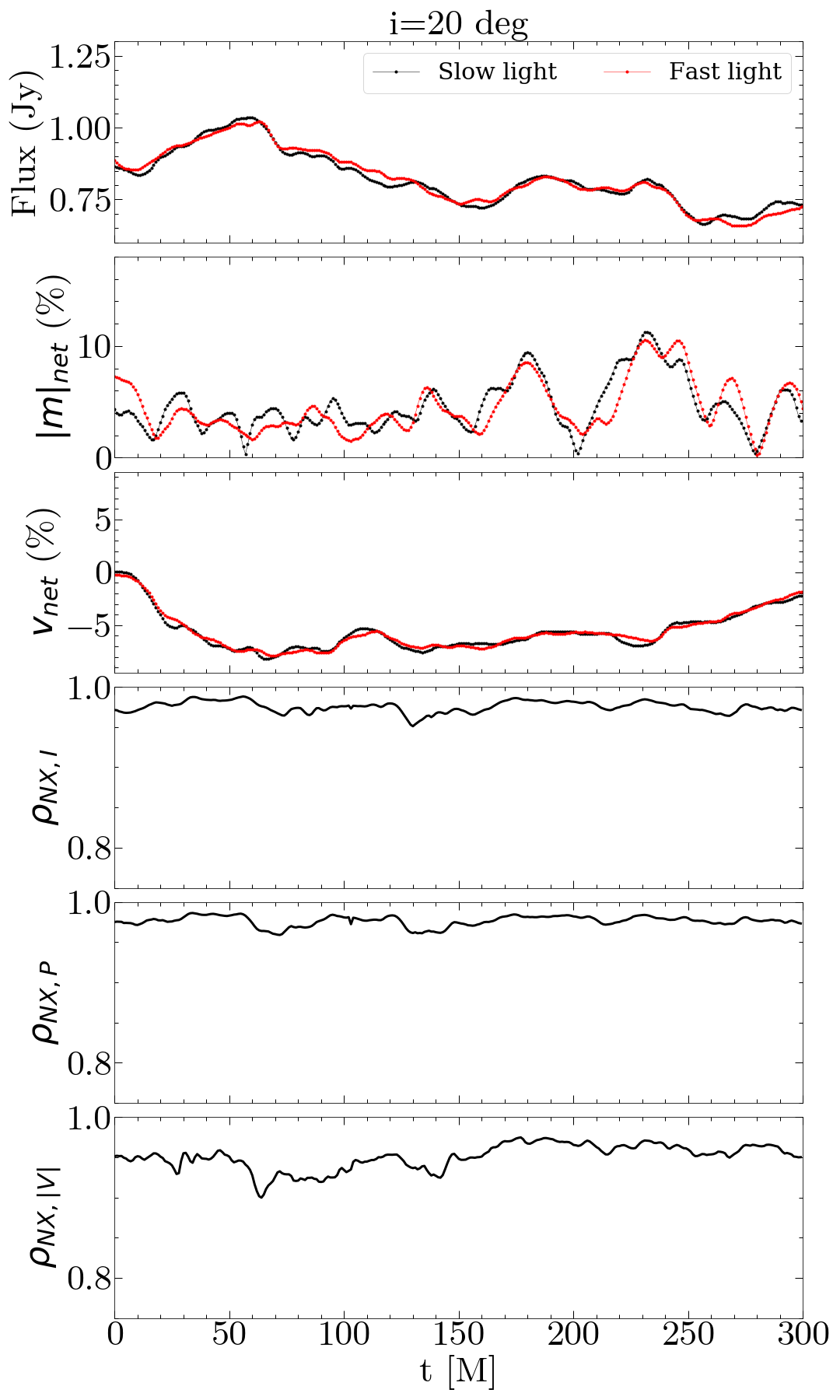}
\includegraphics[width=0.32\linewidth]{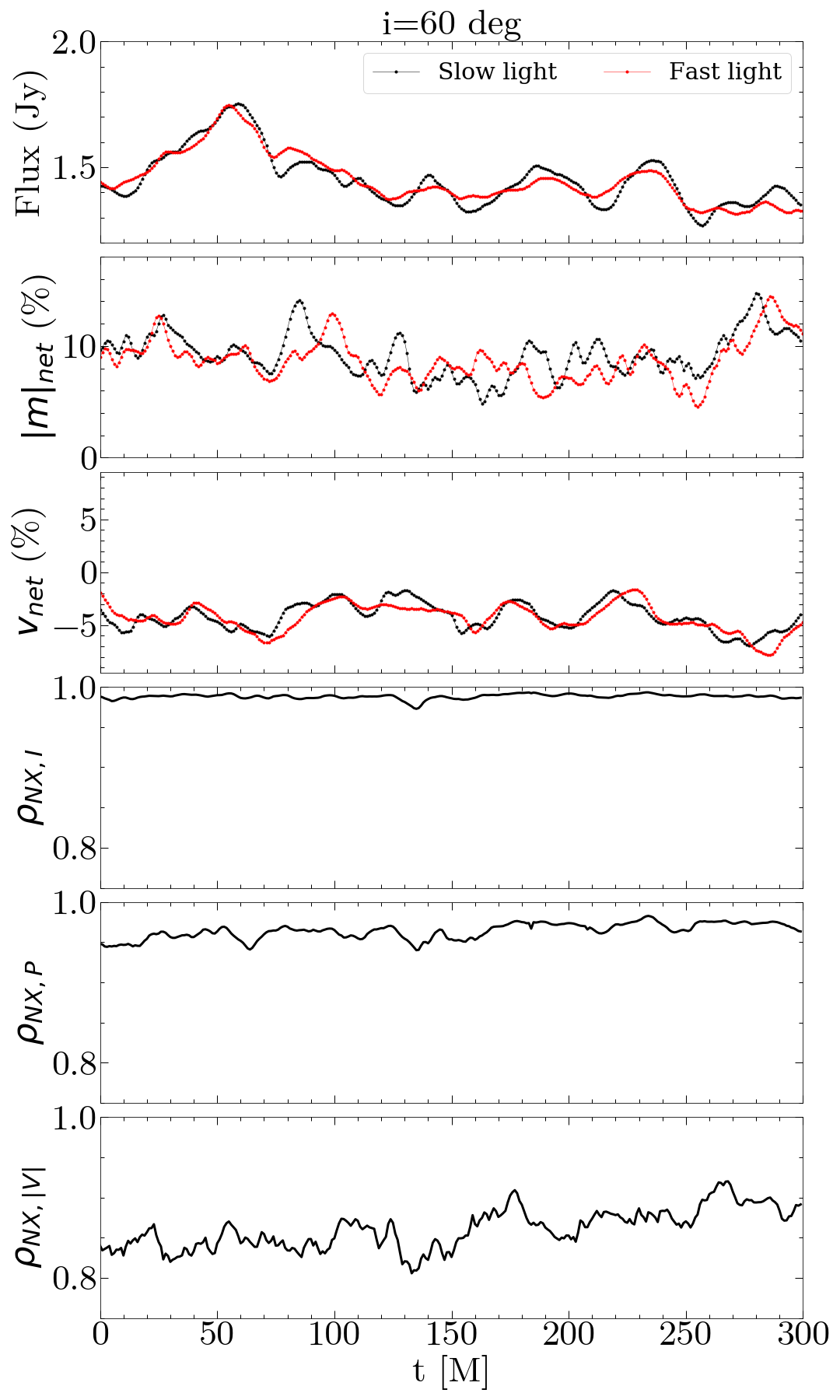}
\includegraphics[width=0.32\linewidth]{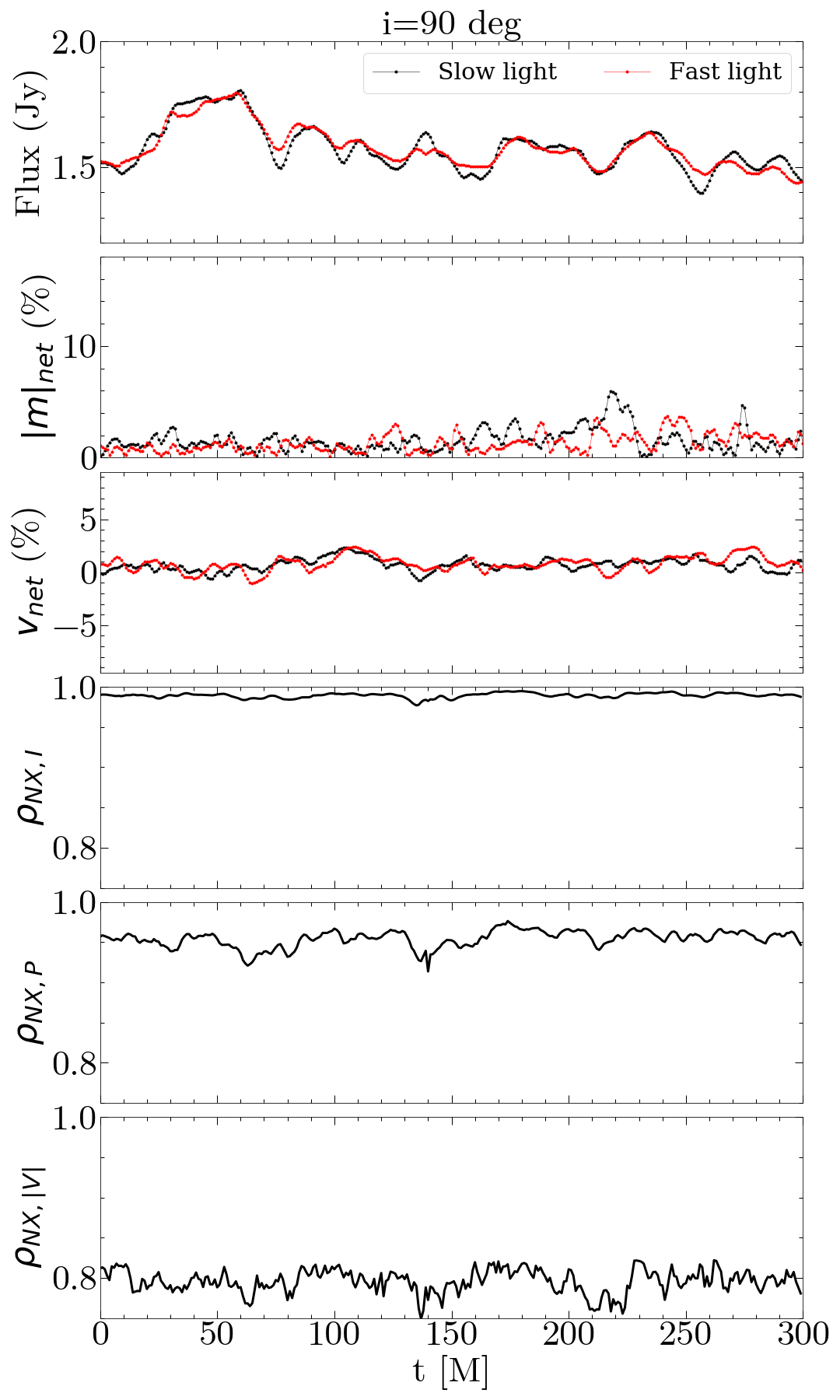}
\caption{Temporal behavior of observables calculated using slow-light and
  fast-light approximation for three viewing angles
  ($i=20, 60$ and $90 \deg$ from left to right). Top three panels display flux, fractional linear
  and circular polarization as a function of time. Bottom three panels display normalized
  cross-correlation between images computed using slow-light
  and fast-light approximation. Cross-correlation coefficient
  is computed separately for total intensity images,
  $\rho_{NX,I}$, linear polarimetric images (where
  $P=\sqrt{Q^2+U^2}$), $\rho_{NX,P}$ and circular polarimetric images,
  $\rho_{NX,|V|}$. The slow-light and fast-light light curves
  in the top three panels are aligned so that the cross-correlation
  coefficients in the bottom panels have maximum mean value within the shown
  time interval.}\label{fig:fast_slow_comparison}
\end{figure*}

\begin{figure*}
\def\arraystretch{0.0}
\centering
\setlength{\tabcolsep}{0pt}
\begin{tabular}{cc}
  \multicolumn{2}{c}{$i=20 \deg$}\\
\raisebox{0.09\linewidth}[0pt][0pt]{\rotatebox[origin=c]{90}{FAST
    LIGHT}}\phantom{.} &
\includegraphics[width=0.8\linewidth,trim=0.cm 12.cm 0.cm 12.cm,clip]{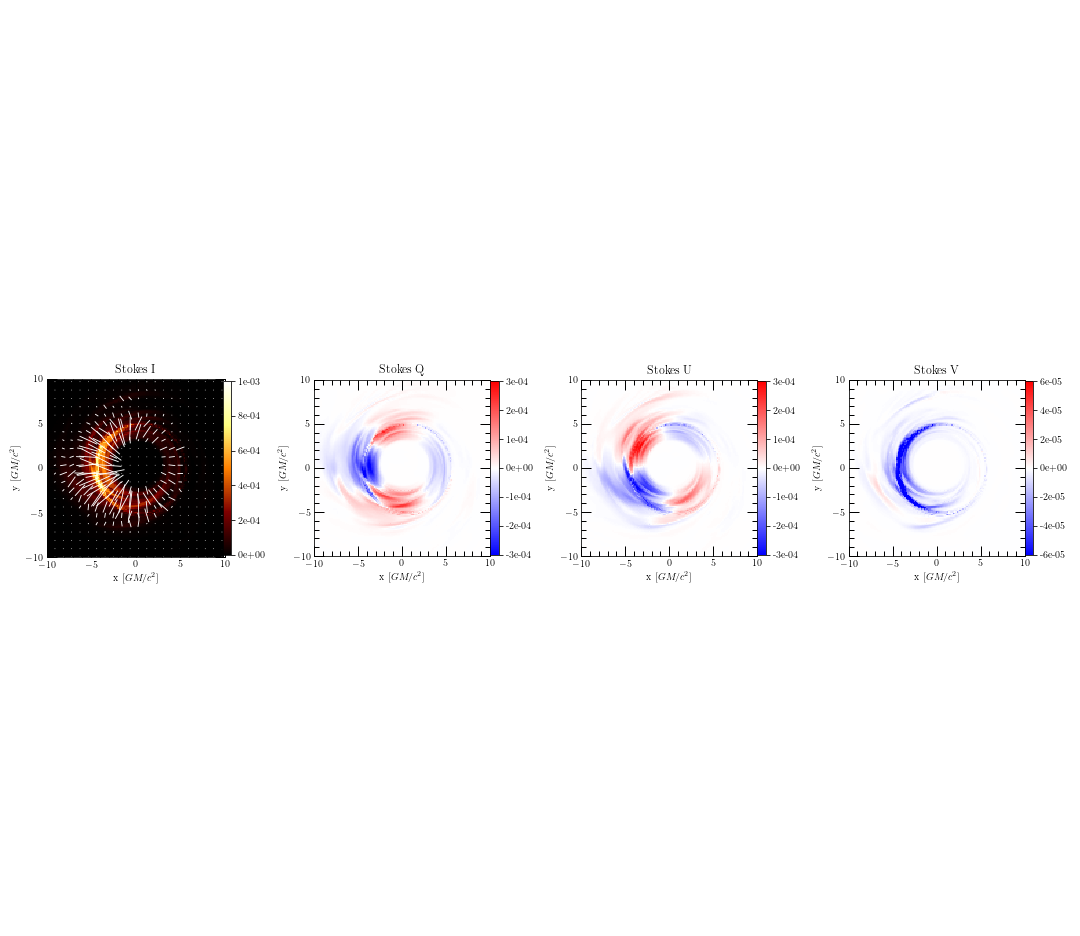}\\
\raisebox{0.09\linewidth}[0pt][0pt]{\rotatebox[origin=c]{90}{SLOW LIGHT}}\phantom{.} &
\includegraphics[width=0.8\linewidth,trim=0.cm 12.cm 0.cm 12.cm,clip]{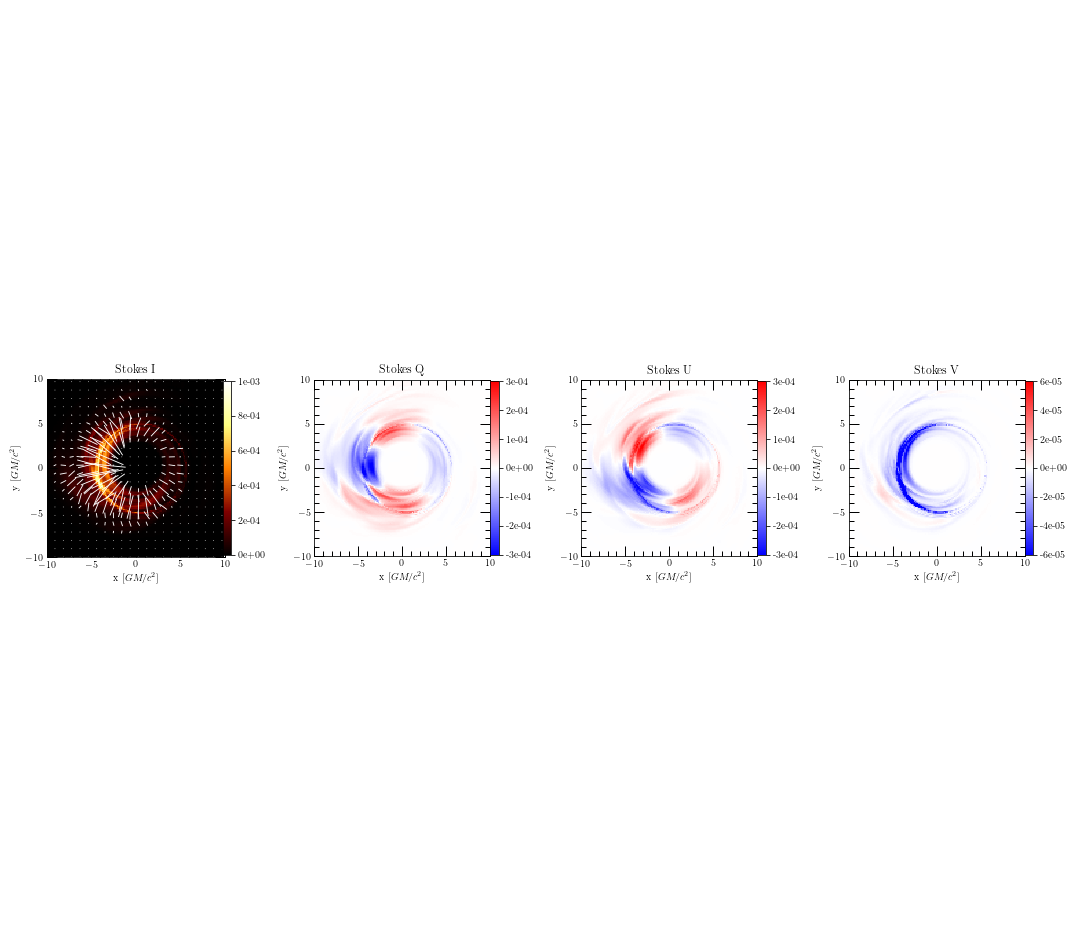}\\
\multicolumn{2}{c}{$i=60 \deg$}\\
\raisebox{0.09\linewidth}[0pt][0pt]{\rotatebox[origin=c]{90}{FAST LIGHT}}\phantom{.} &
\includegraphics[width=0.8\linewidth,trim=0.cm 12.cm 0.cm 12.cm,clip]{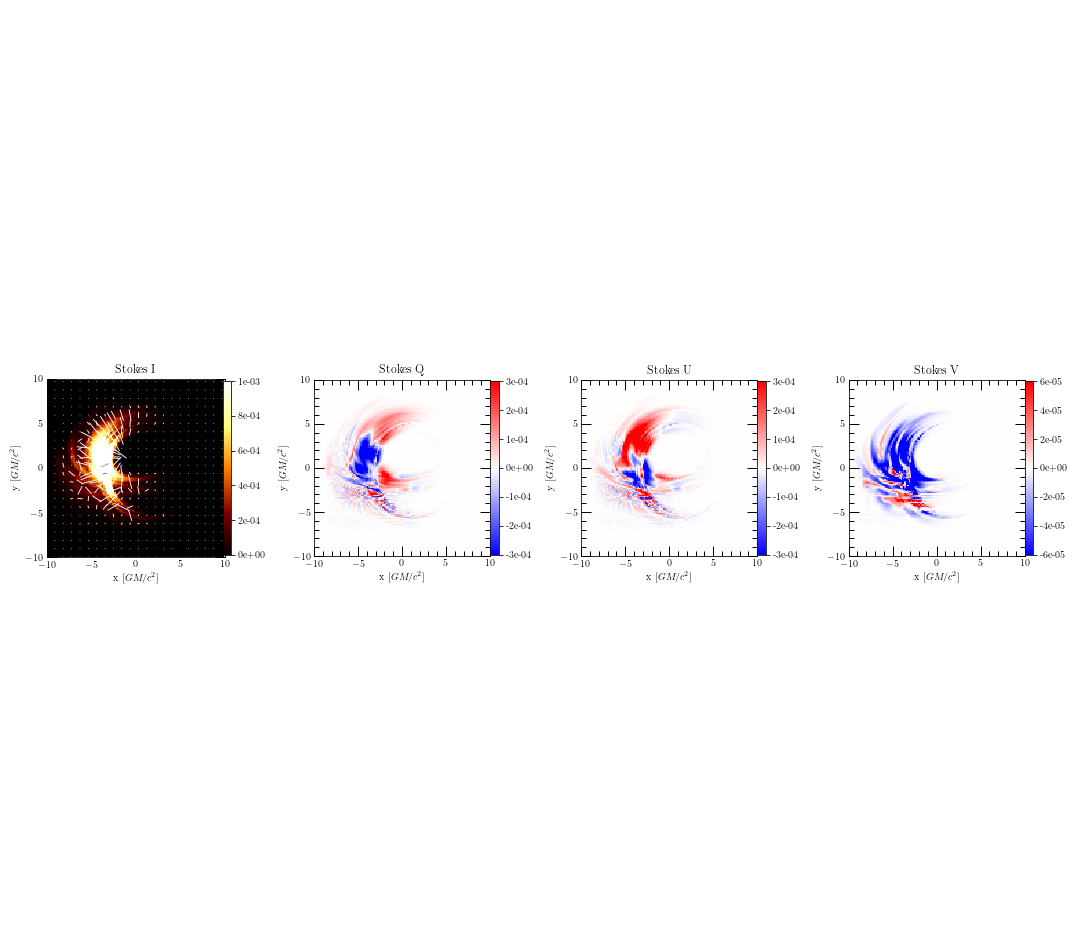}\\
\raisebox{0.09\linewidth}[0pt][0pt]{\rotatebox[origin=c]{90}{SLOW LIGHT}}\phantom{.} &
\includegraphics[width=0.8\linewidth,trim=0.cm 12.cm 0.cm 12.cm,clip]{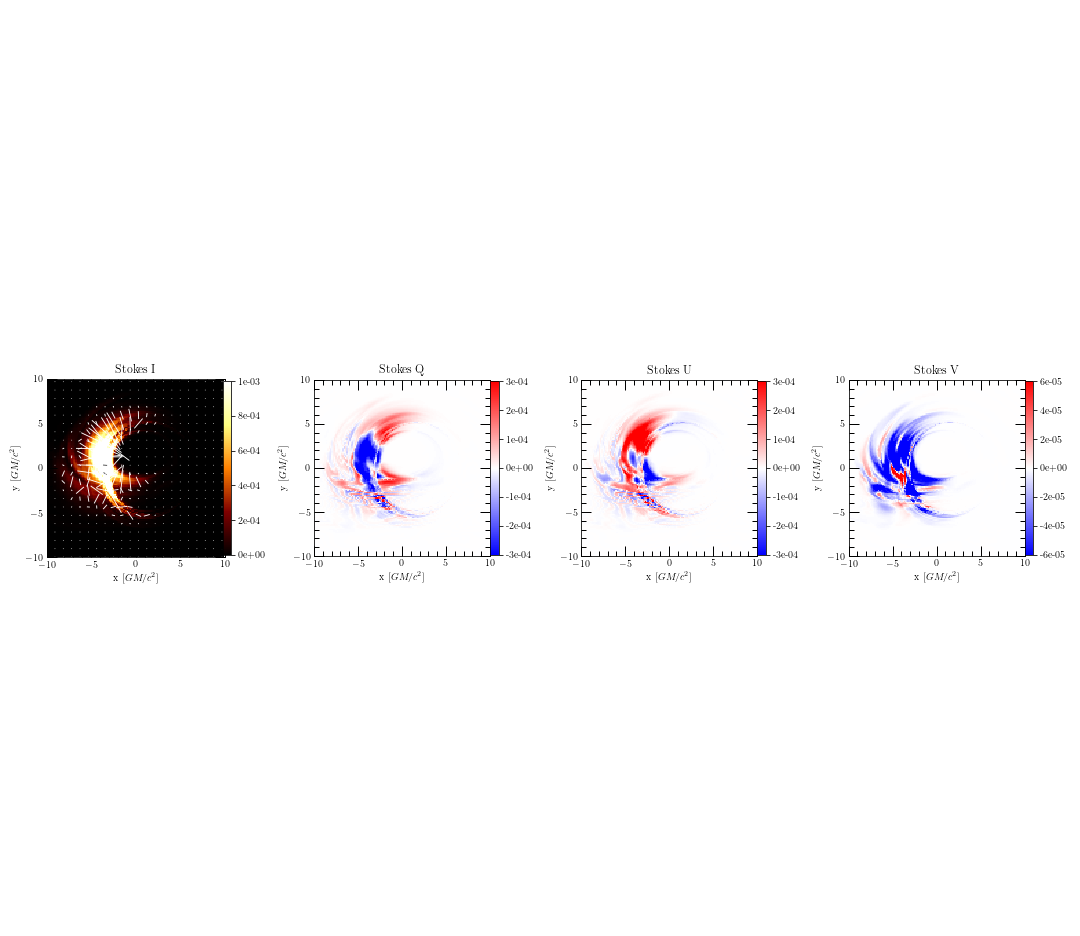}\\
\multicolumn{2}{c}{$i=90 \deg$}\\
\raisebox{0.09\linewidth}[0pt][0pt]{\rotatebox[origin=c]{90}{FAST LIGHT}}\phantom{.} &
\includegraphics[width=0.8\linewidth,trim=0.cm 12.cm 0.cm 12.cm,clip]{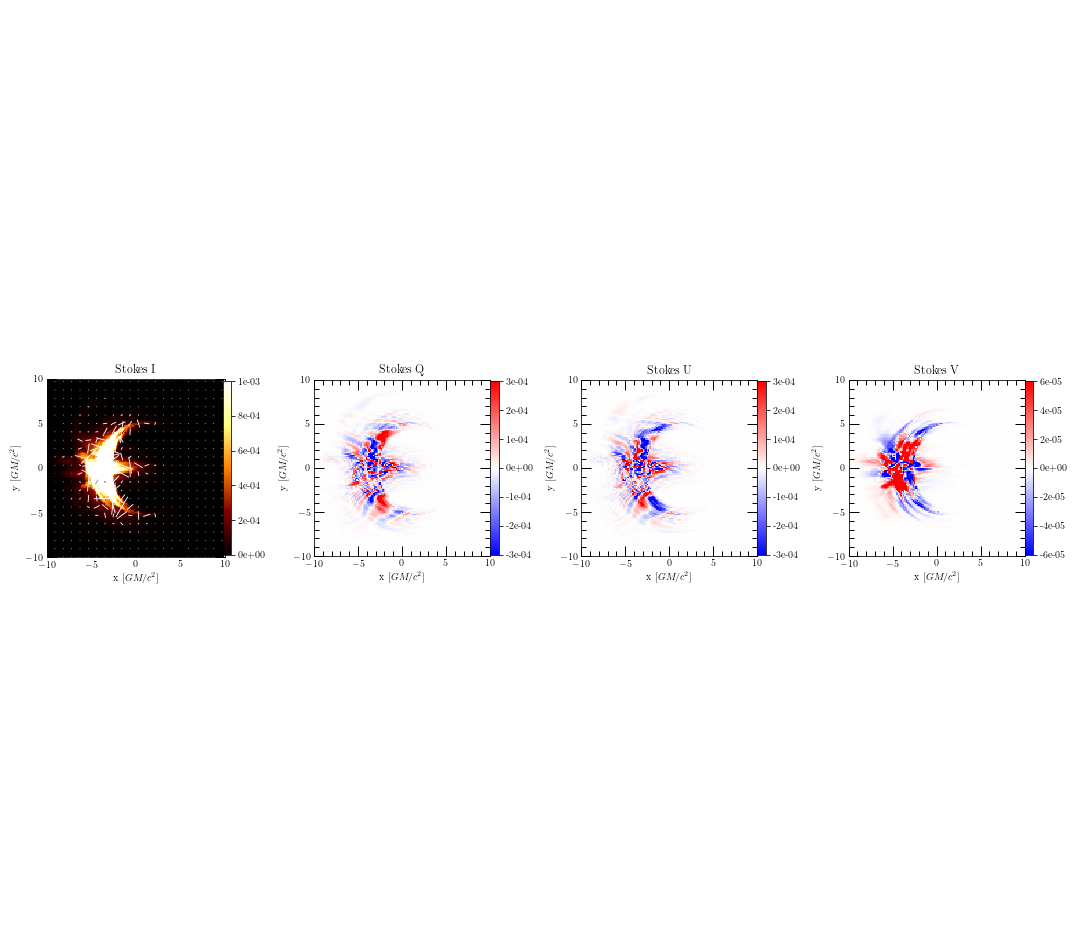}\\
\raisebox{0.09\linewidth}[0pt][0pt]{\rotatebox[origin=c]{90}{SLOW LIGHT}}\phantom{.} &
\includegraphics[width=0.8\linewidth,trim=0.cm 12.cm 0.cm 12.cm,clip]{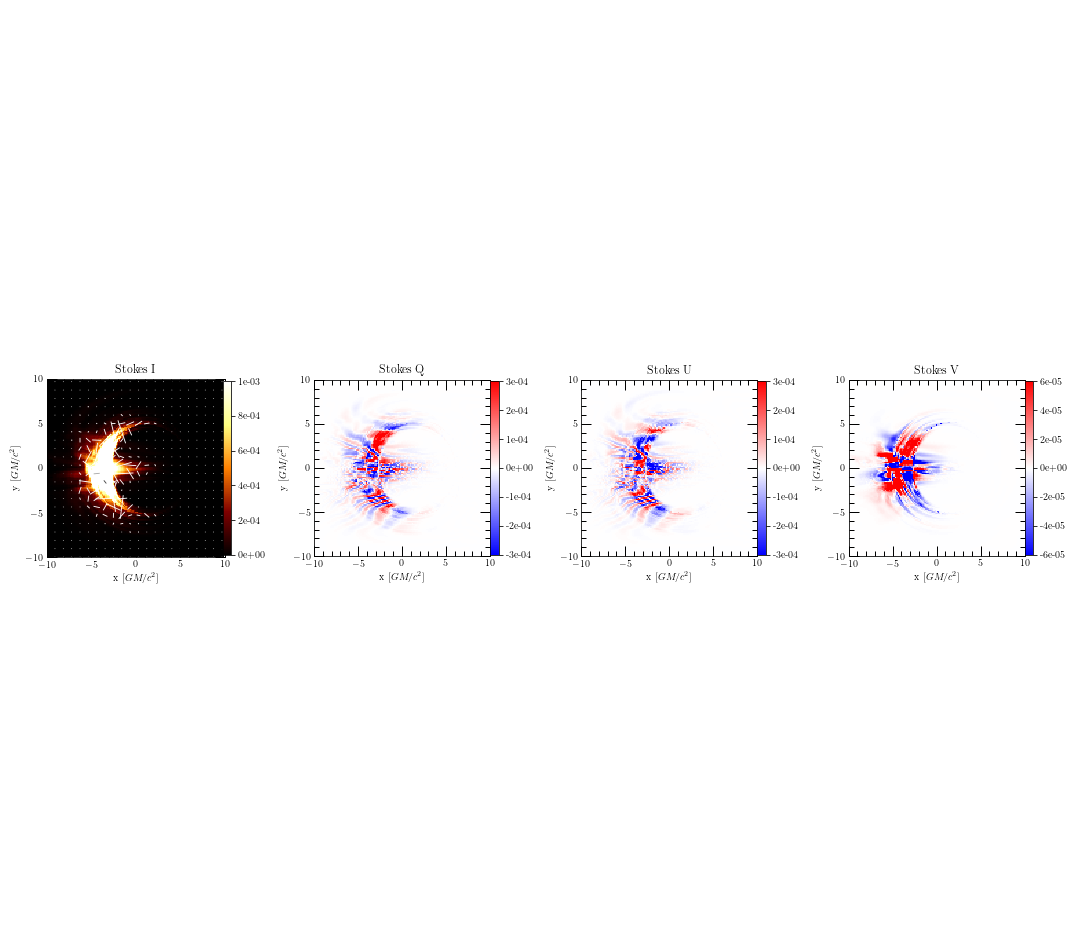}\\
\end{tabular}
  \centering
\caption{Examples of Stokes images constructed using slow-light
and fast-light radiative transfer models for three chosen viewing angles.
For these examples, the normalized cross-correlations between slow- and
fast-light polarimetric images, $(\rho_{NX,I},\rho_{NX,P},\rho_{NX,|V|})$ are (0.98,0.98,0.96) for $i=20 \deg$,
(0.98,0.97,0.9) for $i=60\deg$ and (0.99,0.96,0.8) for $i=90\deg$.}\label{fig:fast_slow_comparison_images}
\end{figure*}

In the radiative transfer simulations we neglect the light travel time delay
effects and post-process each simulation snapshot separately,
as if photons propagated at infinite speed (so called ``fast-light''
approach). Here we explicitly show how well this commonly-used
approximation holds for total intensity and light linear and circular polarizations.

We carry out several radiative transfer simulations and compare light curves and images generated using
the fast-light approach to those generated by propagating photons through
time-evolving simulation (refered as to ``slow-light''). To carry
out the tests we utilize another 3-D GRMHD simulation of a torus
where the cadence of the dump files is
$\Delta t=1$M (the simulation is SANE with black hole spin parameter $a_*=0.9375$).
High cadence is needed for accurate interpolations of fluid variables between
the time-snapshots \citep{dexter:2010}. All fast-light and slow-light transfer calculations are
carried out in modified Kerr-Schild coordinates.  

In Figure~\ref{fig:fast_slow_comparison}, in three top panels, we show
temporal behaviour of total flux and fractional linear and circular
polarizations as measured by a distant observer at three chosen viewing
angles. The time is shown in units of M, where $t=0$~M corresponds to about
$t=1500$~M in evolution of the torus, hence the turbulence in the torus is
well evolved for this test.  The two types of light curves shown correspond to results of
slow- and fast-light models. The difference in total flux is less than 10 per cent.
This result is consistent with results and
discussion by \citealt{dexter:2010} who compared slow- and fast-light approach in
unpolarized radiative transfer models. Notice that here we do not see the time
shift between two light curves that is visible in Figs.~9 and 10 of \citet{dexter:2010}
because the light curves produced in slow and fast-light
calculations are aligned in time by maximizing mean normalized
cross-correlation between the individual frames used to construct the light
curves. In Figure~\ref{fig:fast_slow_comparison}, in three bottom panels, we
show these normalized cross-correlations between slow- and fast-light images,
computed separately for Stokes ${\mathcal I}$, Stokes ${\mathcal
  P}\equiv \sqrt{{\mathcal Q}^2+\mathcal{U}^2}$, and absolute value of Stokes ${\mathcal V}$.
The cross-correlation coefficients, $\rho_{NX}$, are
calculated using {\tt nxcorr} function in ${\tt eht-imaging}$ library
\citep{chael:2016}. The images computed in slow light
model are highly correlated with those from fast-light approximation both in
total intensity and both polarizations.
In Figure~\ref{fig:fast_slow_comparison_images}, we show examples of fast and
slow light images in all Stokes parameters. Images with more scrambled polarization
(e.g., with viewing angle of $i=90 \deg$) the cross-correlation for ${\mathcal
  P}$ and $|{\mathcal V}|$ is slightly worse compared to Stokes ${\mathcal
  I}$, because polarization maps are more sensitive 
to small changes in the fluid conditions. What follows, one should be cautious
when interpreting any observed time
variations in Stokes ${\mathcal Q}$ and ${\mathcal U}$ with fast-light models that have
scrambled linear polarization (where scrambled polarization can be due
to either turbulence or higher Faraday depth).

The slow-light models are computationally far more expensive compared to
the fast-light models. Given small differences between two approaches
we choose the cheaper approach to predict radiative properties of the long duration simulations.

\section{B-field polarity in GRMHD simulations}\label{app:polarityB}

\begin{figure*}
\def\arraystretch{0.0}
\centering
\setlength{\tabcolsep}{0pt}
\begin{tabular}{cccccc}
  \multicolumn{3}{c}{B} & \multicolumn{3}{c}{-B} \\
  \\
${\bf \mathcal{I}}$ & ${\bf \mathcal{I+LP}}$ & ${\bf \mathcal{CP}}$ & ${\bf \mathcal{I}}$ & ${\bf \mathcal{I+LP}}$ & ${\bf \mathcal{CP}}$ \\
\raisebox{0.06\linewidth}[0pt][0pt]{\rotatebox[origin=c]{90}{Emission+abs}}\phantom{.} 
 \includegraphics[width=0.15\linewidth]{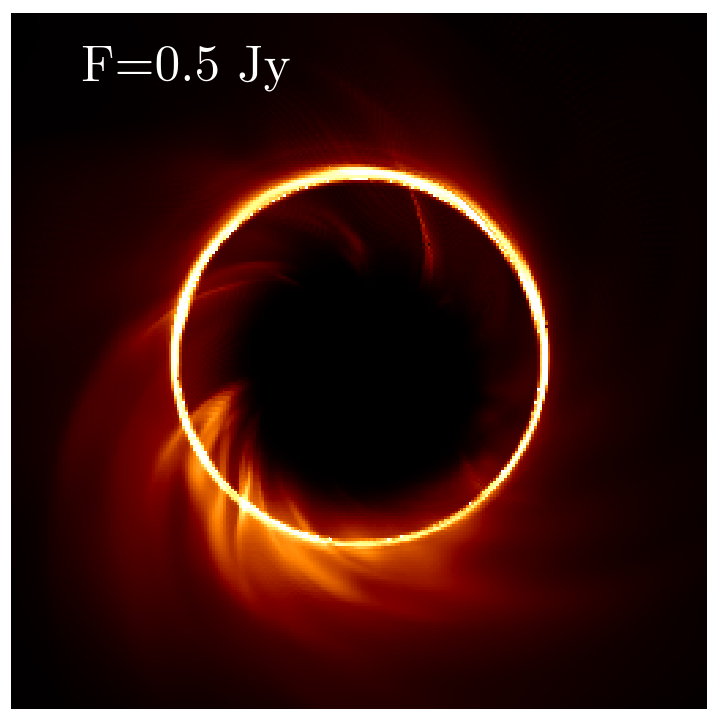}
& \includegraphics[width=0.15\linewidth]{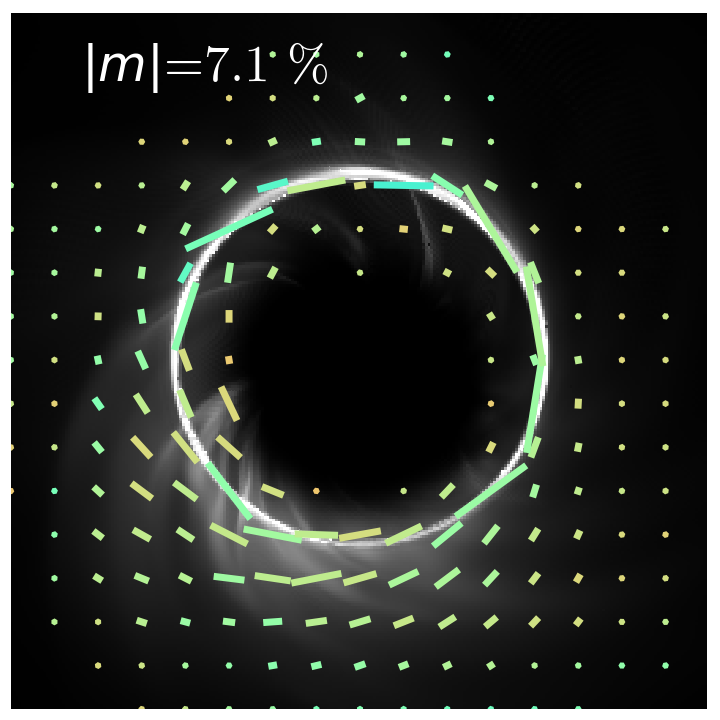}
& \includegraphics[width=0.15\linewidth]{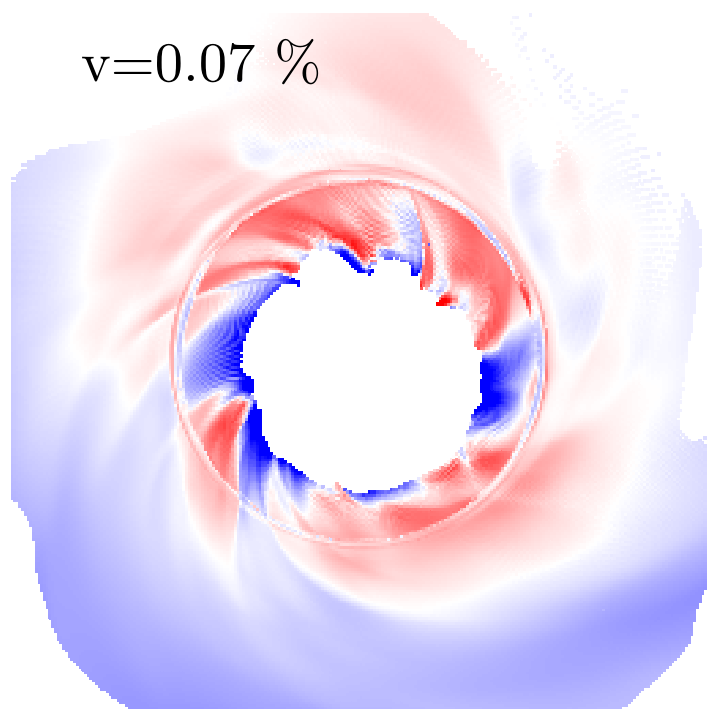}
& \includegraphics[width=0.15\linewidth]{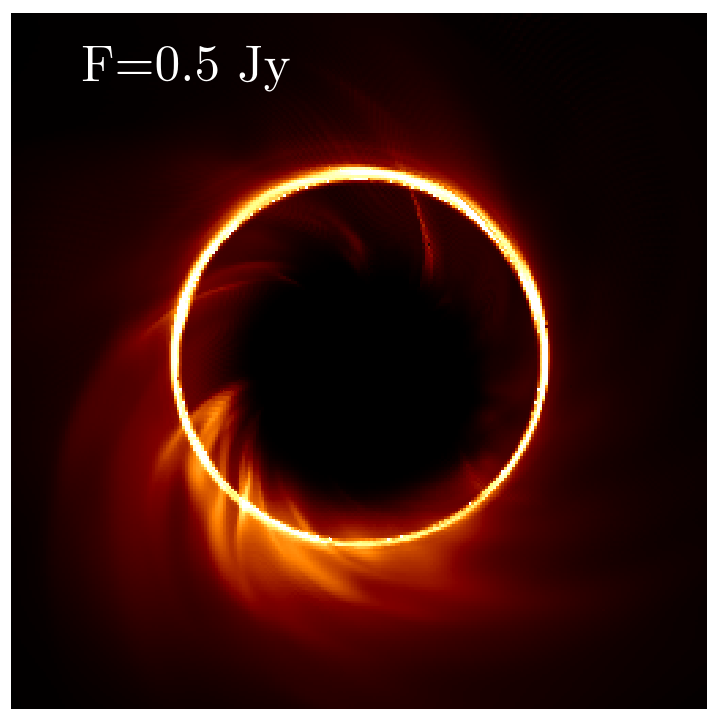}
& \includegraphics[width=0.15\linewidth]{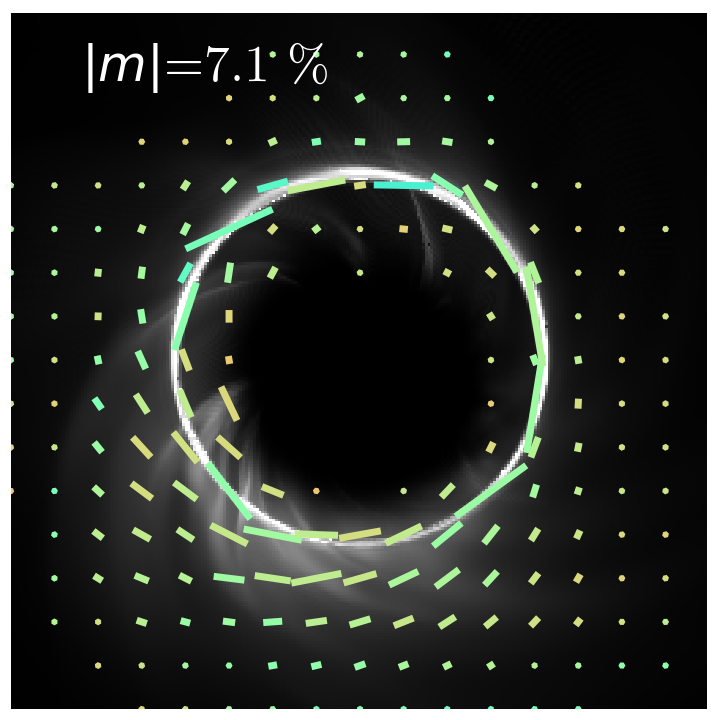}
& \includegraphics[width=0.15\linewidth]{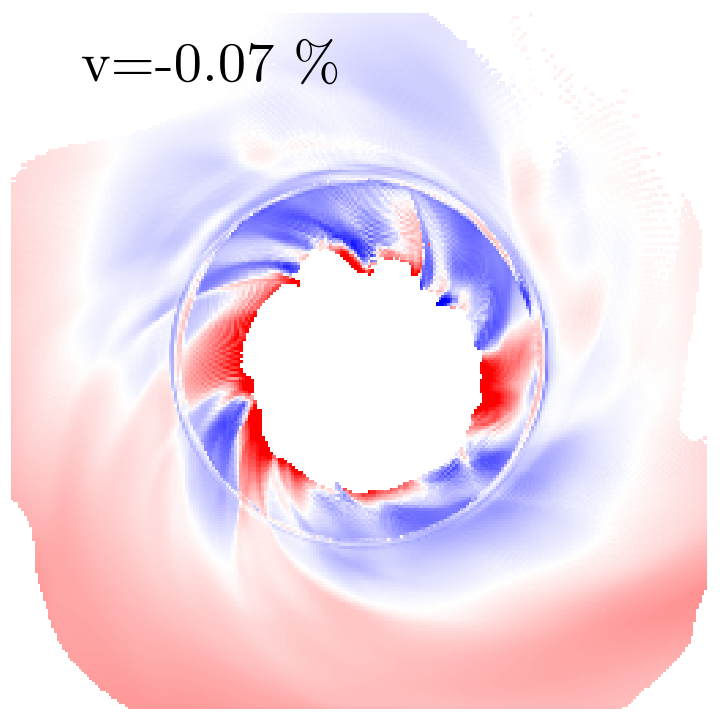}\\
\raisebox{0.06\linewidth}[0pt][0pt]{\rotatebox[origin=c]{90}{Full RT}}\phantom{.}
 \includegraphics[width=0.15\linewidth]{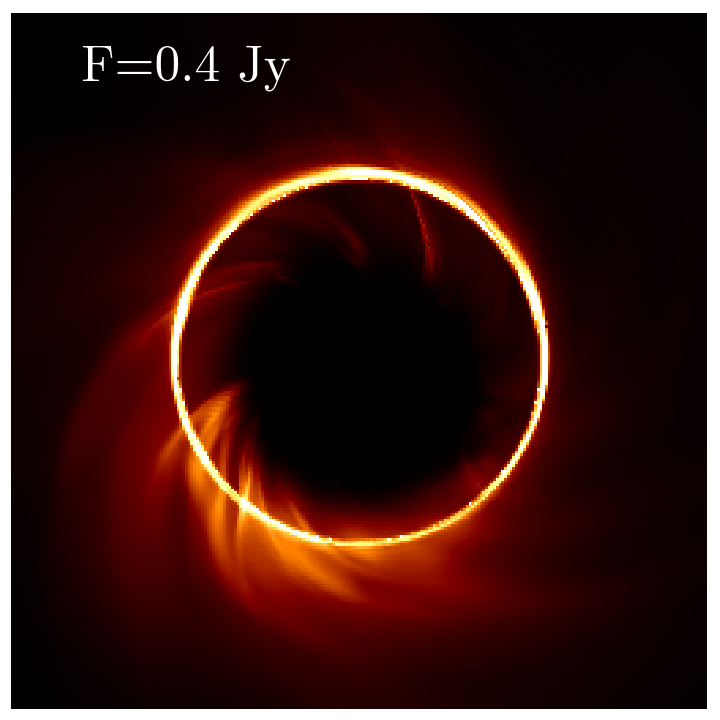}
& \includegraphics[width=0.15\linewidth]{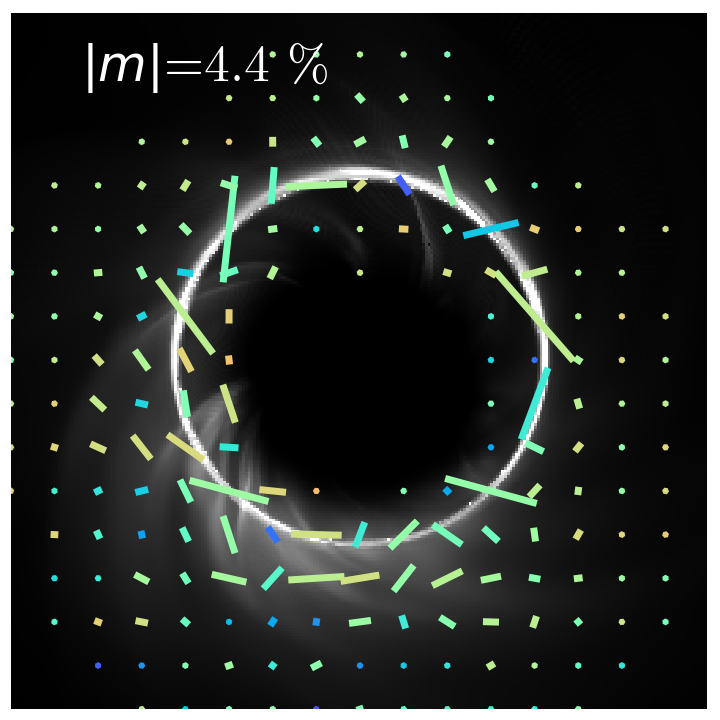}
& \includegraphics[width=0.15\linewidth]{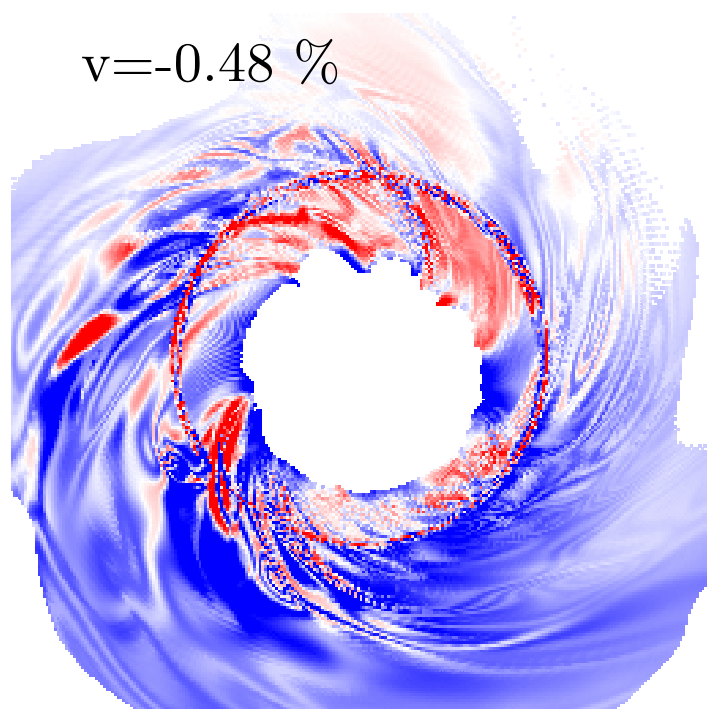}
& \includegraphics[width=0.15\linewidth]{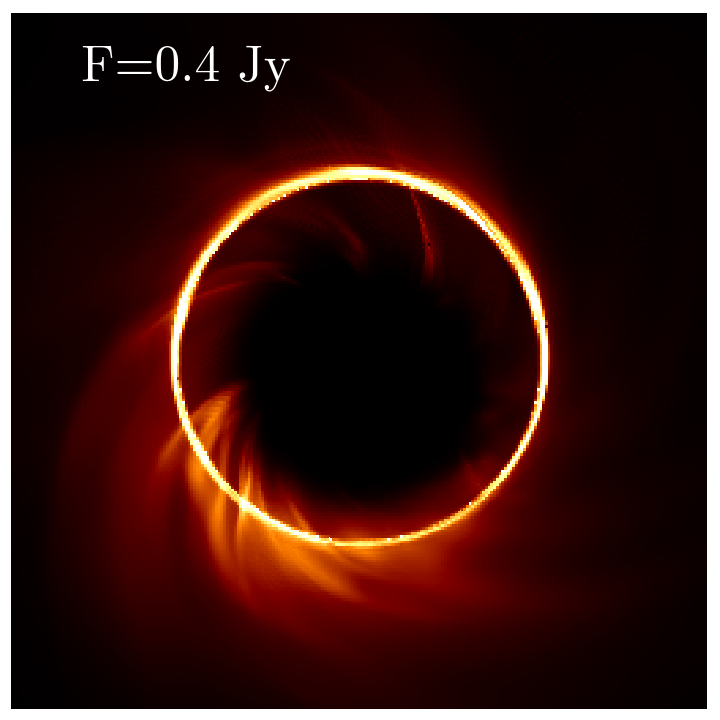}
& \includegraphics[width=0.15\linewidth]{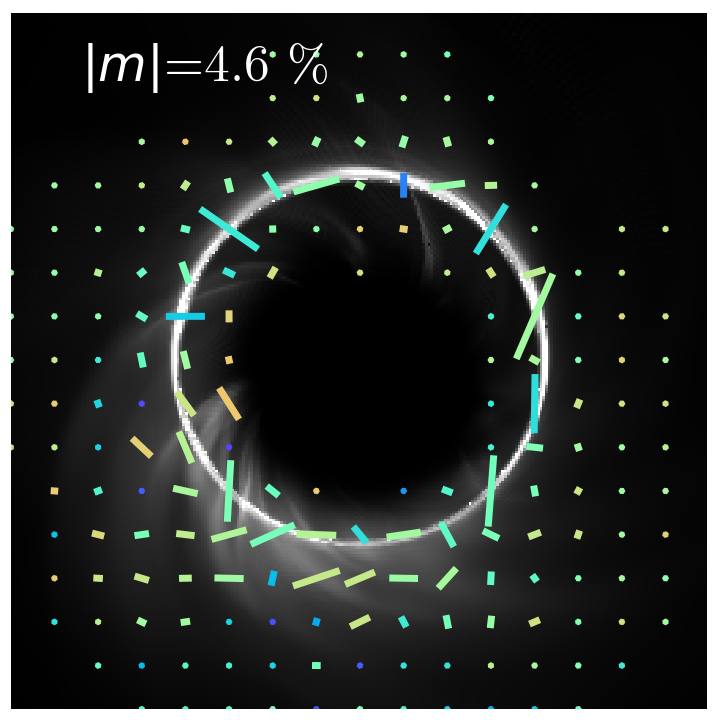}
& \includegraphics[width=0.15\linewidth]{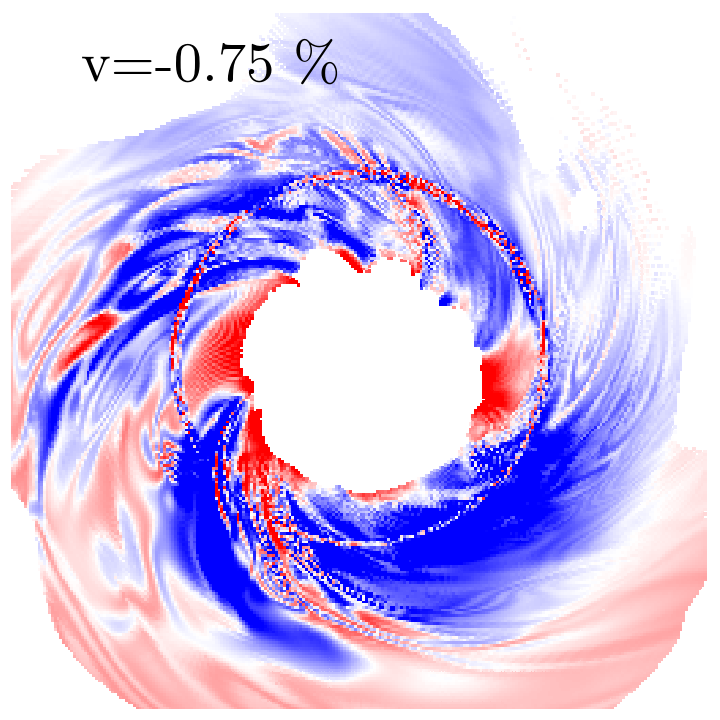}\\
\includegraphics[width=0.14\linewidth,trim=0 18cm 0 0,clip]{plots/Model1a_beta30_a0/cbar_I_hor_2.png} &
\includegraphics[width=0.14\linewidth,trim=0 18cm 0 0,clip]{plots/Model1a_beta30_a0/cbar_LP_hor.png} &
\includegraphics[width=0.14\linewidth,trim=0 18cm 0 0,clip]{plots/Model1a_beta30_a0/cbar_CP.png} &
\includegraphics[width=0.14\linewidth,trim=0 18cm 0 0,clip]{plots/Model1a_beta30_a0/cbar_I_hor_2.png} &
\includegraphics[width=0.14\linewidth,trim=0 18cm 0 0,clip]{plots/Model1a_beta30_a0/cbar_LP_hor.png} &
\includegraphics[width=0.14\linewidth,trim=0 18cm 0 0,clip]{plots/Model1a_beta30_a0/cbar_CP.png} \\
\end{tabular}
\caption{Impact of the global magnetic field polarity on the linear and circular
  polarization maps of model A200. Top panels show the maps computed without
  Faraday effects accounting only for emission and self-absorption effects. As
  expected linear polarization remains unaffected with circular polarization
  maps swap sign when magnetic field sign is reversed.
  Bottom panels show the same model when solving full RT equations including
  Faraday effects, as expected a significant change is visible in both linear and
  circular polarization maps.}\label{fig:invertB}
\end{figure*}

We start GRMHD simulations with a single loop of magnetic field inside of a
torus. The polarity of the loop is set arbitrarily because the GRMHD model
evolution is insensitive to the polarity of the magnetic fields. However, the
polarized radiative transfer equations in the plasma frame are not symmetric
when changing the B-field sign which may result in different polarimetric images.

The asymmetry of polarized radiative transfer with respect to B-field polarity
becomes evident when examining the properties of Equation~\ref{eq:delin2}.
Let's consider two cases. In the first case let's assume that there are no Faraday effects
($\vv{\rho} \times \vv{p}=0$). Then Stokes ${\mathcal Q}$ (linear polarization)
will not depend on B-field polarity but the sign of Stokes ${\mathcal V}$ will
reflect B-field polarity. This is because the sign of the normalized synchrotron
emissivity $\epsilon_Q$, absorptivity $\alpha_Q$, as well as 
$\vv{\alpha} \cdot \vv{p}$ do not depend on the
sign of B-field while the sign of $\epsilon_V$ and $\alpha_V$ do. Notice that in this
first case the magnitude of circular polarization remains unchanged when
changing the sign of the B-field.
In the second case, we consider full RT model with all Faraday effects included. In
this case the polarity of the magnetic field does make a difference to both
linear and circular polarization and, to a very small degree, can affect
Stokes ${\mathcal I}$. To understand this effect one has to understand the behaviour of
$(\vv{\rho} \times \vv{p})$ when the sign of B-field changes. It is useful to write this term
separately for Stokes ${\mathcal Q,U,V}$:
\begin{equation}
(\vv{\rho} \times \vv{p})_Q=-\rho_V p_U, 
\end{equation}
\begin{equation}
(\vv{\rho} \times \vv{p})_U=\rho_V p_Q - \rho_Q p_V,
\end{equation}
\begin{equation}
(\vv{\rho} \times \vv{p})_V=\rho_Q p_U,
\end{equation}
where sign of Faraday conversion coefficient, $\rho_Q$, does not depend on
B-field sign while sign of Faraday rotativity, $\rho_V$, does.
It is evident that the change in B-field sign will cause
change in the Stokes ${\mathcal Q}$ evolution via Faraday rotation (only one term
in Equation~\ref{eq:delin2} will change sign), which will
be followed by change in evolution of Stokes 
${\mathcal U}$ and then ${\mathcal V}$, via Faraday conversion.
In this case we also expect to see small
difference in Stokes ${\mathcal I}$ because
the magnitude of term $\vv{\alpha}\cdot \vv{p}$ in Equation~\ref{eq:delin1}
can be altered.

In Figure~\ref{fig:invertB} we show that our numerical scheme is in agreement
with the above considerations. We show a snapshot of model A200 (the same one as
shown in Figure~\ref{fig:invert}) computed with and without the inversion of
B-field sign. We present two aforementioned cases in which emission and
absorption only and emission, absorption and Faraday effects are included
in the computation. As expected in the first case, the inversion of B-fields makes
no difference for Stokes ${\mathcal I,Q,U}$ and the magnitude of Stokes
${\mathcal V}$ but the sign of Stokes ${\mathcal  V}$
is reversed. The first case additionally demonstrates that our numerical scheme is accurate. 
In the second case, in model with Faraday effects included (lower panels in
Figure~\ref{fig:invertB}), the linear and circular polarizations changed
between models of different B-field polarity. Interestingly in maps with circular
polarization only regions that are Faraday-thin (within the photon ring in the
central regions of the frame and in
the outer regions of the frame) swap the polarity, while
Faraday-thick regions in the most bright part of the frame - retain the Stokes
${\mathcal V}$ signature. These results support the idea that the handedness of
Stokes ${\mathcal V}$ is more sensitive to the direction of the rotation of the
flow (or in other words: the handedness of the magnetic field helix), rather then the polarity of the magnetic fields \citep{enslin:2003}.

\section{Observing Frequency Effects}\label{app:freq}

In our analysis we study polarimetric maps at $f=$230~GHz corresponding to the EHT observing frequency. Future observations by this instrument will image M87 black hole near horizon emission at $f=$345~GHz. Planned space VLBI projects will aim to observe black hole horizons at even higher frequencies $f$=690~GHz. At these higher frequencies the images of black holes are expected to have higher resolution enabling to resolve better the finer structures around the black hole shadow such as e.g. the lensing ring. Here we re-calculate model A200 images at two higher frequencies to investigate which frequency is optimal for circular polarization observations or how observations at different frequencies can complement each other. 

\begin{figure}
\def\arraystretch{0.0}
\centering
\setlength{\tabcolsep}{0pt}
\begin{tabular}{ccc}
${\bf \mathcal{I}}$ & ${\bf \mathcal{I+LP}}$ & ${\bf \mathcal{CP}}$ \\
\raisebox{0.12\linewidth}[0pt][0pt]{\rotatebox[origin=c]{90}{A200 230\,GHz}}\phantom{.}
 \includegraphics[width=0.3\linewidth]{plots/Model1a_beta30_a0/m87_cp_new_256/ipole_ang160_freq230_Munit1_3e+29_tratj1_tratd200_datI.png} & \includegraphics[width=0.3\linewidth]{plots/Model1a_beta30_a0/m87_cp_new_256/ipole_ang160_freq230_Munit1_3e+29_tratj1_tratd200_datIP.png}
& \includegraphics[width=0.3\linewidth]{plots/Model1a_beta30_a0/m87_cp_new_256/ipole_ang160_freq230_Munit1_3e+29_tratj1_tratd200_datV.png}\\
\raisebox{0.12\linewidth}[0pt][0pt]{\rotatebox[origin=c]{90}{A200 345\,GHz }}\phantom{.} 
 \includegraphics[width=0.3\linewidth]{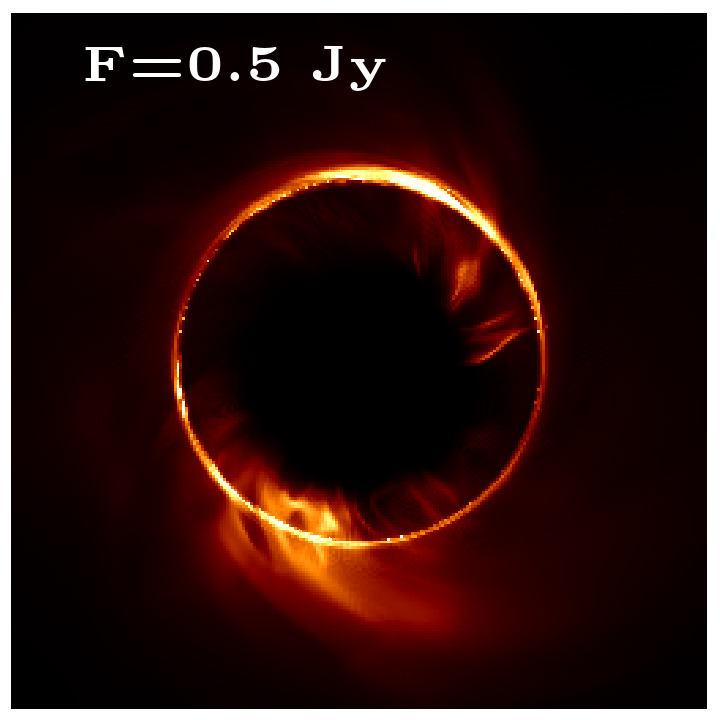}
& \includegraphics[width=0.3\linewidth]{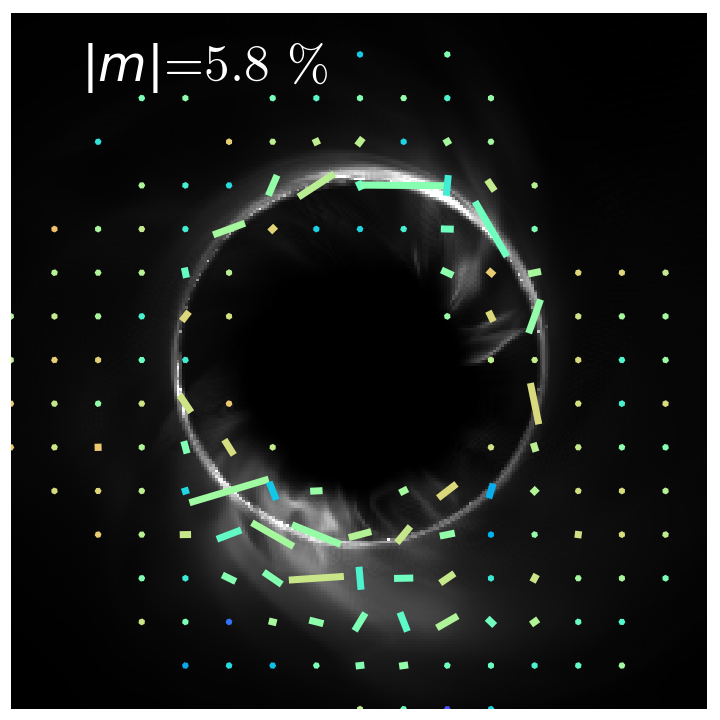}
& \includegraphics[width=0.3\linewidth]{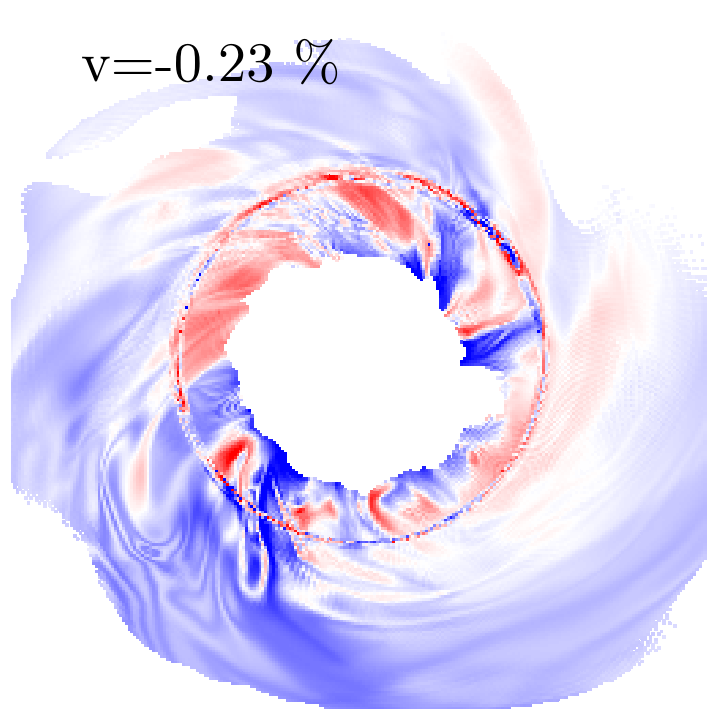}\\
\raisebox{0.12\linewidth}[0pt][0pt]{\rotatebox[origin=c]{90}{A200 690\,GHz }}\phantom{.}
 \includegraphics[width=0.3\linewidth]{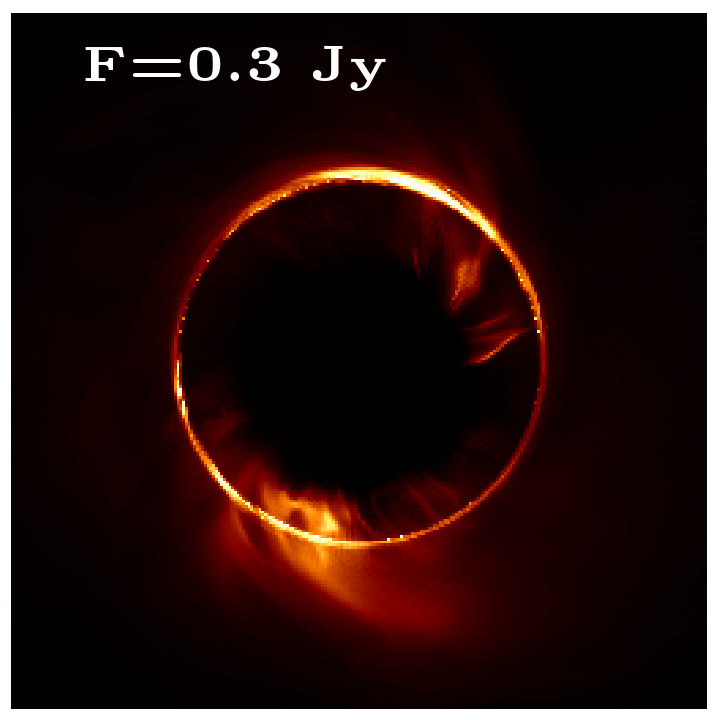}
& \includegraphics[width=0.3\linewidth]{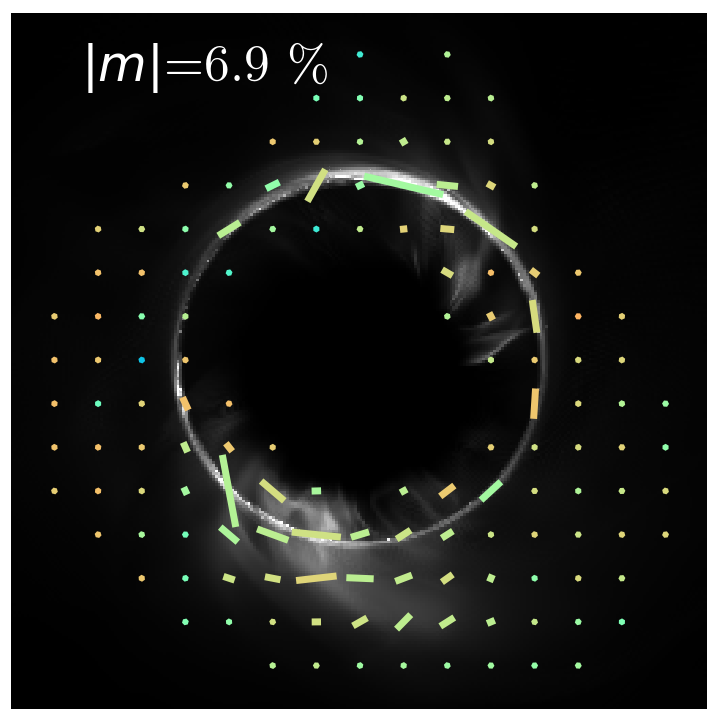}
& \includegraphics[width=0.3\linewidth]{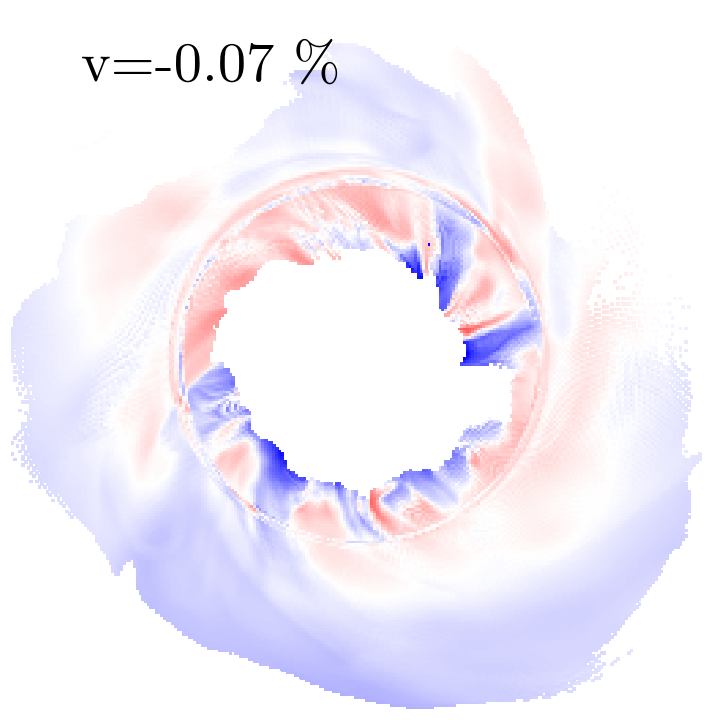}\\
\includegraphics[width=0.29\linewidth,trim=0 18cm 0 0,clip]{plots/Model1a_beta30_a0/cbar_I_hor_2.png} &
\includegraphics[width=0.29\linewidth,trim=0 18cm 0 0,clip]{plots/Model1a_beta30_a0/cbar_LP_hor.png} &
\includegraphics[width=0.29\linewidth,trim=0 18cm 0 0, clip]{plots/Model1a_beta30_a0/cbar_CP.png} \\
\end{tabular}
\caption{Frequency dependency of polarimetric images of model A200 observed at
  $i=160 \deg$. The time moment is the same as shown in Figure~\ref{fig:Rh} and images are displayed in the same manner as in Figure~\ref{fig:Rh}.}\label{fig:freq}
\end{figure}

In Figure~\ref{fig:freq} we show example snapshot of model A200 at $f=$230, 345, and 690 GHz. The Faraday conversion is decreasing with observing frequency as $f^{-3}$ and so the fractional circular polarization decreases significantly for higher energies. The inversion of polarization in the lensed emission is hardly visible already at 345~GHz. However, also the Faraday rotations decrease with the frequency and one can observe the intrinsic polarization patterns which are not scrambled and reflect better the geometry of magnetic fields in the emission region.

\bsp
\label{lastpage}
\end{document}